\documentclass[aps,prd,amsmath,amssymb,nofootinbib]{revtex4-2}
\usepackage{graphicx}
\usepackage{amsmath}
\usepackage{amsfonts,amsbsy}
\usepackage{amssymb}
\usepackage{xcolor}
\usepackage{hyperref}

\usepackage{slashed}

\def\d{\partial}
\def\p{{\mathbf p}}
\def\q{{\mathbf q}}
\def\k{{\mathbf k}}
\def\x{{\mathbf x}}
\def\y{{\mathbf y}}

\def\z{{\mathbf z}}
\def\v{{\mathbf v}}

\def\w{{\mathbf w}}
\def\ux{{\underline{x}}}
\def\uy{{\underline{y}}}
\def\uz{{\underline{z}}}

\def\d{\partial}
\def\p{{\mathbf p}}
\def\q{{\mathbf q}}
\def\k{{\mathbf k}}
\def\x{{\mathbf x}}
\def\y{{\mathbf y}}

\def\z{{\mathbf z}}
\def\v{{\mathbf v}}

\def\ux{{\underline{x}}}
\def\uy{{\underline{y}}}
\def\uz{{\underline{z}}}
\def\uw{{\underline{w}}}

\newcommand{\be}{\begin{equation}}
\newcommand{\ee}{\end{equation}}
\newcommand{\ba}{\begin{eqnarray}}
\newcommand{\ea}{\end{eqnarray}}
\newcommand{\ban}{\begin{eqnarray*}}
\newcommand{\ean}{\end{eqnarray*}}

\begin{document}

\title{\bf Quark and scalar propagators at next-to-eikonal accuracy in the CGC through a dynamical background gluon field}

\author{Tolga Altinoluk$^{a}$ and Guillaume Beuf$^{a}$}

\affiliation{$^a$Theoretical Physics Division, National Centre for Nuclear Research,
Pasteura 7, Warsaw 02-093, Poland}


\begin{abstract}
We revisit the calculation of quark and scalar propagators in a gluon background field at Next-to-Eikonal accuracy, which will provide
the first power correction to scattering processes in the high-energy limit, including gluon saturation effects.
The two main results of our study are as follows.
First, we derive
a very general Next-to-Eikonal expansion of the quark propagator in terms of the scalar propagator, which allows us to isolate spinor-specific Next-to-Eikonal corrections from the universal ones.
Second, we present
a new method allowing, for the first time, a systematic calculation
of the Next-to-Eikonal corrections coming from the $x^-$ dependence of the background field. This opens the possibility to study high-energy scattering processes on a target with non-trivial dynamics instead of on a static one, by going beyond infinite Lorentz time dilation of the target.
\end{abstract}

\maketitle

\section{Introduction}
\label{sec:intro}

The effective theory that describes the high-energy hadronic collisions in the dilute-dense systems, such as proton-nucleus (pA) collisions, is known as the Color Glass Condensate (CGC) (see \cite{Gelis:2010nm,Albacete:2014fwa,Blaizot:2016qgz} for recent reviews and references therein).  This effective theory relies on the phenomenon of gluon saturation that is encountered in the Regge-Gribov limit where the high scattering energy is achieved due to the decrease of the longitudinal momentum fraction (Bjorken $x$) carried by the interacting partons. In the Regge-Gribov limit, with increasing energy, gluon density rapidly increases in the scattering hadrons. At sufficiently high energies, or equivalently sufficiently small values of $x$, this rapid increase in the gluon density is tamed by the non-linear interactions of the emitted gluons which causes the aforementioned phenomenon gluon saturation and it is characterized  by a new perturbative scale $Q_s$, known as the saturation scale.

These saturation ideas were initially studied through the so-called dipole model in \cite{Nikolaev:1990ja,Nikolaev:1991et,Mueller:1993rr,Mueller:1994jq,Mueller:1994gb}. This model considers color neutral dipoles as most convenient degrees of freedom at high energy. When boosted a color dipole emits gluons. Since in the limit of large number of colors $N_c$, a gluon can be treated as a quark-antiquark pair, and the process can be considered a dipole emitting another dipole at each step of the evolution, forming a dipole cascade which finally interacts with the target. Later, it was noted by McLerran and Venugopalan that a convenient  approach to study gluon saturation is provided by the nonlinearities of the classical Yang-Mills theory \cite{McLerran:1993ni,McLerran:1993ka,McLerran:1994vd}. With these developments, the nonlinear rapidity evolution equation, Balitsky-Kovchegov/ Jalilian-Marian-Iancu-McLerran-Weigert-Leonidov-Kovner (BK-JIMWLK) functional evolution equation was derived \cite{Balitsky:1995ub,Kovchegov:1999yj,Kovchegov:1999ua, Jalilian-Marian:1996mkd,Jalilian-Marian:1997qno,Jalilian-Marian:1997jhx,Jalilian-Marian:1997ubg,Kovner:2000pt,Weigert:2000gi,Iancu:2000hn,Iancu:2001ad,Ferreiro:2001qy}. In recent years, these developments have become the basis for the phenomenological studies of saturation physics applied to high-energy collision data.

The key approximation routinely adopted within the CGC framework is the eikonal approximation. Effectively, it amounts to describing the high energy scattering processes by the leading contribution in energy and systematically neglecting the power suppressed corrections. High scattering energies in dilute-dense systems is achieved by boosting the dense target from its rest frame to the scattering energies. In such a case, eikonal approximation amounts to the following three assumptions. First, the highly boosted background field that describes the dense target experiences Lorentz contraction and therefore it is assumed to localize in the longitudinal direction (around $x^+=0$). Second, the dense target is described by the leading component of the background field which in our setup corresponds to ''-" component, whereas the other components (transverse and "+" components) are neglected due to the fact that they are suppressed by the Lorentz boost factor compared to the leading one. Third, the background field of the target is assumed to be independent of the light-cone coordinate $x^-$ due to Lorentz time dilation which effectively amounts to neglecting the internal dynamics of the target. With these three assumptions, the background field of the target is given by
\be
A^{\mu}_a(x^-,x^+,\x) \approx \delta^{\mu -} \, \delta(x^+) \, A^-_a(\x) \, ,
\ee
which is also referred to as the shockwave limit.

While the eikonal approximation is an extremely powerful and reliable tool to study the scattering processes occurring at asymptotically high energies, one should relax this approximation
in order to study the processes at moderate energies, and calculate the power suppressed corrections to the leading energy contribution. In recent years, saturation-based computations that are performed at strict eikonal order have been quite successful to describe the high-energy collision data from the Large Hadron Collider (LHC). However, neither the energies probed at the Relativistic Heavy Ion Collider (RHIC) nor the ones at the future Electron-Ion Collider (EIC) are extremely high. Therefore, power corrections to the leading energy contribution, i.e. subeikonal corrections, are expected to be more relevant for phenomenological studies of RHIC and EIC than at LHC. The computation of these corrections and the analysis of their impact on observables will provide an opportunity to extend the applicability region of the CGC framework from high to moderate energies.
Moreover, the subeikonal corrections are expected to provide insight on new physics effects such as spin and quark flavor at small-x which are absent in the standard eikonal CGC framework.

Over the last decade, computations of the first power corrections to the leading energy contribution, namely next-to-eikonal (NEik) corrections, have advanced considerably. The first studies performed in \cite{Altinoluk:2014oxa, Altinoluk:2015gia}, focused on the computation of the subeikonal corrections to the gluon propagator due to the finite longitudinal width of the target. These corrections are understood to be associated with the transverse motion of the gluon during its interaction with the target. The results were applied to the single inclusive gluon production and various spin asymmetries in proton-nucleus collisions at mid-rapidity. Later on, in \cite{Altinoluk:2015xuy}, the effects of these corrections were studied in the weak field limit and it was shown that subeikonal corrections associated with the transverse motion can be interpreted as subeikonal corrections to the Lipatov vertex. Moreover, the effects of these corrections on the two particle correlations and azimuthal harmonics were studied in \cite{Agostini:2019avp, Agostini:2019hkj} for the proton-proton collisions at mid-rapidity. Finally, in a recent study \cite{Altinoluk:2020oyd}, NEik corrections are computed for the quark propagator in a gluon background field. Apart from the corrections due to finite longitudinal width of the target, the study performed  in \cite{Altinoluk:2020oyd} includes the NEik corrections due to interaction of the quark with the transverse component of the background field.
NEik corrections have also been studied in many other contexts in the literature. In series of papers \cite{Kovchegov:2015pbl,Kovchegov:2016weo,Kovchegov:2016zex,Kovchegov:2017jxc,Kovchegov:2017lsr,Kovchegov:2018znm,Kovchegov:2018zeq,Kovchegov:2020hgb,Adamiak:2021ppq,Kovchegov:2021lvz}, quark and gluon helicity (and transversity) distributions were studied in detail at NEik accuracy. The helicity-dependent generalization of the nonlinear rapidity evolution was studied in \cite{Cougoulic:2019aja,Cougoulic:2020tbc} at NEik order. In a recent study \cite{Kovchegov:2021iyc}, the quark Sivers function was studied with subeikonal evolution. In \cite{Chirilli:2018kkw,Chirilli:2021lif}, NEik corrections to both quark and gluon propagators were derived and used to study the high-energy operator product expansion (OPE) of two electromagnetic currents. Apart from the aforementioned studies, NEik effects were also studied in the context of the transverse momentum dependent distribution functions (TMDs) in \cite{Balitsky:2015qba,Balitsky:2016dgz,Balitsky:2017flc,Balitsky:2017gis,Balitsky:2019ayf,Balitsky:2020jzt,Balitsky:2021fer} where evolution equations for TMDs that interpolate between moderate and high energies were derived. A similar idea was proposed recently in \cite{Boussarie:2020vzf,Boussarie:2020fpb} where a new formulation for the unintegrated gluon distributions that interpolate between the moderate and high values of energy is introduced. A different approach to study NEik corrections was pursued in \cite{Jalilian-Marian:2017ttv,Jalilian-Marian:2018iui,Jalilian-Marian:2019kaf} by including the longitudinal momentum exchange between the projectile and the target during the interaction. Last but not least, in \cite{Hatta:2016aoc,Boussarie:2019icw}, NEik effects were studied in the context of orbital angular momentum.

The quark propagator at NEik accuracy is one of the main building blocks in order to study many observables at NEik accuracy, and it was computed in \cite{Chirilli:2018kkw} and \cite{Altinoluk:2020oyd} by using completely different methods.
In these studies, the results for the NEik quark propagator are obtained in a different form, which complicates their direct comparison.
The common feature of most NEik studies in the literature is that the light-cone coordinate $x^-$ dependence of the background field is still neglected. Motivated by these two issues, the main purposes of the present manuscript are: (i) to develop a systematic procedure to account for the $x^-$ dependence of the background field of the target at NEik accuracy for the quark propagator on a gluon background field and (ii) to revisit the derivation of the NEik quark propagator to further study the apparent differences in the results obtained
in \cite{Chirilli:2018kkw} and \cite{Altinoluk:2020oyd}.

The plan for the rest of the paper is as follows. In Sec. \ref{sec:quark_vs_scal_prop}, we study the details of how in-medium quark and scalar propagators are related to each other at NEik accuracy.  In Sec. \ref{sec:scal_prop}, we derive the complete NEik corrections to the scalar propagator in a gluon background, including the effects that are originating from the $x^-$ dependence of the background field. Sec. \ref{sec:quark_prop} is devoted to combining the results of the two previous sections in order to obtain the quark propagator at NEik accuracy. This result is then compared with the existing ones in the literature. Finally, conclusions are given in Sec. \ref{sec:Discussions}.  Technical details related with the derivation performed in Sec.  \ref{sec:quark_vs_scal_prop} are provided in Appendices \ref{app:L_vs_R_cov_deriv_on_GF}, \ref{app:LR_equiv_other_blocks}, and \ref{app:BG_block_exp}. A detailed comparison between the results from Sec. \ref{sec:scal_prop}
and from Ref.~\cite{Chirilli:2018kkw} for the scalar propagator at NEik accuracy is provided in Appendix \ref{app:scalar_comparison_Giovanni}.


\section{Relating in-medium quark and scalar propagators at NEik accuracy\label{sec:quark_vs_scal_prop}}


The aim of this section is to improve our understanding of the interplay between spin effects and the Eikonal approximation and subleading corrections to it. More precisely, we study the quark propagator in a gluon background field at high energy and express it in terms of the scalar propagator in that same gluon background field, as an expansion beyond the Eikonal approximation. In such a way, universal non-Eikonal effects stay contained within the scalar propagator, whereas quark-specific NEik corrections are isolated and derived explicitly. 
Moreover, the study in this section is performed in a completely general setup, without imposing any gauge fixing condition. 
Our results for the quark propagator at NEik accuracy in this section are valid without any constraints on the endpoints of the propagator, which can thus be either inside or outside of the target.


\subsection{Integral equations for the quark propagator in a gluon background field}

Our starting point is the Dirac equations obeyed by the Feynman quark propagator in a gluon background field, namely
\begin{align}
\left(i\, \slashed{D}_x\!-\!m\right) S_F(x,y) = &\, i \delta^{(D)}(x\!-\!y)
\label{Dirac_eq_L}
\\
 S_F(x,y)\left(-i\, \overleftarrow{\slashed{D}}_y\!-\!m\right) = &\, i \delta^{(D)}(x\!-\!y)
\label{Dirac_eq_R}
\, ,
\end{align}
with the right and left covariant derivatives defined as\footnote{Hereafter, we consider the gauge field and the field strength to be fundamental color matrices, defined as $A_{\mu}(x)\equiv t^a\, A^a_{\mu}(x)$ and $F_{\mu\nu}(x)\equiv t^a\, F^a_{\mu\nu}(x)$, in terms of the corresponding fields with adjoint indices.}
\begin{align}
D_{x^{\mu}}\equiv  &\, \d_{x^{\mu}}+ig\, A_{\mu}(x)
\label{Cov_deriv}
\\
 \overleftarrow{D}_{x^{\mu}}\equiv  &\, \overleftarrow{\d}_{x^{\mu}}-ig\, A_{\mu}(x)
\label{Cov_deriv_back}
\, ,
\end{align}
and the relation between
the field strength and the covariant derivative is given by
\begin{align}
ig\, F_{\mu\nu}(x)  =&\, [D_{x^{\mu}},D_{x^{\nu}}]= [\overleftarrow{D}_{x^{\mu}},\overleftarrow{D}_{x^{\nu}}]
\label{Fmunu_from_commutator}
\, .
\end{align}

In order to solve the left-evolution equation~\eqref{Dirac_eq_L} for the quark propagator, it is convenient to introduce an auxiliary propagator $L_F(x,y)$, defined as
\begin{align}
S_F(x,y) = &\, \left(i\, \slashed{D}_x\!+\!m\right) L_F(x,y)
\label{from_Dirac_prop_to_aux_L}
\, .
\end{align}
Since
\begin{align}
\left(i\, \slashed{D}_x\!-\!m\right)\left(i\, \slashed{D}_x\!+\!m\right)
=&\, - \gamma^{\mu}D_{x^{\mu}} \gamma^{\nu}D_{x^{\nu}} -m^2
= -\frac{1}{2}\, \big\{\gamma^{\mu},\gamma^{\nu}\big\}\, D_{x^{\mu}}D_{x^{\nu}}
  -\frac{1}{2}\, \big[\gamma^{\mu},\gamma^{\nu}\big]\, D_{x^{\mu}}D_{x^{\nu}}
 -m^2
\label{op_for_aux_prop}
\, ,
\end{align}
the auxiliary propagator obeys the following equation
\begin{align}
\left[-g^{\mu\nu}\, D_{x^{\mu}}D_{x^{\nu}}-m^2
-\frac{i}{4}\, [\gamma^{\mu},\gamma^{\nu}]\, g F_{\mu\nu}(x)
\right]
L_F(x,y) = &\, i \delta^{(D)}(x\!-\!y)
\label{aux_prop_eq_L}
\, .
\end{align}

On the other hand, the scalar Feynman propagator in the same background field is defined through the following equations
\begin{align}
\left[-g^{\mu\nu}\, D_{x^{\mu}}D_{x^{\nu}}-m^2
\right]
G_F(x,y) = &\, i \delta^{(D)}(x\!-\!y)
\label{scalar_prop_eq_L}
\\
G_F(x,y)
\left[-g^{\mu\nu}\, \overleftarrow{D}_{y^{\mu}}\overleftarrow{D}_{y^{\nu}}-m^2
\right]
 = &\, i \delta^{(D)}(x\!-\!y)
 \label{scalar_prop_eq_R}
 \, .
\end{align}
Since the same differential operator appears in both Eqs.~\eqref{aux_prop_eq_L} and \eqref{scalar_prop_eq_L}, the scalar propagator can be used as a Green's function in order to write the integral equation
\begin{align}
L_F(x,y) = &\,  G_F(x,y)
-i\int d^Dz\, G_F(x,z)\, \frac{i}{4}\, [\gamma^{\mu},\gamma^{\nu}]\, g F_{\mu\nu}(z)\, L_F(z,y)
\label{sol_aux_prop_L_1}
\end{align}
for the auxiliary propagator. That equation can be iterated multiple times, and leads to an explicit solution for $L_F(x,y)$ as a geometric series, summing arbitrary number of insertions of $(1/4)\, [\gamma^{\mu},\gamma^{\nu}]\, g F_{\mu\nu}(z)$ into the scalar propagator. Hence, Eq.~\eqref{sol_aux_prop_L_1} is equivalent to
\begin{align}
L_F(x,y) = &\,  G_F(x,y)
-i\int d^Dz\, L_F(x,z)\, \frac{i}{4}\, [\gamma^{\mu},\gamma^{\nu}]\, g F_{\mu\nu}(z)\, G_F(z,y)
\label{sol_aux_prop_L_2}
\, ,
\end{align}
which would lead to the same solution as geometric series. Thanks to Eq.~\eqref{from_Dirac_prop_to_aux_L}, one thus obtains the following equation for the quark propagator
\begin{align}
S_F(x,y) = &\,  \left(i\, \slashed{D}_x\!+\!m\right) G_F(x,y)
+\int d^Dz\, S_F(x,z)\, \frac{1}{4}\, [\gamma^{\mu},\gamma^{\nu}]\, g F_{\mu\nu}(z)\, G_F(z,y)
\label{sol_Dirac_prop_L}
\, .
\end{align}

In a similar way, one can study the right-evolution equation~\eqref{Dirac_eq_R} for the quark propagator, by introducing an auxiliary propagator $R_F(x,y)$ defined by
\begin{align}
S_F(x,y) = &\,  R_F(x,y) \left(-i\, \overleftarrow{\slashed{D}}_y\!+\!m\right)
\label{from_Dirac_prop_to_aux_R}
\, ,
\end{align}
which obeys
\begin{align}
R_F(x,y)
\left[-g^{\mu\nu}\, \overleftarrow{D}_{y^{\mu}}\overleftarrow{D}_{y^{\nu}}-m^2
-\frac{i}{4}\, [\gamma^{\mu},\gamma^{\nu}]\, g F_{\mu\nu}(y)
\right]
 = &\, i \delta^{(D)}(x\!-\!y)
\label{aux_prop_eq_R}
\, .
\end{align}
Following the same method, one finds the following integral equation for the quark propagator
\begin{align}
S_F(x,y) = &\,  G_F(x,y)\left(-i\, \overleftarrow{\slashed{D}}_y\!+\!m\right)
+\int d^Dz\, G_F(x,z)\, \frac{1}{4}\, [\gamma^{\mu},\gamma^{\nu}]\, g F_{\mu\nu}(z)\, S_F(z,y)
\label{sol_Dirac_prop_R}
\, .
\end{align}


\subsection{Projection into blocks}

From the fundamental relation $\{\gamma^{\mu}, \gamma^{\nu}\}=2g^{\mu\nu}$, it is clear that $\gamma^+\gamma^+=\gamma^-\gamma^-=0$ and that both $\gamma^+$ and $\gamma^-$ anticommute with the transverse $\gamma^i$. Moreover, ${\cal P}_G$ and ${\cal P}_B$ defined as
\begin{align}
{\cal P}_G \equiv &\,  \frac{\gamma^{-}\gamma^{+}}{2}
\label{G_projection}
\\
{\cal P}_B \equiv &\,  \frac{\gamma^{+}\gamma^{-}}{2}
\label{B_projection}
\end{align}
are projections, with ${\cal P}_G+{\cal P}_B= \mathbf{1}$ and ${\cal P}_G{\cal P}_B= 0$.
These projections are used notably in Light-Front quantization, with ${\cal P}_G$ restricting Dirac fields to their so-called good components, which are independent degrees of freedom in that context. By contrast, ${\cal P}_B$ projects Dirac fields into their so-called bad components, which are obeying constraint equations in Light-Front quantization.

Coming back to Eqs.~\eqref{sol_Dirac_prop_L} and \eqref{sol_Dirac_prop_R}, we can now expand the field strength insertion as
\begin{align}
\frac{1}{4}\, [\gamma^{\mu},\gamma^{\nu}]\, g F_{\mu\nu}
= &\,  \frac{1}{4}\, [\gamma^{i},\gamma^{j}]\, g F_{ij}
+  \frac{1}{2}\, [\gamma^{+},\gamma^{j}]\, g F_{+j}
+  \frac{1}{2}\, [\gamma^{-},\gamma^{j}]\, g F_{-j}
+  \frac{1}{2}\, [\gamma^{+},\gamma^{-}]\, g F_{+-}
\nonumber\\
= &\,  \frac{1}{4}\, [\gamma^{i},\gamma^{j}]\, g F_{ij}
+  \gamma^{+}\gamma^{j}\, g F_{+j}
+  \gamma^{-}\gamma^{j}\, g F_{-j}
+  ({\cal P}_B\!-\! {\cal P}_G)\, g F_{+-}
\label{sigmamunu_Fmunu_decomp}
\, .
\end{align}
In order to understand the effect of the insertion of each component of the field strength, it is convenient to use the projections ${\cal P}_G$ and ${\cal P}_B$ to split the quark propagator (seen as a Dirac matrix) into four blocks.
Then, projecting the left evolution equation, Eq.~\eqref{sol_Dirac_prop_L}, with either ${\cal P}_G$ or ${\cal P}_B$ on each side, one obtains the following four equations for the blocks in $S_F(x,y)$
\begin{align}
[{\cal P}_G S_F(x,y) {\cal P}_B] = &\,  i\, \gamma^- D_{x^-} G_F(x,y)
+\int d^Dz\, [{\cal P}_G S_F(x,z) {\cal P}_B]\,
\left\{\frac{1}{4}\, [\gamma^{i},\gamma^{j}]\, g F_{ij}(z) + g F_{+-}(z)
\right\}\, G_F(z,y)
\nonumber\\
&\, +\int d^Dz\, [{\cal P}_G S_F(x,z) {\cal P}_G]\, \gamma^{-}\gamma^{j}\, g F_{-j}(z)\, G_F(z,y)
\label{GB_block_L_exact}
\, ,
\end{align}
\begin{align}
[{\cal P}_G S_F(x,y) {\cal P}_G] = &\,  {\cal P}_G\left(i\, \gamma^j D_{x^j}+m\right) G_F(x,y)
+\int d^Dz\, [{\cal P}_G S_F(x,z) {\cal P}_G]\,
\left\{\frac{1}{4}\, [\gamma^{i},\gamma^{j}]\, g F_{ij}(z) - g F_{+-}(z)
\right\}\, G_F(z,y)
\nonumber\\
&\, +\int d^Dz\, [{\cal P}_G S_F(x,z) {\cal P}_B]\, \gamma^{+}\gamma^{j}\, g F_{+j}(z)\, G_F(z,y)
\label{GG_block_L_exact}
\, ,
\end{align}
\begin{align}
[{\cal P}_B S_F(x,y) {\cal P}_B] = &\,  {\cal P}_B\left(i\, \gamma^j D_{x^j}+m\right) G_F(x,y)
+\int d^Dz\, [{\cal P}_B S_F(x,z) {\cal P}_B]\,
\left\{\frac{1}{4}\, [\gamma^{i},\gamma^{j}]\, g F_{ij}(z) + g F_{+-}(z)
\right\}\, G_F(z,y)
\nonumber\\
&\, +\int d^Dz\, [{\cal P}_B S_F(x,z) {\cal P}_G]\, \gamma^{-}\gamma^{j}\, g F_{-j}(z)\, G_F(z,y)
\label{BB_block_L_exact}
\end{align}
and
\begin{align}
[{\cal P}_B S_F(x,y) {\cal P}_G] = &\,  i\, \gamma^+ D_{x^+} G_F(x,y)
+\int d^Dz\, [{\cal P}_B S_F(x,z) {\cal P}_G]\,
\left\{\frac{1}{4}\, [\gamma^{i},\gamma^{j}]\, g F_{ij}(z) - g F_{+-}(z)
\right\}\, G_F(z,y)
\nonumber\\
&\, +\int d^Dz\, [{\cal P}_B S_F(x,z) {\cal P}_B]\, \gamma^{+}\gamma^{j}\, g F_{+j}(z)\, G_F(z,y)
\label{BG_block_L_exact}
\, .
\end{align}
Similarly, projecting the right evolution equation, Eq.~\eqref{sol_Dirac_prop_R}, one finds the following four equations
\begin{align}
[{\cal P}_G S_F(x,y) {\cal P}_B] = &\,   G_F(x,y) (-i)\, \gamma^-  \overleftarrow{D}_{y^-}
+\int d^Dz\, \,  G_F(x,z)
\left\{\frac{1}{4}\, [\gamma^{i},\gamma^{j}]\, g F_{ij}(z) - g F_{+-}(z)
\right\}\, [{\cal P}_G S_F(z,y) {\cal P}_B]
\nonumber\\
&\, +\int d^Dz\, G_F(x,z)\, \gamma^{-}\gamma^{j}\, g F_{-j}(z)\, [{\cal P}_B S_F(z,y) {\cal P}_B]
\label{GB_block_R_exact}
\, ,
\end{align}
\begin{align}
[{\cal P}_G S_F(x,y) {\cal P}_G] = &\,   G_F(x,y) {\cal P}_G \left(-i\, \gamma^j  \overleftarrow{D}_{y^j}+m\right)
+\int d^Dz\, \,  G_F(x,z)
\left\{\frac{1}{4}\, [\gamma^{i},\gamma^{j}]\, g F_{ij}(z) - g F_{+-}(z)
\right\}\, [{\cal P}_G S_F(z,y) {\cal P}_G]
\nonumber\\
&\, +\int d^Dz\, G_F(x,z)\, \gamma^{-}\gamma^{j}\, g F_{-j}(z)\, [{\cal P}_B S_F(z,y) {\cal P}_G]
\label{GG_block_R_exact}
\, ,
\end{align}
\begin{align}
[{\cal P}_B S_F(x,y) {\cal P}_B] = &\,   G_F(x,y) {\cal P}_B \left(-i\, \gamma^j  \overleftarrow{D}_{y^j}+m\right)
+\int d^Dz\, \,  G_F(x,z)
\left\{\frac{1}{4}\, [\gamma^{i},\gamma^{j}]\, g F_{ij}(z) + g F_{+-}(z)
\right\}\, [{\cal P}_B S_F(z,y) {\cal P}_B]
\nonumber\\
&\, +\int d^Dz\, G_F(x,z)\, \gamma^{+}\gamma^{j}\, g F_{+j}(z)\, [{\cal P}_G S_F(z,y) {\cal P}_B]
\label{BB_block_R_exact}
\end{align}
and
\begin{align}
[{\cal P}_B S_F(x,y) {\cal P}_G] = &\,   G_F(x,y) (-i)\, \gamma^+  \overleftarrow{D}_{y^+}
+\int d^Dz\, \,  G_F(x,z)
\left\{\frac{1}{4}\, [\gamma^{i},\gamma^{j}]\, g F_{ij}(z) + g F_{+-}(z)
\right\}\, [{\cal P}_B S_F(z,y) {\cal P}_G]
\nonumber\\
&\, +\int d^Dz\, G_F(x,z)\, \gamma^{+}\gamma^{j}\, g F_{+j}(z)\, [{\cal P}_G S_F(z,y) {\cal P}_G]
\label{BG_block_R_exact}
\, .
\end{align}


\subsection{Target-space Eikonal expansion\label{sec:taget_space_eik_expansion}}

So far, we have obtained the exact equations (\ref{GB_block_L_exact}-\ref{BG_block_R_exact}) for the blocks inside $S_F(x,y)$. In order to understand the high-energy limit of $S_F(x,y)$ in the Eikonal approximation and beyond, let us discuss the power counting relevant in this limit.
The high-energy limit can be achieved by performing a large Lorentz boost along the $x^-$ direction of the background field representing the target. The corresponding power counting can then be formulated in terms of the Lorentz boost parameter, $\gamma_T$.
Under a large boost along the $x^-$ direction, the components of the background field strength scale as
\begin{align}
&\, F_{+j} 
 \propto \gamma_T \gg 1
\label{F_enhanced}
\\
&\, F_{ij}  \propto (\gamma_T)^0 =1
\label{F_ij_order_one}
\\
&\, F_{+-}  \propto (\gamma_T)^0 =1
\label{F_pm_order_one}
\\
&\, F_{-j} 
  \propto \frac{1}{\gamma_T} \ll 1
\label{F_suppressed}
\, .
\end{align}
Large Lorentz boosts along the $x^-$ direction contract the target along the $x^+$ direction. Hence, for a target of finite extent, the field strength has a small support along the $x^+$ direction after performing the large boost.\footnote{Actually, a strictly finite support of $F_{\mu\nu}(x)$ along $x^+$ is not necessary. It is sufficient that $F_{\mu\nu}(x)$ decays faster than a power at $x^+\rightarrow \pm \infty$ in order to avoid the appearance of extra power suppressed corrections by mixing between the tails of the $F_{\mu\nu}(x)$ distribution and the non-Eikonal effects. In that case of faster-than-power decay for $F_{\mu\nu}(x)$, a typical scale $L^+$ arises as the width of the distribution, which can be considered as an approximate support of $F_{\mu\nu}(x)$, and the power counting introduced in this section stays valid. Note that due to confinement, it is expected that the field strength $F_{\mu\nu}(x)$ of a hadron decays exponentially at large distance, in its rest frame. Our power counting is thus pertinent in the case of hadronic or nuclear targets.}
This small support is noted as $-L^+/2<x^+<L^+/2$, with the longitudinal width scaling as
 $L^+\propto 1/\gamma_T$.
Hence, in Eqs.~(\ref{GB_block_L_exact}-\ref{BG_block_R_exact}), the integration over the $z^+$ of each field strength insertion leads to a suppression of order $L^+\propto 1/\gamma_T$. This suppression is an extra effect on top of the suppression or the enhancement of the field strength components under a large boost, given in Eqs.~(\ref{F_enhanced}-\ref{F_suppressed}).

Let us now consider the scalar propagator $G_F(x,y)$. By definition, it describes the propagation of a scalar probe through the background gauge field. As stated earlier, the high-energy limit is achieved in our setup by boosting the target but not the projectile. Here, this amounts to boosting the background gauge field, but keeping the coordinates $x$ and $y$ fixed, since they are associated with the trajectory of the probe playing the role of the projectile.
Due to the fact that $G_F(x,y)$ is a scalar propagator, it stays of order one under such large boost, namely  
\begin{align}
G_F(x,y)\propto (\gamma_T)^0 =1
\label{G_F_power_count}
\, .
\end{align}
We also need to discuss the power counting for covariant derivatives acting on $G_F(x,y)$, like in  $D_{x^{\mu}} G_F(x,y)$.  Somewhat counterintuitively, the discussion of the power counting for such object is non-trivial even in the vacuum limit, because it might contain a factor of $\delta(x^+\!-\!y^+)$. Such a factor formally provides an enhancement as  $\gamma_T$ in our power counting, which can be understood as follows. Let us consider a situation in which we have two insertions of background field strength components, at points $z$ and $w$, and a factor $D_{z^{\mu}} G_F(z,w)$ linking the two insertions. On top of the behavior of the field strength components themselves, we have a suppression by $1/{\gamma_T}^2$, with one $1/{\gamma_T}$ factor coming from the integration over $z^+$ and another $1/{\gamma_T}$ factor from the integration over $w^+$. Then, if $D_{z^{\mu}} G_F(z,w)$ contains a factor of $\delta(z^+\!-\!w^+)$, one of these two integrals can be performed trivially, which will remove one suppression factor of  $1/\gamma_T$. In this sense, $\delta(z^+\!-\!w^+)$ can be considered as an enhancement by a factor of  ${\gamma_T}$.

With this discussion in mind, let us first focus on the vacuum scalar propagator $G_{0,F}(x,y)$ and study which components of 
 $\partial_{x^{\mu}} G_{0,F}(x,y)$ do contain a factor of $\delta(x^+\!-\!y^+)$. This vacuum propagator obeys the vacuum limit of Eq.~\eqref{scalar_prop_eq_L}, namely
\begin{align}
\left[-2\, {\partial}_{x^{-}}  {\partial}_{x^{+}}  +{\delta}^{ij}\, {\partial}_{x^{i}}{\partial}_{x^{j}}  -m^2
\right]
G_{0,F}(x,y) = &\, i \delta^{(D)}(x\!-\!y)
\label{scalar_prop_eq_L_vac}
 \, .
\end{align}
The right hand side of Eq.~\eqref{scalar_prop_eq_L_vac} is of order $\gamma_T$ since it contains a factor of $\delta(x^+\!-\!y^+)$. By contrast, the vacuum propagator $G_{0,F}(x,y)$ is of order $(\gamma_T)^0$, since it is the vacuum limit of Eq.~\eqref{G_F_power_count}, so that the term in $m^2$ is subleading in Eq.~\eqref{scalar_prop_eq_L_vac}. Moreover, the action of transverse derivatives on $G_{0,F}(x,y)$ is not expected to change the power counting, so that ${\partial}_{x^{j}}G_{0,F}(x,y) \propto (\gamma_T)^0$ and ${\partial}_{x^{i}}{\partial}_{x^{j}}G_{0,F}(x,y) \propto (\gamma_T)^0$.
Eq.~\eqref{scalar_prop_eq_L_vac} thus becomes 
\begin{align}
 {\partial}_{x^{-}}  {\partial}_{x^{+}}
G_{0,F}(x,y) = &\, -\frac{i}{2}\, \delta^{(D)}(x\!-\!y) + O({\gamma_T}^0)
\label{scalar_prop_eq_L_vac_approx}
 \, .
\end{align}
Integrating Eq.~\eqref{scalar_prop_eq_L_vac_approx} over $x^-$ leaves the factor of $\delta(x^+\!-\!y^+)$ on the right hand side unchanged, so that 
\begin{align}
  {\partial}_{x^{+}}
G_{0,F}(x,y) 
\propto &\, \delta(x^+\!-\!y^+)
\propto  {\gamma_T} \gg 1
\label{xplus_deriv_vac_scalar_prop}
 \, .
\end{align}
By constrast, integrating Eq.~\eqref{scalar_prop_eq_L_vac_approx} over $x^+$ transforms the factor of $\delta(x^+\!-\!y^+)$ on the right hand side into a Heaviside or a sign function, depending on the integration constant, which does not correspond to an enhancement. Hence
\begin{align}
  {\partial}_{x^{-}}
G_{0,F}(x,y) 
\propto &\, ({\gamma_T})^0
\label{xminus_deriv_vac_scalar_prop}
 \, .
\end{align}
In summary, in the vacuum case, only the component $ {\partial}_{x^{+}} G_{0,F}(x,y) $ is formally enhanced as $\gamma_T$ since it contains a $\delta(x^+\!-\!y^+)$ factor, whereas the other components  ${\partial}_{x^{-}} G_{0,F}(x,y) $ and  ${\partial}_{x^{j}} G_{0,F}(x,y) $ are of order $ ({\gamma_T})^0$, like the propagator $ G_{0,F}(x,y) $ itself.

It remains now to discuss to what extent these results for $ {\partial}_{x^{\mu}} G_{0,F}(x,y) $ can be generalized for $ {D}_{x^{\mu}} G_{F}(x,y) $, in the presence of the background field. If the interactions with the background field are treated in perturbation theory, $ {\partial}_{x^{\mu}} G_{0,F}(x,y) $ is the zeroth order term in the perturbative series for $ {D}_{x^{\mu}} G_{F}(x,y) $. It is expected that a subset of the higher order interaction terms can be resummed into gauge links dressing this zeroth order term. Such gauge links cannot modify the power counting found for the components of $ {\partial}_{x^{\mu}} G_{0,F}(x,y) $. In addition, $ {D}_{x^{\mu}} G_{F}(x,y) $ can also contain extra terms in which the Lorentz vector structure is provided by the background field.\footnote{ At first, one would consider terms proportional to the background gauge field $A_{\mu}(x)$ in $D_{x^{\mu}} G_F(x,y)$. However, taking gauge symmetry into account, it is more natural to consider instead Lorentz vectors with a simpler gauge transformation, like $[D_{x^\nu},{F_{\mu}}^{\nu}(x)]$.}
The behavior of such terms in $ {D}_{x^{\mu}} G_{F}(x,y) $  under a large boost of the target is driven by the behavior of the background field itself. In that case, the component $\mu=+$ is enhanced as $\gamma_T$, the component $\mu=-$ is suppressed as  $1/\gamma_T$ whereas the transverse components are of order $(\gamma_T)^0$. For each component, the behavior of this extra medium contribution to  $ {D}_{x^{\mu}} G_{F}(x,y) $ is the same as the one of the vacuum term $ {\partial}_{x^{\mu}} G_{0,F}(x,y) $, except for the $\mu=-$ component in which the extra contribution is suppressed compared to $ {\partial}_{x^{-}} G_{0,F}(x,y) $. Hence, the power counting found for the components of $ {\partial}_{x^{\mu}} G_{0,F}(x,y) $ is not modified by interactions with the background field, and stay valid for $ {D}_{x^{\mu}} G_{F}(x,y) $.
The case of $G_F(x,y)\overleftarrow{D}_{y^{\mu}}$ is fully analog to $D_{x^{\mu}} G_F(x,y)$.
Hence, we obtain
\begin{align}
D_{x^-} G_F(x,y)   \sim G_F(x,y)\overleftarrow{D}_{y^-}  \propto &\, (\gamma_T)^0 =1
\label{D_min_power_count}
\\
D_{x^j} G_F(x,y)   \sim G_F(x,y)\overleftarrow{D}_{y^j} \propto &\, (\gamma_T)^0 =1
\label{D_perp_power_count}
\\
D_{x^+} G_F(x,y)   \sim G_F(x,y)\overleftarrow{D}_{y^+} \propto  &\, \gamma_T \gg 1
\label{D_plus_power_count}
\, .
\end{align}

Thanks to the power counting that we have introduced so far, we can analyse Eqs.~(\ref{GB_block_L_exact}-\ref{BG_block_L_exact}) in order to extract the behavior of each block inside $S_F(x,y)$ under a large Lorentz boost of the target.
First, let us note that in each of the Eqs.~(\ref{GB_block_L_exact}-\ref{BG_block_L_exact}), the second term involves the same block of $S_F(x,y)$ as on the left hand side of the equation. Apart from that, it contains $G_F(z,y)$,
 $F_{ij}(z)$ and $F_{+-}(z)$, which are all of order $(\gamma_T)^0$, and the integration over $z^+$, which brings a power of $1/\gamma_T$. Hence, in each of Eqs.~(\ref{GB_block_L_exact}-\ref{BG_block_L_exact}), the second term is overall suppressed as $1/\gamma_T$ compared to the left hand side. Thus, the leading behavior of each block inside $S_F(x,y)$ is determined by either the first or third terms in Eqs.~(\ref{GB_block_L_exact}-\ref{BG_block_L_exact}), whereas this homogeneous second term is always subleading.

 Let us now focus on Eqs.~\eqref{GB_block_L_exact} and \eqref{GG_block_L_exact}. In both of them, the inhomogeneous first term is of order $(\gamma_T)^0$, according to Eqs.~\eqref{D_min_power_count} and \eqref{D_perp_power_count}. By contrast, the third term in both Eqs.~\eqref{GB_block_L_exact} and \eqref{GG_block_L_exact} couple these two equations with each other. In the third term in Eq.~\eqref{GB_block_L_exact}, both the integration over $z^+$ and the field strength component $F_{-j}(z)$ bring a suppression by a factor of $1/\gamma_T$. Hence, the third term in Eq.~\eqref{GB_block_L_exact} is suppressed by $1/{\gamma_T}^2$ compared to the left hand side of Eq.~\eqref{GG_block_L_exact}. By contrast, in the third term in Eq.~\eqref{GG_block_L_exact}, $F_{+j}(z)$ brings an enhancement by a factor of $\gamma_T$ which compensates the suppression of $1/\gamma_T$ due to the integration over $z^+$. The third term in Eq.~\eqref{GB_block_L_exact} is then of the same order as the left hand side of Eq.~\eqref{GB_block_L_exact}. From this analysis, we can conclude that
\begin{align}
[{\cal P}_G S_F(x,y) {\cal P}_B]\propto &\,(\gamma_T)^0 =1
\label{G_B_block_power_count}
\\
[{\cal P}_G S_F(x,y) {\cal P}_G]\propto &\,(\gamma_T)^0 =1
\label{G_G_block_power_count}
\, .
\end{align}

Eqs.~\eqref{BB_block_L_exact} and \eqref{BG_block_L_exact} can be analyzed in the same way.
The first term in Eq.~\eqref{BB_block_L_exact} is of order $(\gamma_T)^0$, according to Eq.~\eqref{D_perp_power_count}, whereas the first term in Eq.~\eqref{BG_block_L_exact} is of order $\gamma_T$,
according to Eq.~\eqref{D_plus_power_count}.
The third term in Eq.~\eqref{BB_block_L_exact} is suppressed by $1/{\gamma_T}^2$ compared to the left hand side of Eq.~\eqref{BG_block_L_exact}, with one power of $1/{\gamma_T}$ coming from the integration over $z^+$ and one from the field strength component $F_{-j}(z)$. Finally, the third term in Eq.~\eqref{BG_block_L_exact} is of the same order as the left hand side of Eq.~\eqref{BB_block_L_exact}, due to the compensation of the enhancement by a factor of $\gamma_T$ coming from $F_{+j}(z)$ and of the suppression by a factor of $1/\gamma_T$ coming from the integration over $z^+$. We can then deduce
\begin{align}
[{\cal P}_B S_F(x,y) {\cal P}_B]\propto &\,(\gamma_T)^0 =1
\label{B_B_block_power_count}
\\
[{\cal P}_B S_F(x,y) {\cal P}_G] \propto &\, \gamma_T \gg 1
\label{B_G_block_power_count}
\, .
\end{align}

%
In ${\cal P}_B S_F(x,y) {\cal P}_G$,  the contribution enhanced as $\gamma_T$ is related via Eqs.~\eqref{BG_block_L_exact} and \eqref{BG_block_R_exact} to the enhanced contributions to $D_{x^+} G_F(x,y)$ and $G_F(x,y)\overleftarrow{D}_{y^+}$ as
\begin{align}
[{\cal P}_B S_F(x,y) {\cal P}_G] = &\,  i\, \gamma^+ D_{x^+} G_F(x,y) +O\big({\gamma_T}^0\big) = G_F(x,y) (-i)\, \gamma^+  \overleftarrow{D}_{y^+}+O\big({\gamma_T}^0\big)
\label{BG_block_enhanced}
\, .
\end{align}

%

Based on this discussion of power counting, we can now derive approximate solutions of Eqs.~(\ref{GB_block_L_exact}-\ref{BG_block_R_exact}), valid up to corrections suppressed as $1/{\gamma_T}^2$ in the limit of large Lorentz boost of the target.
For example, in Eq.~\eqref{GB_block_L_exact}, both the left hand side and the first term on the right hand side are of order $({\gamma_T})^0$, which corresponds to the Eikonal order in the high-energy limit. By contrast, the homogeneous second term is of order $1/{\gamma_T}$, which corresponds to a NEik correction, and the third term is of order $1/{\gamma_T}^2$, corresponding to a next-to-next-to-eikonal correction (NNEik). Hence, an approximate solution to Eq.~\eqref{GB_block_L_exact} valid at NEik accuracy can be obtained by discarding the third term and iterating Eq.~\eqref{GB_block_L_exact} once (in order to make the contribution from the homogeneous term fully explicit), which leads to

\begin{align}
[{\cal P}_G S_F(x,y) {\cal P}_B] = &\,  i\, \gamma^- D_{x^-}\left\{ G_F(x,y)
+\int d^Dz\, G_F(x,z)\,
\left(\frac{1}{4}\, [\gamma^{i},\gamma^{j}]\, g F_{ij}(z) + g F_{+-}(z)
\right)\, G_F(z,y)\right\}
 +\textrm{NNEik}
\label{GB_block_L_NEik}
\, .
\end{align}
Similarly, one finds from Eq.~\eqref{GB_block_R_exact}
\begin{align}
[{\cal P}_G S_F(x,y) {\cal P}_B] = &\,  \left\{ G_F(x,y)
+\int d^Dz\, \,  G_F(x,z)
\left(\frac{1}{4}\, [\gamma^{i},\gamma^{j}]\, g F_{ij}(z) - g F_{+-}(z)
\right)\, G_F(z,y)\right\}
(-i)\, \gamma^-  \overleftarrow{D}_{y^-}
+\textrm{NNEik}
\label{GB_block_R_NEik}
\, .
\end{align}

In Eq.~\eqref{GG_block_L_exact}, all terms are of order $({\gamma_T})^0$ apart from the homogeneous second term on the right hand side, which is of order $1/{\gamma_T}$ and thus a NEik correction. In particular, the third term involves ${\cal P}_G S_F(x,y) {\cal P}_B$. In order to solve Eq.~\eqref{GG_block_L_exact}, we can first insert the expression given in Eq.~\eqref{GB_block_L_NEik} for ${\cal P}_G S_F(x,y) {\cal P}_B$ into this third term, which will provide both a  contribution of order $({\gamma_T})^0$ and a correction of order $1/{\gamma_T}$.
Then, in order to write an explicit approximate solution of Eq.~\eqref{GG_block_L_exact} taking into account the homogeneous term, one has to iterate Eq.~\eqref{GG_block_L_exact} once. Each of the two contributions of order $({\gamma_T})^0$, corresponding to the first and third terms in Eq.~\eqref{GG_block_L_exact}, generates an extra correction of order $1/{\gamma_T}$ via this iteration. All in all, one finds
\begin{align}
& [{\cal P}_G S_F(x,y) {\cal P}_G] =   {\cal P}_G\left(i\, \gamma^j D_{x^j}+m\right) G_F(x,y)
\nonumber\\
&\, +2i\, {\cal P}_G D_{x^-}\int d^Dz\,
\left\{ G_F(x,z) +\int d^Dw\, G_F(x,w)
\left(\frac{1}{4}\, [\gamma^{i},\gamma^{j}]\, g F_{ij}(w) + g F_{+-}(w)\right)\,
G_F(w,z)\right\}\gamma^{l}\, g {F^{-}}_{l}(z)\, G_F(z,y)
\nonumber\\
&\, + {\cal P}_G \int d^Dz\,
\left\{\left(i\, \gamma^l D_{x^l}+m\right) G_F(x,z)
+2i\, D_{x^-}\int d^Dw\, G_F(x,w) \gamma^{l}\, g {F^{-}}_{l}(w)\, G_F(w,z)
\right\}
\nonumber\\
&\, \hspace{2cm} \times\;
\left(\frac{1}{4}\, [\gamma^{i},\gamma^{j}]\, g F_{ij}(z) - g F_{+-}(z)\right)\,
G_F(z,y)
+\textrm{NNEik}
\, .
\label{GG_block_L_NEik}
\end{align}
%
In Eq.~\eqref{GG_block_R_exact}, the third term on the right hand side involves the enhanced block ${\cal P}_B S_F(x,y) {\cal P}_G$. Due to the suppression both by the integration over $z^+$ and by the field strength component, this third term is overall of order $1/{\gamma_T}$. Hence, only the enhanced contribution (given in Eq.~\eqref{BG_block_enhanced}) to ${\cal P}_B S_F(x,y) {\cal P}_G$ is relevant in this third term in order to obtain an approximate solution to Eq.~\eqref{GG_block_R_exact} up to corrections of order $1/{\gamma_T}^2$. After this replacement, Eq.~\eqref{GG_block_R_exact} can be solved at the desired accuracy via an iteration, in order to account for the homogeneous second term. One obtains
\begin{align}
[{\cal P}_G S_F(x,y) {\cal P}_G] = &\,  {\cal P}_G \left\{G_F(x,y)
+\int d^Dz\, \,  G_F(x,z)
\left(\frac{1}{4}\, [\gamma^{i},\gamma^{j}]\, g F_{ij}(z) - g F_{+-}(z)
\right)\, G_F(z,y)\right\} \left(-i\, \gamma^l  \overleftarrow{D}_{y^l}+m\right)
\nonumber\\
&\, +2i\, {\cal P}_G \int d^Dz\, G_F(x,z)\, \gamma^{j}\, g {F^{+}}_{j}(z)\, G_F(z,y)  \overleftarrow{D}_{y^+}
+\textrm{NNEik}
\, .
\label{GG_block_R_NEik}
\end{align}
%
%
%
Concerning the ${\cal P}_B S_F(x,y) {\cal P}_B$ block, Eq.~\eqref{BB_block_L_exact} is analogous to Eq.~\eqref{GG_block_R_exact} and Eq.~\eqref{BB_block_R_exact} is analogous to Eq.~\eqref{GG_block_L_exact}. Following the same method as for the ${\cal P}_G S_F(x,y) {\cal P}_G$ block, one obtains
\begin{align}
[{\cal P}_B S_F(x,y) {\cal P}_B] = &\,  {\cal P}_B\left(i\, \gamma^l D_{x^l}+m\right) \left\{G_F(x,y)
+\int d^Dz\, G_F(x,z)\,
\left(\frac{1}{4}\, [\gamma^{i},\gamma^{j}]\, g F_{ij}(z) + g F_{+-}(z)
\right)\, G_F(z,y)\right\}
\nonumber\\
&\, +2i\,{\cal P}_B\,  D_{x^+}\int d^Dz\,  G_F(x,z)\, \gamma^{j}\, g {F^+}_{j}(z)\, G_F(z,y)
+\textrm{NNEik}
\label{BB_block_L_NEik}
\end{align}
from the left evolution equation \eqref{BB_block_L_exact} and
\begin{align}
& [{\cal P}_B S_F(x,y) {\cal P}_B] =    G_F(x,y) {\cal P}_B \left(-i\, \gamma^j  \overleftarrow{D}_{y^j}+m\right)
\nonumber\\
&+2i\, {\cal P}_B \int d^Dz\, \,  G_F(x,z) \gamma^{l}\, g {F^{-}}_{l}(z)\,
\left\{G_F(z,y)+ \int d^Dw\, G_F(z,w)
\left(\frac{1}{4}\, [\gamma^{i},\gamma^{j}]\, g F_{ij}(w) - g F_{+-}(w)\right) G_F(w,y)
\right\}\overleftarrow{D}_{y^-}
\nonumber\\
& +{\cal P}_B \int d^Dz\, \,  G_F(x,z)
\left(\frac{1}{4}\, [\gamma^{i},\gamma^{j}]\, g F_{ij}(z) + g F_{+-}(z)\right)
\nonumber\\
& \hspace{2cm}\times\; \left\{G_F(z,y) \left(-i\, \gamma^l  \overleftarrow{D}_{y^l}+m\right)
 +2i\int d^Dw\, G_F(z,w) \gamma^{l}\, g {F^{-}}_{l}(w)\, G_F(w,y) \overleftarrow{D}_{y^-}
\right\}
+\textrm{NNEik}
\label{BB_block_R_NEik}
\end{align}
from the right evolution equation \eqref{BB_block_R_exact}.

Note that for each block in $S_F(x,y)$, one obtains two different expressions from the left and right evolution equations. Obviously, for each block, the two expressions should be equivalent 
up to NNEik contributions. This equivalence will be shown explicitly in Sec.~\ref{sec:LR_equi}.

The case of the $[{\cal P}_B S_F(x,y) {\cal P}_G]$ block is more complicated due to the enhancement of the leading contribution, see Eq.~\eqref{B_G_block_power_count}. Therefore, we leave the detailed study of this block to Sec.~\ref{sec:LR_equi}. It will be performed after introducing now in Sec.~\ref{sec:taming_enhancement} the necessary tools to address this complication.


\subsection{Taming the enhanced contribution\label{sec:taming_enhancement}}

The enhanced contribution appearing in the $[{\cal P}_B S_F(x,y) {\cal P}_G]$ block is related to
$D_{x^+} G_F(x,y)$ via Eq.~\eqref{BG_block_enhanced}. Since the light-cone time $x^+$ is playing the role of evolution variable in high-energy context, one should use the Klein-Gordon equation in gluon background given in Eq.~\eqref{scalar_prop_eq_L} in order to obtain a more explicit expression for $D_{x^+} G_F(x,y)$.
Eq.~\eqref{scalar_prop_eq_L} can be equivalently written as
\begin{align}
D_{x^{-}}D_{x^{+}}G_F(x,y) = &\, -\frac{i}{2}\, \delta^{(D)}(x\!-\!y)
+\frac{1}{2} \Big[ \delta^{ij}\, D_{x^{i}}D_{x^{j}} -m^2 -ig F_{+-}(x)
\Big]\, G_F(x,y)
\label{Dxmin_scalar_prop_eq_L}
\, .
\end{align}
In order to obtain the desired expression for $D_{x^+} G_F(x,y)$, we have to understand how to invert $D_{x^{-}}$.

For that purpose let us
consider the equation
\begin{align}
D_{x^{-}} {\cal B}(x) =&\, {\cal C}(x)
\label{Dxmin_eq}
\end{align}
with generic matrix-valued functions ${\cal B}(x)$ and ${\cal C}(x)$ in the fundamental representation.
It is convenient to introduce the gauge link along the $x^-$ direction
\begin{align}
{\cal U}^{[-]}_F(x^-, y^-; \uz)
\equiv &\, {\cal P}_{-} \exp
\left\{
-ig \int_{y^-}^{x^-} \!\!dz^-\, A^+(z)
\right\}
\label{def_minus_Wilson_line}
\end{align}
which satisfies the relations
\begin{align}
\left[{\cal U}^{[-]}_F(x^-, y^-; \uz)\right]^{\dag}=&\, {\cal U}^{[-]}_F(y^-, x^-; \uz)
\label{prop_minus_Wilson_line}
\end{align}
and
\begin{align}
D_{x^{-}} {\cal U}^{[-]}_F(x^-, y^-; \uz)=&\, 0
\label{minus_cov_deriv_x_on_Wilson_line}
\\
 {\cal U}^{[-]}_F(x^-, y^-; \uz) \overleftarrow{D}_{y^{-}}=&\, 0
\label{minus_cov_deriv_yx_on_Wilson_line}
\, .
\end{align}
It is then possible to solve Eq.~\eqref{Dxmin_eq} as
\begin{align}
{\cal B}(x) =&\,  {\cal U}^{[-]}_F(x^-, -\infty; \ux) \, {\cal B}(\ux,-\infty)
+\int_{-\infty}^{x^-} \!\!dz^-\, {\cal U}^{[-]}_F(x^-, z^-; \ux)\, {\cal C}(\ux,z^-)
\label{Dxmin_eq_sol_1}
\end{align}
or as
\begin{align}
{\cal B}(x) =&\,  \left[{\cal U}^{[-]}_F(+\infty, x^-; \ux)\right]^{\dag} \, {\cal B}(\ux,+\infty)
-\int^{+\infty}_{x^-} \!\!dz^-\, \left[{\cal U}^{[-]}_F(z^-, x^-; \ux)\right]^{\dag}\, {\cal C}(\ux,z^-)
\label{Dxmin_eq_sol_2}
\, ,
\end{align}
in terms of the limit of ${\cal B}(x)$ at $x^-\rightarrow -\infty$ or at $x^-\rightarrow +\infty$.

For the particular case given in Eq.~\eqref{Dxmin_scalar_prop_eq_L}, Eqs.~\eqref{Dxmin_eq_sol_1} and \eqref{Dxmin_eq_sol_2} provide the solutions
\begin{align}
D_{x^{+}}G_F(x,y) =&\,  {\cal U}^{[-]}_F(x^-, -\infty; \ux) \, D_{x^{+}}G_F(\ux,-\infty;y)
-\frac{i}{2}\, \delta^{(D-1)}(\ux\!-\!\uy)\, \theta(x^-\!-\!y^-)\, {\cal U}^{[-]}_F(x^-, y^-; \ux)
\nonumber\\
&\,
+\int_{-\infty}^{x^-} \!\!dz^-\, {\cal U}^{[-]}_F(x^-, z^-; \ux)\,
\int d^{D-1}\uz\, \delta^{(D-1)}(\ux\!-\!\uz)\;
\frac{1}{2} \Big[ \delta^{ij}\, D_{z^{i}}D_{z^{j}} -m^2 -ig F_{+-}(z)
\Big]\, G_F(z,y)
\label{Dxmin_scalar_prop_eq_L_sol_1}
\end{align}
and
\begin{align}
D_{x^{+}}G_F(x,y) =&\,  \left[{\cal U}^{[-]}_F(+\infty, x^-; \ux)\right]^{\dag} \, D_{x^{+}}G_F(\ux,+\infty;y)
+\frac{i}{2}\, \delta^{(D-1)}(\ux\!-\!\uy)\, \theta(y^-\!-\!x^-)\,  \left[{\cal U}^{[-]}_F(y^-, x^-; \ux)\right]^{\dag}
\nonumber\\
&\,
-\int^{+\infty}_{x^-} \!\!dz^-\, \left[{\cal U}^{[-]}_F(z^-, x^-; \ux)\right]^{\dag}\,
\int d^{D-1}\uz\, \delta^{(D-1)}(\ux\!-\!\uz)\;
\frac{1}{2} \Big[ \delta^{ij}\, D_{z^{i}}D_{z^{j}} -m^2 -ig F_{+-}(z)
\Big]\, G_F(z,y)
\label{Dxmin_scalar_prop_eq_L_sol_2}
\, .
\end{align}

In general, the scalar Feynman propagator $G_F(x,y)$ should have the limits
\begin{align}
\lim_{x^-\rightarrow -\infty} \theta(x^+\!-\!y^+) D_{x^+} G_F(x,y)
 = &\, 0
 \label{boundary_cond_GF_xmin_min_infinity}
\\
 \lim_{x^-\rightarrow +\infty} \theta(y^+\!-\!x^+) D_{x^+} G_F(x,y)
 = &\, 0
\label{boundary_cond_GF_xmin_plus_infinity}
\end{align}
because such configurations correspond to infinitely large spacelike separations $-(x\!-\!y)^2>0$, for which $G_F(x,y)$ has to be exponentially suppressed.

Taking the linear combination of the two solutions Eqs.~\eqref{Dxmin_scalar_prop_eq_L_sol_1} and \eqref{Dxmin_scalar_prop_eq_L_sol_2} with weights $\theta(x^+\!-\!y^+)$ and $\theta(y^+\!-\!x^+)$ respectively, one obtains a solution of Eq.~\eqref{Dxmin_scalar_prop_eq_L} in which the boundary terms drop thanks to the relations \eqref{boundary_cond_GF_xmin_min_infinity} and \eqref{boundary_cond_GF_xmin_plus_infinity}, namely
\begin{align}
& D_{x^{+}}G_F(x,y) =
\Delta(x,y)
\nonumber\\
&\,
+\theta(x^+\!-\!y^+)\int_{-\infty}^{x^-} \!\!dz^-\, {\cal U}^{[-]}_F(x^-, z^-; \ux)\,
\int d^{D-1}\uz\, \delta^{(D-1)}(\ux\!-\!\uz)\;
\frac{1}{2} \Big[ \delta^{ij}\, D_{z^{i}}D_{z^{j}} -m^2 -ig F_{+-}(z)
\Big]\, G_F(z,y)
\nonumber\\
&\,
-\theta(y^+\!-\!x^+)
\int^{+\infty}_{x^-} \!\!dz^-\, \left[{\cal U}^{[-]}_F(z^-, x^-; \ux)\right]^{\dag}\,
\int d^{D-1}\uz\, \delta^{(D-1)}(\ux\!-\!\uz)\;
\frac{1}{2} \Big[ \delta^{ij}\, D_{z^{i}}D_{z^{j}} -m^2 -ig F_{+-}(z)
\Big]\, G_F(z,y)
\label{Dxmin_scalar_prop_eq_L_sol_final_1}
\, .
\end{align}
In Eq.~\eqref{Dxmin_scalar_prop_eq_L_sol_final_1}, we have used the notation\footnote{Here, we have introduced the sign function, defined as $\textrm{sgn}(X)\equiv \theta(X)-\theta(-X)$.}
\begin{align}
\Delta(x,y)
\equiv &\,
-\frac{i}{4}\, \delta^{(D-1)}(\ux\!-\!\uy)\, \textrm{sgn}(x^-\!-\!y^-)\, {\cal U}^{[-]}_F(x^-, y^-; \ux)
\label{def_Delta}
\, .
\end{align}
Note that $\Delta(x,y)$ satisfies the relations
\begin{align}
D_{x^{-}} \Delta(x,y)=&\, - \Delta(x,y)\overleftarrow{D}_{y^{-}}=  -\frac{i}{2}\, \delta^{(D)}(x\!-\!y)
\label{eq_Delta}
\, .
\end{align}
The terms in the second and third lines of Eq.~\eqref{Dxmin_scalar_prop_eq_L_sol_final_1} can be rearranged, in order to obtain
\begin{align}
& D_{x^{+}}G_F(x,y) =
\Delta(x,y)
+i
\int d^{D}z\, \Delta(x,z)\;
 \Big[ \delta^{ij}\, D_{z^{i}}D_{z^{j}} -m^2 -ig F_{+-}(z)
\Big]\, G_F(z,y)
\nonumber\\
&\,
+\frac{1}{4}\textrm{sgn}(x^+\!-\!y^+)\, \int d^{D}z\, \delta^{(D-1)}(\ux\!-\!\uz)\;
{\cal U}^{[-]}_F(x^-, z^-; \ux)\,
 \Big[ \delta^{ij}\, D_{z^{i}}D_{z^{j}} -m^2 -ig F_{+-}(z)
\Big]\, G_F(z,y)
\label{Dxmin_scalar_prop_eq_L_sol_final_2}
\, .
\end{align}

%
Let us now discuss the power counting of each term in Eq.~\eqref{Dxmin_scalar_prop_eq_L_sol_final_2}. The first term 
$\Delta(x,y)$, defined in Eq.~\eqref{def_Delta}, contains a factor $\delta(x^+\!-\!y^+)$. It is thus formally enhanced as $\gamma_T$, as discussed in the previous subsection.
In the second term in Eq.~\eqref{Dxmin_scalar_prop_eq_L_sol_final_2}, the $\delta(x^+\!-\!z^+)$ factor contained in $\Delta(x,z)$ can be used to perform the integration over $z^+$. Hence, the second term in Eq.~\eqref{Dxmin_scalar_prop_eq_L_sol_final_2} is overall of order $(\gamma_T)^0$.

The case of the last contribution, in the second line of Eq.~\eqref{Dxmin_scalar_prop_eq_L_sol_final_2}, is more complicated, but can be discussed as follows.
In the second line of Eq.~\eqref{Dxmin_scalar_prop_eq_L_sol_final_2}, only the Wilson line ${\cal U}^{[-]}_F(x^-, z^-; \ux)\,$ is dependent on $x^-$. Using the identity
\begin{align}
{\cal U}^{[-]}_F(x^-, z^-; \ux)
= & \,
{\cal U}^{[-]}_F(x^-, 0; \ux)\; \Big[{\cal U}^{[-]}_F(z^-,0; \ux)\Big]^{\dag}
\label{Wilson_line_zmin_split}
\, ,
\end{align}
one can factorize the $x^-$ dependence from the $y^-$ dependence and from the integration over $z^-$. Hence, in second line of Eq.~\eqref{Dxmin_scalar_prop_eq_L_sol_final_2}, the $x^-$ and $y^-$ dependencies separately arise from the dependence of the background field on $z^-$, and not from the propagation of scalar modes from $y$ to $x$.
The distribution in final momentum $q^+$ given by the second line of Eq.~\eqref{Dxmin_scalar_prop_eq_L_sol_final_2} is thus determined by
\begin{align}
f(q^+, \ux)
=&\,
\int dx^-\; e^{iq^+ x^-}\;
{\cal U}^{[-]}_F(x^-, 0; \ux)
\label{FT_of_Umin}
\, .
\end{align}
The gauge field component $A^+(x)$ is expected to be suppressed as $A^+(x)\sim  1/\gamma_T$ in the limit of large Lorentz boost of the target, so that
\begin{align}
f(q^+, \ux)
=&\,
\int dx^-\; e^{iq^+ x^-}\; \bigg[
1 -ig \int_{0}^{x^-}dw^-\;  A^+(\ux,w^-)
\bigg]
+\textrm{NNEik}
\label{FT_of_Umin_2}
\, .
\end{align}
On the other hand,
\begin{align}
\int_{0}^{x^-}dw^-\;  A^+(\ux,w^-)
=&\,
\int_{0}^{x^-}dw^-\; 
\Big[ A^+(\ux,0) +\int_{0}^{w^-}dv^-\,  \d_{-}A^+(\ux,v^-)
\Big]
\nonumber\\
=&\,
x^-\, A^+(\ux,0)
+
\int_{0}^{x^-} dv^-\; (x^-\!-\!v^-) \;  \d_{-}A^+(\ux,v^-)
\label{FT_of_Umin_3}
\, .
\end{align}
Since $\d_{-}A^+(\ux,v^-)\sim 1/{\gamma_T}^2$ under a large Lorentz boost of the target, the second term in Eq.~\eqref{FT_of_Umin_3} is a NNEik correction, so that one finds from Eqs.~\eqref{FT_of_Umin_2} and \eqref{FT_of_Umin_3}
\begin{align}
f(q^+, \ux)
=&\,
2\pi\, \delta(q^+)\,
- g A^+(\ux,0)\, 2\pi\, \delta'(q^+)
+\textrm{NNEik}
\label{FT_of_Umin_4}
\, .
\end{align}
Apart from NNEik contributions, $f(q^+, \ux)$ is thus a singular distribution localized at $q^+=0$.
%
As stated previously, in our approach, the high-energy limit of a collision process is obtained by performing a large boost of the target. However, only the modes with strictly positive light cone momentum from the projectile can contribute to such scattering process. Indeed, no matter how much the target is boosted along the $x^-$ direction, it will never catch up with the zero modes ($p^+=0$) from the projectile and cross them. Therefore, such zero mode contribution from the projectile should be discarded in our study of the propagators beyond the Eikonal approximation, anticipating their applications to scattering processes.
%
The second factor in the second line of Eq.~\eqref{Dxmin_scalar_prop_eq_L_sol_final_2}, containing the $z^-$ integration and the $y^-$ dependence but no $x^-$ dependence is expected to behave in a similar way, namely to contain zero modes and power suppressed corrections, even though it is more complicated to show explicitly in the presence of the background field. In any case, Eq.~\eqref{FT_of_Umin_4} is sufficient to conclude that the second line of Eq.~\eqref{Dxmin_scalar_prop_eq_L_sol_final_2} is a combination of zero modes and of contributions which are NNEik or smaller. All of these are beyond the scope of the present study. Hence, Eq.~\eqref{Dxmin_scalar_prop_eq_L_sol_final_2} can be written as
\begin{align}
& D_{x^{+}}G_F(x,y) =
\Delta(x,y)
+i
\int d^{D}z\, \Delta(x,z)\;
 \Big[ \delta^{ij}\, D_{z^{i}}D_{z^{j}} -m^2 -ig F_{+-}(z)
\Big]\, G_F(z,y)
+\textrm{NNEik}
\label{Dxmin_scalar_prop_eq_L_sol_final_3}
\, .
\end{align}
A similar analysis can be performed for $G_F(x,y)\overleftarrow{D}_{y^{+}}$, leading to
\begin{align}
G_F(x,y)\overleftarrow{D}_{y^{+}}
=&\,
-\Delta(x,y)
-i
\int d^{D}z\, G_F(x,z)\;
 \Big[ \delta^{ij}\, D_{z^{i}}D_{z^{j}} -m^2 +ig F_{+-}(z)
\Big]\, \Delta(z,y)
+\textrm{NNEik}
\label{Dxmin_scalar_prop_eq_R_sol_final}
\, .
\end{align}


\subsection{Equivalence of left and right evolution for each block in $S_F(x,y)$\label{sec:LR_equi}}

In this section, for each of the blocks in $S_F(x,y)$, we show how the NEik expressions obtained from either the left or the right evolution equation are equivalent. We will in particular make use of the identity
\begin{align}
D_{x^{\rho}} G_F(x,y)
 = &\, -G_F(x,y) \overleftarrow{D}_{y^{\rho}}
 +\int d^Dz\, G_F(x,z)\, \Big[
 g {F_{\rho}}^{\mu}(z)\, \overrightarrow{D}_{z^{\mu}} -\overleftarrow{D}_{z^{\mu}} g {F_{\rho}}^{\mu}(z)
 \Big]G_F(z,y)
\label{deriv_scalar_prop_flip_generic}
\, ,
\end{align}
derived in Appendix~\ref{app:L_vs_R_cov_deriv_on_GF}.

\subsubsection{${\cal P}_G S_F(x,y) {\cal P}_B$ block:}

Let us consider the $\rho=-$ component of Eq.~\eqref{deriv_scalar_prop_flip_generic}, namely
\begin{align}
D_{x^{-}} G_F(x,y)
 = &\, -G_F(x,y) \overleftarrow{D}_{y^{-}}
 \nonumber\\
 &
 +\int d^Dz\, G_F(x,z)\, \Big[
 -g F_{+-}(z)\, \overrightarrow{D}_{z^{-}} +\overleftarrow{D}_{z^{-}} g F_{+-}(z)
 +g F^{+j}(z)\, \overrightarrow{D}_{z^{j}} -\overleftarrow{D}_{z^{j}} g F^{+j}(z)
 \Big]G_F(z,y)
\label{deriv_scalar_prop_flip_minus}
\, .
\end{align}
From the power counting rules introduced in Sec.~\ref{sec:taget_space_eik_expansion}, $D_{x^{-}} G_F(x,y)$ and $G_F(x,y) \overleftarrow{D}_{y^{-}}$ are of order $(\gamma_T)^0$. By contrast the terms involving $F_{+-}(z)$ in Eq.~\eqref{deriv_scalar_prop_flip_minus} are NEik corrections (of order $1/\gamma_T$), and the ones involving $F^{+j}(z)$ are NNEik corrections  (of order $1/{\gamma_T}^2$). Hence, iterating Eq.~\eqref{deriv_scalar_prop_flip_minus}, one can obtain
the expansion
\begin{align}
D_{x^{-}} G_F(x,y)
 = &\, -G_F(x,y) \overleftarrow{D}_{y^{-}}
 +2\int d^Dz\, G_F(x,z)\, g F_{+-}(z)\,  G_F(z,y) \overleftarrow{D}_{y^{-}}
 +\textrm{NNEik}
\label{deriv_scalar_prop_flip_minus_NEik_1}
\end{align}
or equivalently
\begin{align}
 G_F(x,y)\overleftarrow{D}_{y^{-}}
 = &\, -D_{x^{-}}G_F(x,y)
 -2D_{x^{-}}\int d^Dz\, G_F(x,z)\, g F_{+-}(z)\,  G_F(z,y)
 +\textrm{NNEik}
\label{deriv_scalar_prop_flip_minus_NEik_2}
\, .
\end{align}
Eqs.~\eqref{deriv_scalar_prop_flip_minus_NEik_1} or \eqref{deriv_scalar_prop_flip_minus_NEik_2} show the equivalence between the expansions \eqref{GB_block_L_NEik} and \eqref{GB_block_R_NEik} obtained from the left and right evolution equations for the $[{\cal P}_G S_F(x,y) {\cal P}_B]$ block, up to the terms involving $F_{ij}(z)$.

The discussion of these $F_{ij}(z)$ terms in Eqs.~\eqref{GB_block_L_NEik} and \eqref{GB_block_R_NEik} goes as follows. First, these terms are suppressed by the integration over $z^+$, making them NEik corrections. Hence, in order to compare them between
Eqs.~\eqref{GB_block_L_NEik} and \eqref{GB_block_R_NEik}, we can truncate further the relation~\eqref{deriv_scalar_prop_flip_minus_NEik_1}, and use $D_{x^{-}} G_F(x,y)=-G_F(x,y) \overleftarrow{D}_{y^{-}}+\textrm{NEik}$ in order to move the covariant derivative through the scalar propagators. Moreover, one has by power counting
\begin{align}
\big[D_{z^{-}} , F_{ij}(z)\big] \propto &\,\frac{1}{\gamma_T} \ll 1
\, .
\end{align}
It is then easy to see that the terms involving $F_{ij}(z)$ in \eqref{GB_block_L_NEik} and \eqref{GB_block_R_NEik} are equivalent to each other, up to NNEik corrections which are beyond our scope.

All in all, we have checked that the expansions \eqref{GB_block_L_NEik} and \eqref{GB_block_R_NEik} for the ${\cal P}_G S_F(x,y) {\cal P}_B$ block are indeed equivalent to each other, up to NNEik corrections.
In particular, we can as well choose the average between the two expressions, which is
\begin{align}
[{\cal P}_G S_F(x,y) {\cal P}_B] = &\,  i\, \frac{\gamma^-}{2}\, D_{x^-}\left\{ G_F(x,y)
+\int d^Dz\, G_F(x,z)\,
\left(\frac{1}{4}\, [\gamma^{i},\gamma^{j}]\, g F_{ij}(z) + g F_{+-}(z)
\right)\, G_F(z,y)\right\}
\nonumber\\
&
+\left\{ G_F(x,y)
+\int d^Dz\, \,  G_F(x,z)
\left(\frac{1}{4}\, [\gamma^{i},\gamma^{j}]\, g F_{ij}(z) - g F_{+-}(z)
\right)\, G_F(z,y)\right\}
(-i)\, \frac{\gamma^-}{2}\,  \overleftarrow{D}_{y^-}
 +\textrm{NNEik}
\nonumber\\
 = &\,  i\, \frac{\gamma^-}{2}\, D_{x^-}\left\{ G_F(x,y)
+\int d^Dz\, G_F(x,z)\,
\frac{1}{4}\, [\gamma^{i},\gamma^{j}]\, g F_{ij}(z)
\, G_F(z,y)\right\}
\nonumber\\
&
+\left\{ G_F(x,y)
+\int d^Dz\, \,  G_F(x,z)
\frac{1}{4}\, [\gamma^{i},\gamma^{j}]\, g F_{ij}(z)
\, G_F(z,y)\right\}
(-i)\, \frac{\gamma^-}{2}\,  \overleftarrow{D}_{y^-}
 +\textrm{NNEik}
\label{GB_block_averaged_NEik}
\, .
\end{align}
The terms involving $F_{+-}(z)$ indeed cancel each other in the symmetrization, up to NNEik corrections, which can be justified by using the same arguments as for the earlier discussion of the $F_{ij}(z)$ terms.


\subsubsection{${\cal P}_G S_F(x,y) {\cal P}_G$ and ${\cal P}_B S_F(x,y) {\cal P}_B$ blocks}

Inserting the relation \eqref{Dxmin_scalar_prop_eq_R_sol_final} into Eq.~\eqref{GG_block_R_NEik} leads to the expansion
\begin{align}
[{\cal P}_G S_F(x,y) {\cal P}_G] = &\,  {\cal P}_G \left\{G_F(x,y)
+\int d^Dz\, \,  G_F(x,z)
\left(\frac{1}{4}\, [\gamma^{i},\gamma^{j}]\, g F_{ij}(z) - g F_{+-}(z)
\right)\, G_F(z,y)\right\} \left(-i\, \gamma^l  \overleftarrow{D}_{y^l}+m\right)
\nonumber\\
&\, -2i\, {\cal P}_G \int d^Dz\, G_F(x,z)\, \gamma^{j}\, g {F^{+}}_{j}(z)\, \Delta(z,y)
+\textrm{NNEik}
\label{GG_block_R_NEik_2}
\end{align}
for the ${\cal P}_G S_F(x,y) {\cal P}_G$ block obtained from the evolution on the right. We thus have to show the equivalence between Eq.~\eqref{GG_block_R_NEik_2} and the expansion~\eqref{GG_block_L_NEik}, obtained from the evolution on the left. That demonstration is performed in Appendix~\ref{app:LR_equiv_other_blocks}.

For the ${\cal P}_B S_F(x,y) {\cal P}_B$ block, inserting the relation \eqref{Dxmin_scalar_prop_eq_L_sol_final_3} into Eq.~\eqref{BB_block_L_NEik} leads to the expansion
\begin{align}
[{\cal P}_B S_F(x,y) {\cal P}_B] = &\,  {\cal P}_B\left(i\, \gamma^l D_{x^l}+m\right) \left\{G_F(x,y)
+\int d^Dz\, G_F(x,z)\,
\left(\frac{1}{4}\, [\gamma^{i},\gamma^{j}]\, g F_{ij}(z) + g F_{+-}(z)
\right)\, G_F(z,y)\right\}
\nonumber\\
&\, +2i\,{\cal P}_B\,  \int d^Dz\,  \Delta(x,z)\, \gamma^{j}\, g {F^+}_{j}(z)\, G_F(z,y)
+\textrm{NNEik}
\label{BB_block_L_NEik_2}
\end{align}
from the evolution on the left. It should be compared to the expansion~\eqref{BB_block_R_NEik}, obtained from the evolution on the right for the ${\cal P}_B S_F(x,y) {\cal P}_B$ block. This case is entirely symmetric to the one of the ${\cal P}_G S_F(x,y) {\cal P}_G$ block, so that the demonstration of equivalence of Eqs.~\eqref{BB_block_R_NEik} and \eqref{BB_block_L_NEik_2} would follow the same steps as in Appendix~\ref{app:LR_equiv_other_blocks}, and does not need to be repeated.


\subsubsection{${\cal P}_B S_F(x,y) {\cal P}_G$ block}

The case of the ${\cal P}_B S_F(x,y) {\cal P}_G$ block is complicated by its enhancement, already discussed in Sec.~\ref{sec:taget_space_eik_expansion}. From the results of Sec.~\ref{sec:taming_enhancement}, we know that the enhanced contribution in ${\cal P}_B S_F(x,y) {\cal P}_G$ can be written as $i\, \gamma^+ \Delta(x,y)$. It is thus convenient to subtract that term, in order to recover a quantity of order $(\gamma_T)^0$.

The evolution equation Eq.~\eqref{BG_block_L_exact} on the left side for ${\cal P}_B S_F(x,y) {\cal P}_G$ is then rewritten as
\begin{align}
& \Big[{\cal P}_B S_F(x,y) {\cal P}_G - i\, \gamma^+ \Delta(x,y)\Big]
= \,  i\, \gamma^+ \Big[D_{x^+} G_F(x,y) -\Delta(x,y)\Big]
\nonumber\\
&\, \hspace{2cm}
+\int d^Dz\, \bigg\{i\, \gamma^+ \Delta(x,z)+\Big[{\cal P}_B S_F(x,z) {\cal P}_G- i\, \gamma^+ \Delta(x,z)\Big]\bigg\}\,
\left\{\frac{1}{4}\, [\gamma^{i},\gamma^{j}]\, g F_{ij}(z) - g F_{+-}(z)
\right\}\, G_F(z,y)
\nonumber\\
&\, \hspace{2cm}
+\int d^Dz\, [{\cal P}_B S_F(x,z) {\cal P}_B]\, \gamma^{+}\gamma^{j}\, g {F^{-}}_{j}(z)\, G_F(z,y)
\label{BG_block_L_exact_subtr}
\, .
\end{align}
This evolution equation is studied in Appendix~\ref{app:BG_block_exp}, where it is shown that it leads to the expansion
\begin{align}
& \Big[{\cal P}_B S_F(x,y) {\cal P}_G - i\, \gamma^+ \Delta(x,y)\Big]
\nonumber\\
= \, &
\big( i\gamma^l\,{D}_{x^{l}} +m \big)\frac{\gamma^+}{2}\int d^{D}w\, \Delta(x,w)\,
\bigg\{
G_F(w,y)
\nonumber\\
&\hspace{3cm}
+\int d^{D}z\, G_F(w,z)\bigg[
\frac{1}{4}\, [\gamma^{i},\gamma^{j}]\, g F_{ij}(z)
-g F_{+-}(z)
\bigg]G_F(z,y)
\bigg\}\big( -i\gamma^{m}\,\overleftarrow{D}_{y^j} +m \big)
\nonumber\\
&
+\big( i\gamma^l\,{D}_{x^{l}} +m \big)\frac{\gamma^+}{2}\int d^{D}w\,
\bigg\{G_F(x,w)\,
\nonumber\\
&\hspace{3cm}
+\int d^{D}z\, G_F(x,z)\bigg[
\frac{1}{4}\, [\gamma^{i},\gamma^{j}]\, g F_{ij}(z)
+g F_{+-}(z)
\bigg]G_F(z,w)
\bigg\}
\Delta(w,y)\big( -i\gamma^{m}\,\overleftarrow{D}_{y^j} +m \big)
\nonumber\\
&\,
-2i\big( i\gamma^i\,{D}_{x^{i}} +m \big)\gamma^+\gamma^{j}\int d^{D}z\, \int d^{D}w\, \Delta(x,z)\,
  G_F(z,w)\, g {F^+}_{j}(w)\, \Delta(w,y)
\nonumber\\
&\, +2i\gamma^i\gamma^+\int d^{D}z\, \int d^{D}w\, \Delta(x,w)\, g {F^+}_{i}(w)\,  G_F(w,z)
\, \Delta(z,y)\, \big( -i\gamma^j\,\overleftarrow{D}_{y^{j}} +m \big)
+\textrm{NNEik}
\label{BG_block_L_subtr_NEik_result_2}
\end{align}
for the ${\cal P}_B S_F(x,y) {\cal P}_G$ block.
That expression is fully symmetric, and could also be obtained from the evolution equation Eq.~\eqref{BG_block_R_exact} on the right side for ${\cal P}_B S_F(x,y) {\cal P}_G$, using the same steps as in Appendix~\ref{app:BG_block_exp}.


\subsection{Reconstructing the NEik quark propagator from the scalar propagator}

So far, we have obtained relatively compact expressions for the expansion at NEik for each of the four blocks inside $S_F(x,y)$, and checked that these expressions could be obtained both from the evolution equation on the left and on the right. However, in future applications, it will not always be convenient to manipulate the four blocks of $S_F(x,y)$ separately. Fortunately, the expressions~\eqref{GB_block_averaged_NEik}, \eqref{GG_block_R_NEik_2}, \eqref{BB_block_L_NEik_2} and \eqref{BG_block_L_subtr_NEik_result_2} for the blocks can be recombined into
\begin{align}
  S_F(x,y)
= &\, i\, \gamma^+ \Delta(x,y)
\nonumber\\
&\,
+\big( i\,\slashed{D}_{x} +m \big)\frac{\gamma^+}{2}\int d^{D}w\, \Delta(x,w)\,
\bigg\{
G_F(w,y)
\nonumber\\
&\hspace{3cm}
+\int d^{D}z\, G_F(w,z)\bigg[
\frac{1}{4}\, [\gamma^{i},\gamma^{j}]\, g F_{ij}(z)
-g F_{+-}(z)
\bigg]G_F(z,y)
\bigg\}\big( -i\,\overleftarrow{\slashed{D}}_{y} +m \big)
\nonumber\\
&
+\big( i\,\slashed{D}_{x} +m \big)\frac{\gamma^+}{2}\int d^{D}w\,
\bigg\{G_F(x,w)\,
\nonumber\\
&\hspace{3cm}
+\int d^{D}z\, G_F(x,z)\bigg[
\frac{1}{4}\, [\gamma^{i},\gamma^{j}]\, g F_{ij}(z)
+g F_{+-}(z)
\bigg]G_F(z,w)
\bigg\}
\Delta(w,y)\big( -i\,\overleftarrow{\slashed{D}}_{y} +m \big)
\nonumber\\
&\,
-2i\big( i\,\slashed{D}_{x} +m \big)\gamma^+\gamma^{j}\int d^{D}z\, \int d^{D}w\, \Delta(x,z)\,
  G_F(z,w)\, g {F^+}_{j}(w)\, \Delta(w,y)
\nonumber\\
&\, +2i\gamma^j\gamma^+\int d^{D}z\, \int d^{D}w\, \Delta(x,w)\, g {F^+}_{j}(w)\,  G_F(w,z)
\, \Delta(z,y)\, \big( -i\,\overleftarrow{\slashed{D}}_{y} +m \big)
+\textrm{NNEik}
\label{quark_prop_NEik_from_scalar_result}
\, .
\end{align}
Eq.~\eqref{quark_prop_NEik_from_scalar_result} is our final result for the NEik expansion of the quark propagator $S_F(x,y)$ in gluon background field, written in terms of the scalar propagator $G_F(x,y)$ in the same gluon background field.


\section{Scalar propagator through a shockwave at NEik accuracy\label{sec:scal_prop}}


In this section, we derive the NEik corrections to the scalar propagator through a gluon shockwave, using a similar approach as the one introduced for the quark propagator in Ref.~\cite{Altinoluk:2020oyd}. NEik corrections to the scalar propagator were also derived in Ref.~\cite{Chirilli:2018kkw}, using a different approach.  An important effect, which has never been systematically taken into account in the CGC framework, is the $x^-$ dependence of the background field $A^{\mu}(x)$ of the target. As a major improvement, it is introduced in this work beyond the strict eikonal approximation. With this improvement,
we are finally able to address non-eikonal corrections of all types, and we obtain the complete set of NEik corrections to the scalar propagator.

\subsection{Setup of the calculation}

The scalar Feynman propagator in background field is determined by the Green's equation, Eq.~\eqref{scalar_prop_eq_L}.
Expanding the covariant derivatives, that equation can also be written as
\begin{align}
\Big[-g^{\mu\nu}\, \d_{x^{\mu}}\d_{x^{\nu}}-m^2
-2ig\, A^{\mu}(x)\, \d_{x^{\mu}}
-ig \left(\d_{x^{\mu}} A^{\mu}(x)\right)
+g^2\, A_{\mu}(x)\, A^{\mu}(x)
\Big]
G_F(x,y) = &\, i \delta^{(D)}(x\!-\!y)
\label{scalar_prop_eq_L_Amu}
\, ,
\end{align}
with its vacuum version being
\begin{align}
\left[-g^{\mu\nu}\, \d_{x^{\mu}}\d_{x^{\nu}}-m^2
\right]
G_{0,F}(x,y) = &\, i \delta^{(D)}(x\!-\!y)
\label{free_scalar_prop_eq_L}
\, .
\end{align}
Thanks to the free Feynman propagator solving Eq.~\eqref{free_scalar_prop_eq_L},
the relevant solution of Eq.~\eqref{scalar_prop_eq_L_Amu} obeys the following integral equation
\begin{align}
G_F(x,y) = &\,  G_{0,F}(x,y)
+i\int d^Dz\,\, G_{0,F}(x,z)\,
\Big[-2ig\, A^{\mu}(z)\, \d_{z^{\mu}}
-ig \left(\d_{z^{\mu}} A^{\mu}(z)\right)
+g^2\, A_{\mu}(z)\, A^{\mu}(z)
\Big]
\, G_F(z,y)
\label{sol_scalar_prop_L_1}
\, ,
\end{align}
which can equivalently be written as
\begin{align}
G_F(x,y) = &\,  G_{0,F}(x,y)
+\int d^Dz\,\, G_{0,F}(x,z)\,
\Big[g\, A^{\mu}(z)\, \overrightarrow{\d}_{z^{\mu}}
- \overleftarrow{\d}_{z^{\mu}}\,  g\, A^{\mu}(z)
+ig^2\, A_{\mu}(z)\, A^{\mu}(z)
\Big]
\, G_F(z,y)
\label{sol_scalar_prop_L_2}
\, .
\end{align}
By iterating this equation, one generates a geometric series expression for $G_F(x,y)$. The same expression can be obtained from the Feynman rules for a colored scalar in a non-Abelian background field. Note that this geometric series is also equivalent to the following expression
\begin{align}
G_F(x,y)
= &\,  G_{0,F}(x,y)
+\int d^Dz\,\, G_F(x,z)\,
\Big[g\, A^{\mu}(z)\, \overrightarrow{\d}_{z^{\mu}}
- \overleftarrow{\d}_{z^{\mu}}\,  g\, A^{\mu}(z)
+ig^2\, A_{\mu}(z)\, A^{\mu}(z)
\Big]
\, G_{0,F}(z,y)
\label{sol_scalar_prop_R_2}
\, .
\end{align}
For future convenience, let us introduce the notation
\begin{align}
\delta G_F(x,y) \equiv &\, G_F(x,y)-G_{0,F}(x,y)
\label{def_delta_G_F}
\end{align}
for the total interaction contribution to the propagator in the background field.
Separating the different components of the background field $A^{\mu}(x)$, Eq.~\eqref{sol_scalar_prop_L_2} can be organized as
\begin{align}
\delta G_F(x,y) = &\, \int d^Dz\,\, G_{0,F}(x,z)\,
\bigg\{\Big[g\, A^{-}(z)\, \overrightarrow{\d}_{z^{-}}
- \overleftarrow{\d}_{z^{-}}\,  g\, A^{-}(z)\Big]
\nonumber\\
&\, +\Big[-g\, A_{j}(z)\, \overrightarrow{\d}_{z^{j}}
+ \overleftarrow{\d}_{z^{j}}\,  g\, A_{j}(z)
-ig^2\, A_{j}(z)\, A_{j}(z)
\Big]
\nonumber\\
&\,+\Big[
g\, A^{+}(z)\, \overrightarrow{\d}_{z^{+}}
- \overleftarrow{\d}_{z^{+}}\,  g\, A^{+}(z)
+ig^2\, A^{-}(z)\, A^{+}(z)
+ig^2\, A^{+}(z)\, A^{-}(z)
\Big]
\bigg\}
\, G_F(z,y)
\, ,
\label{sol_scalar_prop_L_3}
\end{align}
or equivalently as
\begin{align}
\delta G_F(x,y) = &\, \int d^Dz\,\, G_{0,F}(x,z)\,
\bigg\{\Big[g\, A^{-}(z)\, \overrightarrow{\d}_{z^{-}}
- \overleftarrow{\d}_{z^{-}}\,  g\, A^{-}(z)\Big]
\nonumber\\
&\,
 +\Big[-i\, \overleftarrow{D}_{z^{j}}\, \overrightarrow{D}_{z^{j}}
+i\, \overleftarrow{\d}_{z^{j}}\, \overrightarrow{\d}_{z^{j}}
\Big]
+\Big[
g\, A^{+}(z)\, \overrightarrow{D}_{z^{+}}
- \overleftarrow{D}_{z^{+}}\,  g\, A^{+}(z)
\Big]
\bigg\}
\, G_F(z,y)
\label{sol_scalar_prop_L_4}
\, .
\end{align}

Let us now explain the power counting relevant in the high-energy limit.
As in Sec.~\ref{sec:taget_space_eik_expansion}, this power counting can be formulated in terms of the Lorentz boost parameter, $\gamma_T$, when performing a large Lorentz boost of the target along the $x^-$ direction.
In addition to the high-energy behavior of the background field strength $F_{\mu\nu}(x)$ introduced in Eqs.~(\ref{F_enhanced}-\ref{F_suppressed}), we need to specify the behavior of the background gauge field $A^{\mu}(x)$.
Under such large boost, the components of the background gauge field scale as
\begin{align}
&\, A^{-} \propto     \gamma_T \gg 1
\label{A_enhanced}
\\
&\, A^{i} \propto   (\gamma_T)^0 =1
\label{A_order_one}
\\
&\, A^{+}  \propto   \frac{1}{\gamma_T} \ll 1
\label{A_suppressed}
\end{align}
in a generic gauge.
Moreover, $A^{\mu}(x)$ becomes independent of $x^-$ in the limit of infinite boost, $\gamma_T\rightarrow +\infty$, with subleading corrections scaling as inverse powers of $\gamma_T$.
Let us recall that the background field strength $F_{\mu\nu}(x)$ is assumed to vanish outside of a finite support\footnote{As mentioned earlier our results generalize to the case of an approximately finite support for $F_{\mu\nu}(x)$, meaning $F_{\mu\nu}(x)$ decaying faster than any power for $x^+\rightarrow \pm \infty$. In this section, we will however restrict ourselves to the case of a strictly finite support, for clarity of the derivation and notations.} in $x^+$,
of length $L^+$. This support is then Lorentz contracted under the boost, as $L^+ \sim 1/\gamma_T$.
The background gauge field $A^{\mu}(x)$ is a pure gauge field for $F_{\mu\nu}(x)=0$, outside of this support corresponding to the target.
For simplicity, we assume that $A^{\mu}(x)=0$ as well outside of the target. This assumption can be understood as a partial gauge fixing.

We are now in position to discuss the power counting of each term in Eq.~\eqref{sol_scalar_prop_L_4}. Within our assumptions, each interaction with the background field in Eq.~\eqref{sol_scalar_prop_L_4} occurs within the target, and the corresponding integration over $z^+$ is confined to the Lorentz contracted support of $A^{\mu}(z)$, thus bringing a power of $L^+\propto 1/\gamma_T$.

In the first square bracket in Eq.~\eqref{sol_scalar_prop_L_4}, the enhancement as $\gamma_T$ of the field component $A^{-}(z)$ is compensating this suppression as $1/\gamma_T$ coming from the integration over  $z^+$.
Note that the derivatives in $z^{-}$ act on propagators and not on $A^{-}(z)$, so that they do not bring a further suppression, as discussed in Sec.~\ref{sec:taget_space_eik_expansion} (see in particular Eqs.~\eqref{xminus_deriv_vac_scalar_prop}  and \eqref{D_min_power_count}).
 Hence, insertions of the first square bracket in Eq.~\eqref{sol_scalar_prop_L_4} are of order $(\gamma_T)^0$ and should be resummed to all order already in the Eikonal approximation.

The second square bracket in Eq.~\eqref{sol_scalar_prop_L_4} is of order $(\gamma_T)^0$ and thus cannot compensate the $1/\gamma_T$ suppression coming from the integration over $z^+$. To NEik accuracy (order $1/\gamma_T$), it is thus sufficient to consider contributions to $G_F(x,y)$ with at most one insertion of the second square bracket in Eq.~\eqref{sol_scalar_prop_L_4}.

In the third square bracket in Eq.~\eqref{sol_scalar_prop_L_4}, the component $A^{+}(z)$ is power-suppressed according to Eq.~\eqref{A_suppressed}. However, the $z^+$ covariant derivatives acting on propagators can bring an enhancement as $\gamma_T$ (see Eq.~\eqref{D_plus_power_count}), so that the third square bracket should be considered of order $(\gamma_T)^0$ as well. Hence, each insertion of this third square bracket brings a suppression by a factor of $1/\gamma_T$ due to the integration over $z^+$.

All in all, in order calculate $G_F(x,y)$ at NEik accuracy, one should include three types of contributions: either with a single insertion of the second square bracket in Eq.~\eqref{sol_scalar_prop_L_4} or with a single insertion of the third square bracket, or without any of them. In all cases, insertions of the first square bracket should be resummed to all orders.
These three contributions will be calculated separately in the following subsections.

\subsection{Scalar propagator in a pure $A^{-}(z)$ background : $z^-$-dependence effects\label{sec:zmin_dep_scalar_prop}}

Let us first consider the contribution to $G_F(x,y)$ due to the interactions with a pure $A^{-}(z)$ background. Dropping the second and third square brackets in~\eqref{sol_scalar_prop_L_4} one writes
\begin{align}
\delta G_F(x,y)\bigg|_{\textrm{pure }A^{-}} = &\,  \int d^Dz\,\, G_{0,F}(x,z)\,
\Big[g\, A^{-}(z)\, \overrightarrow{\d}_{z^{-}}
- \overleftarrow{\d}_{z^{-}}\,  g\, A^{-}(z)\Big]
\, G_F(z,y)\bigg|_{\textrm{pure }A^{-}}
\label{sol_scalar_prop_pure_A_minus}
\, .
\end{align}
Iterating this equation into a geometric series expression, and introducing the Fourier representation of the free Feynman propagators, one finds
\begin{align}
\delta G_F(x,y)\bigg|_{\textrm{pure }A^{-}}
= &\,
\sum_{N=1}^{+\infty} \int \bigg[\prod_{n=1}^{N} d^D z_n\bigg]
G_{0,F}(x,z_N)\,
\bigg\{{\cal P}_n\, \prod_{n=1}^{N}\Big[ \Big(
g\, A^{-}(z_n)\, \overrightarrow{\d}_{z_n^{-}}
- \overleftarrow{\d}_{z_n^{-}}\,  g\, A^{-}(z_n)
\Big)\, G_{0,F}(z_n,z_{n-1})
\Big]
\bigg\}
\nonumber\\
= &\,
\sum_{N=1}^{+\infty} \int \bigg[\prod_{n=1}^{N} d^D z_n\bigg]
\int \bigg[\prod_{n=0}^{N} \frac{d^D p_n}{(2\pi)^D}\: \frac{i}{[p_n^2\!-\!m^2 +i\epsilon]} \bigg]\,
e^{-i p_N \cdot x}\, e^{i p_0 \cdot y}
\nonumber\\
&\, \hspace{2cm} \times\;
\bigg\{{\cal P}_n\, \prod_{n=1}^{N}\Big[
 e^{i p_{n} \cdot z_n}
 \Big(
g\, A^{-}(z_n)\, \overrightarrow{\d}_{z_n^{-}}
- \overleftarrow{\d}_{z_n^{-}}\,  g\, A^{-}(z_n)
\Big)\, e^{-i p_{n-1} \cdot z_n}
\Big]
\bigg\}
\nonumber\\
= &\,
\sum_{N=1}^{+\infty} \int \bigg[\prod_{n=1}^{N} d^D z_n\bigg]\,
\bigg[{\cal P}_n\, \prod_{n=1}^{N}\Big(-i g\, A^{-}(z_n)\Big)\bigg]\,
\nonumber\\
&\, \hspace{2cm} \times\;
\int \bigg[\prod_{n=0}^{N} \frac{d^D p_n}{(2\pi)^D}\:
\frac{i}{[p_n^2\!-\!m^2 +i\epsilon]} \,
e^{-i p_{n} \cdot (z_{n+1}-z_n)}
\bigg]\,
\bigg[\prod_{n=1}^{N} \big(p_n^+ + p_{n-1}^+  \big)\bigg]
\label{expansion_scalar_prop_pure_A_minus_0}
\, ,
\end{align}
using the notations $z_0\equiv y$ and $z_{N+1}\equiv x$. Performing the integrations over $p_n^-$, one arrives at
\begin{align}
\delta G_F(x,y)\bigg|_{\textrm{pure }A^{-}}
= &\,
\sum_{N=1}^{+\infty} \int \bigg[\prod_{n=1}^{N} d^D z_n\bigg]\,
\bigg[{\cal P}_n\, \prod_{n=1}^{N}\Big(-i g\, A^{-}(z_n)\Big)\bigg]\,
\nonumber\\
&\, 
\times\;
\int \bigg[\prod_{n=0}^{N} \frac{d^{D-1} \underline{p_n}}{(2\pi)^{D-1}}\:
\frac{1}{2p_n^+} \,
e^{-i p_{n}^+  (z_{n+1}^- -z_n^-)}\,
e^{i \p_{n} \cdot (\z_{n+1}-\z_n)}
\bigg]\,
\bigg[\prod_{n=1}^{N} \big(p_n^+ + p_{n-1}^+  \big)\bigg]
\nonumber\\
&\, 
\,\times\;
\bigg\{\prod_{n=0}^{N}
e^{-i\, \frac{(\p_n^2+m^2)}{2p_n^+}\, (z_{n+1}^+-z_n^+)}\,
\Big[
\theta(z_{n+1}^+\!-\!z_n^+)\, \theta(p_n^+) - \theta(z_n^+\!-\!z_{n+1}^+)\, \theta(-p_n^+)
\Big]
\bigg\}
\label{expansion_scalar_prop_pure_A_minus_1}
\, .
\end{align}

This expression is the exact solution of Eq.~\eqref{sol_scalar_prop_pure_A_minus}. In the third line of Eq.~\eqref{expansion_scalar_prop_pure_A_minus_1}, the phases are non-Eikonal effects since $z_n^+\propto L^+\propto 1/\gamma_T$, with the exception of the contributions from $x^+$ and $y^+$, which are not constrained to be inside the target. These non-Eikonal phases are associated with transverse motion of the scalar particle within the width of the target.
The other non-Eikonal feature present in Eq.~\eqref{expansion_scalar_prop_pure_A_minus_1} is the $z^-$ dependence of $A^-(z)$. In appendix~\ref{app:NEik_from_zmin_dep_scalar_prop}, the NEik corrections arising from the $z^-$ dependence of $A^-(z)$ are studied in full detail, including their interplay with the non-Eikonal phases mentioned earlier. For simplicity, in this subsection, we will instead present a heuristic derivation allowing us to arrive at the same result in a faster way. 
In the calculation of the NEik corrections arising from the $z^-$ dependence of $A^-(z)$, the non-Eikonal phases are not expected to have a relevant impact. We thus neglect these non-Eikonal phases for now, and we will study them instead in the next subsection. 

After neglecting the non-Eikonal phases in the third line of Eq.~\eqref{expansion_scalar_prop_pure_A_minus_1}, the integrals over $\p_n$ from $n=1$ to $N-1$ become trivial, which leads to
\begin{align}
&\delta G_F(x,y)\bigg|_{\textrm{Eik }+\, z^{-}\textrm{ dep.}}
= \, \int d^{D-2} \z
\sum_{N=1}^{+\infty} \int \bigg[\prod_{n=1}^{N} d z_n^+\: d z_n^-\bigg]\,
\bigg[{\cal P}_n\, \prod_{n=1}^{N}\Big(-i g\, A^{-}(z_n^+,z_n^-,\z)\Big)\bigg]\,
\nonumber\\
&\, 
\times\;
\int \frac{d^{D-1} \underline{q}}{(2\pi)^{D-1}}\:
\frac{1}{2q^+} \,
e^{-i q^+  (x^- -z_N^-)}\,
e^{-i \check{q}^- x^+}\,
e^{i \q \cdot (\x -\z)}\,
\Big[
\theta(x^+\!-\!z_N^+)\, \theta(q^+) - \theta(z_N^+\!-\!x^+)\, \theta(-q^+)
\Big]\,
\nonumber\\
&\, 
\times\;
\int \frac{d^{D-1} \underline{k}}{(2\pi)^{D-1}}\:
\frac{1}{2k^+} \,
e^{-i k^+  (z_1^- -y^-)}\,
e^{i \check{k}^-  y^+}\,
e^{i \k \cdot (\z -\y)}\,
\Big[
\theta(z_{1}^+\!-\!y^+)\, \theta(k^+) - \theta(y^+\!-\!z_{1}^+)\, \theta(-k^+)
\Big]
\nonumber\\
&\, 
\times\;
\int \bigg[\prod_{n=1}^{N-1} \frac{d p_n^+}{2\pi}\:
\frac{1}{2p_n^+} \,
e^{-i p_{n}^+  (z_{n+1}^- -z_n^-)}\,
\Big[
\theta(z_{n+1}^+\!-\!z_n^+)\, \theta(p_n^+) - \theta(z_n^+\!-\!z_{n+1}^+)\, \theta(-p_n^+)
\Big]
\bigg]\,
\bigg[\prod_{n=1}^{N} \big(p_n^+ + p_{n-1}^+  \big)\bigg]
\label{expansion_scalar_prop_pure_A_minus_2}
\, ,
\end{align}
using the notations $p_0\equiv k$ and $p_{N}\equiv q$. Like in Ref.~\cite{Altinoluk:2020oyd}, we use the notation
\begin{align}
\check{p}^- \equiv
& \frac{\p^2+m^2}{2p^+}
\label{def_onshell_p_minus}
\end{align}
for the onshell value of the $-$ component of a momentum $p$, and by extension $\check{p}\equiv (p^+,\p,\check{p}^-)$.

As discussed earlier, the dependence on $z^-$ of $A^{\mu}(z)$ is parametrically  slow in the high-energy limit, so that $A^{\mu}(z)$ is usually taken to be independent of $z^-$ in the Eikonal approximation. In order to go beyond that, we note that the dependence on $z^-$ of each $A^{-}(z)$ insertion is associated, via Fourier transformation, to the difference between two successive $p_n^+$. At high energy, each insertion should then correspond to a parametrically small change of the light-cone momentum along the propagator. It is thus natural to define the typical light-cone momentum
\begin{align}
p^+ \equiv &\, \frac{q^++k^+}{2}
\label{def_p_plus}
\end{align}
for the propagator at high energy.
At each step in Eq.~\eqref{expansion_scalar_prop_pure_A_minus_2}, the momentum $p_n^+$ deviates from this typical value only by a small amount
\begin{align}
\Delta p_n^+ \equiv &\, p_n^+- p^+
\label{def_Delta_p_n_plus}
\, ,
\end{align}
so that
\begin{align}
& p_n^+\sim p^+\sim q^+ \sim k^+  \gg \Delta p_n^+
\label{hyp_p_plus_power_counting}
\, ,
\end{align}
and contributions proportional to the small quantity $(\Delta p_n^+)/p^+$ will be considered of NEik order. Indeed, $(\Delta p_n^+)/p^+$ would identically vanish in the Eikonal approximation, in which $A^{\mu}(z)$ is considered to be independent of $z^-$.

In Eq.~\eqref{expansion_scalar_prop_pure_A_minus_2}, it is thus justified to perform the following Taylor expansion
\begin{align}
\bigg[\prod_{n=0}^{N}
\frac{1}{2p_n^+}\bigg]\:
\bigg[\prod_{n=1}^{N} \big(p_n^+ + p_{n-1}^+  \big)\bigg]
=&\,
\bigg[\prod_{n=0}^{N}
\frac{1}{2(p^++ \Delta p_n^+)}\bigg]\:
\bigg[\prod_{n=1}^{N} \big(2p^+ +\Delta p_n^+ +\Delta p_{n-1}^+\big)\bigg]
\nonumber\\
=&\,
\frac{1}{(2p^+)^{N+1}}\bigg[1-  \sum_{n=0}^{N}\frac{\Delta p_n^+}{p^+} +\dots\bigg]\:
(2p^+)^{N}
\bigg[1 + \sum_{n=1}^{N} \Big(\frac{\Delta p_n^+}{2p^+}
 +\frac{\Delta p_{n-1}^+}{2p^+}\Big)
 +\dots\bigg]
 \nonumber\\
=&\,
\frac{1}{2p^+}\bigg[1- \frac{\Delta p_0^+}{2p^+} - \frac{\Delta p_N^+}{2p^+}+O\left(\left(\frac{\Delta p_n^+}{p^+}\right)^2\right)
\bigg]
 \nonumber\\
=&\,
\frac{1}{(q^++k^+)}\bigg[1+O\left(\left(\frac{\Delta p_n^+}{p^+}\right)^2\right)
\bigg]
\label{p_n_plus_fraction_expansion}
\end{align}
noting that $\Delta p_N^+= (q^+\!-\!k^+)/2= -\Delta p_0^+$.
Moreover, we will assume that the deviations $\Delta p_n^+$ are negligible compared to the typical value $p^+$ within the theta functions, as\footnote{It turns out that corrections with respect to this approximation do not appear at NEik accuracy. Even though it is not a mathematically rigorous argument, an intuitive way to see that is to consider Eq.~\eqref{approx_theta_p_n_plus} as a formal Taylor expansion. At the next order, we would then get $\theta(p^++\Delta p_n^+) \simeq  \theta(p^+) +\Delta p_n^+ \, \delta(p^+)$. The correction is a zero-mode $p^+=0$ in the projectile, which cannot contribute to scattering processes even at NEik accuracy, as already discussed. The  rigorous derivation presented in appendix~\ref{app:NEik_from_zmin_dep_scalar_prop}, which leads to the same final result, does not require to perform such Taylor expansion of a $\theta$ function.  }
\begin{align}
\theta(p_n^+) = \theta(p^++\Delta p_n^+) \simeq  \theta(p^+)
\label{approx_theta_p_n_plus}
\end{align}
for example.
Within these approximations, Eq.~\eqref{expansion_scalar_prop_pure_A_minus_2} becomes
\begin{align}
\delta G_F(x,y)\bigg|_{\textrm{Eik }+\, z^{-}\textrm{ dep.}}
=& \, \int d^{D-2} \z
\sum_{N=1}^{+\infty} \int \bigg[\prod_{n=1}^{N} d z_n^+\: d z_n^-\bigg]\,
\bigg[{\cal P}_n\, \prod_{n=1}^{N}\Big(-i g\, A^{-}(z_n^+,z_n^-,\z)\Big)\bigg]\,
\nonumber\\
&\, 
\times\;
\int \frac{d^{D-1} \underline{q}}{(2\pi)^{D-1}}\:
e^{-i q^+  (x^- -z_N^-)}\,
e^{-i \check{q}^- x^+}\,
e^{i \q \cdot (\x -\z)}\,
\nonumber\\
&\, 
\times\;
\int \frac{d^{D-1} \underline{k}}{(2\pi)^{D-1}}\:
e^{-i k^+  (z_1^- -y^-)}\,
e^{i \check{k}^-  y^+}\,
e^{i \k \cdot (\z -\y)}\,
\frac{1}{(q^++k^+)}\:
\nonumber\\
&\, 
\times\;
\bigg[\theta(q^+)\, \theta(k^+) \prod_{n=0}^{N} \theta(z_{n+1}^+\!-\!z_n^+)
+(-1)^{N+1}\theta(-q^+)\, \theta(-k^+) \prod_{n=0}^{N} \theta(z_n^+\!-\!z_{n+1}^+)
\bigg]\,
\nonumber\\
&\, 
\times\;
\int \bigg[\prod_{n=1}^{N-1} \frac{d \Delta p_n^+}{2\pi}\:
e^{-i (p^++\Delta p_{n}^+)  (z_{n+1}^- -z_n^-)}\,
\bigg]\,
\bigg[
1+O\left(\left(\frac{\Delta p_n^+}{p^+}\right)^2\right)\bigg]
\label{expansion_scalar_prop_pure_A_minus_3}
\, .
\end{align}
The integrations over $\Delta p_{n}^+$ result in $\delta(z_{n+1}^- -z_n^-)$, so that
\begin{align}
\delta G_F(x,y)\bigg|_{\textrm{Eik }+\, z^{-}\textrm{ dep.}}
=& \,
\int \frac{d^{D-1} \underline{q}}{(2\pi)^{D-1}}\: e^{-i \check{q}\cdot x}\,
\int \frac{d^{D-1} \underline{k}}{(2\pi)^{D-1}}\: e^{i \check{k}\cdot y}\,
\frac{1}{(q^++k^+)}\:
\int d^{D-2} \z\: e^{-i \z \cdot (\q -\k)}\,
\int d z^-\: e^{i z^-   (q^+ -k^+)}\,
\nonumber\\
&\, 
\times\;
\sum_{N=1}^{+\infty} \int \bigg[\prod_{n=1}^{N} d z_n^+\bigg]\,
\bigg[{\cal P}_n\, \prod_{n=1}^{N}\Big(-i g\, A^{-}(z_n^+,z^-,\z)\Big)\bigg]\,
\nonumber\\
&\, 
\times\;
\bigg[\theta(q^+)\, \theta(k^+) \prod_{n=0}^{N} \theta(z_{n+1}^+\!-\!z_n^+)
+(-1)^{N+1}\theta(-q^+)\, \theta(-k^+) \prod_{n=0}^{N} \theta(z_n^+\!-\!z_{n+1}^+)
\bigg]\,
+\textrm{NNEik}
\nonumber\\
=& \,
\int \frac{d^{D-1} \underline{q}}{(2\pi)^{D-1}}\: e^{-i \check{q}\cdot x}\,
\int \frac{d^{D-1} \underline{k}}{(2\pi)^{D-1}}\: e^{i \check{k}\cdot y}\,
\frac{1}{(q^++k^+)}\:
\int d^{D-2} \z\: e^{-i \z \cdot (\q -\k)}\,
\int d z^-\: e^{i z^-   (q^+ -k^+)}\,
\nonumber\\
&\, 
\times\;
\bigg\{\theta(q^+)\, \theta(k^+)\, \theta(x^+\!-\!y^+)\,  \Big[{\cal U}_F(x^+,y^+;\z,z^-)-\mathbf{1}\Big]
\nonumber\\
&\, \hspace{2cm}
-\theta(-q^+)\, \theta(-k^+)\, \theta(y^+\!-\!x^+)\,  \Big[{\cal U}_F(y^+,x^+;\z,z^-)^{\dag}-\mathbf{1}\Big]\bigg\}
+\textrm{NNEik}
\label{expansion_scalar_prop_pure_A_minus_4}
\, ,
\end{align}
where we have introduced the Wilson line along the $x^+$ direction, defined as\footnote{${\cal P}_{+}$ denotes the ordering of the $A^-(z)$ as color matrices along the $x^+$ direction.}
\begin{align}
{\cal U}_F(x^+,y^+;\z,z^-)
\equiv\, {\cal P}_{+} \exp\bigg\{-ig \int_{y^+}^{x^+} dz^+\; A^-(z)\bigg\}
\label{def_Wilson_line_along_plus}
\, .
\end{align}

Adding the vacuum propagator contribution to Eq.~\eqref{expansion_scalar_prop_pure_A_minus_4}, one obtains the following expression for the scalar propagator in a pure $A^{-}(z)$ background, including $z^-$ dependence effects at NEik accuracy but no finite $z^+$ width effects yet:
\begin{align}
G_F(x,y)\bigg|_{\textrm{Eik }+\, z^{-}\textrm{ dep.}}
=& \,
\int \frac{d^{D-1} \underline{q}}{(2\pi)^{D-1}}\: e^{-i \check{q}\cdot x}\,
\int \frac{d^{D-1} \underline{k}}{(2\pi)^{D-1}}\: e^{i \check{k}\cdot y}\,
\frac{1}{(q^++k^+)}\:
\int d^{D-2} \z\: e^{-i \z \cdot (\q -\k)}\,
\int d z^-\: e^{i z^-   (q^+ -k^+)}\,
\nonumber\\
&\, 
\times\;
\bigg\{\theta(q^+)\, \theta(k^+)\, \theta(x^+\!-\!y^+)\,  {\cal U}_F(x^+,y^+;\z,z^-)
\nonumber\\
&\, \hspace{2cm}
-\theta(-q^+)\, \theta(-k^+)\, \theta(y^+\!-\!x^+)\,  {\cal U}_F(y^+,x^+;\z,z^-)^{\dag}\bigg\}
+\textrm{NNEik}
\label{expansion_scalar_prop_pure_A_minus_5}
\, .
\end{align}

%
The $z^-$ dependence of the Wilson lines in Eq.~\eqref{expansion_scalar_prop_pure_A_minus_5} goes beyond the strict Eikonal approximation usually performed in high-energy QCD studies, in particular in the context of gluon saturation.
In order to obtain the expression in Eq.~\eqref{expansion_scalar_prop_pure_A_minus_5}, we have neglected terms of order $(q^+\!-\!k^+)^2/(q^+\!+\!k^+)^2$, which scale as $1/(\gamma_T)^2$ under a large boost of the target (see Eq.~\eqref{qplus_min_kplus_as_power_suppressed}), and are thus of NNEik order. Hence, any possible NEik effect due to the $z^-$ dependence of the background field is still present in Eq.~\eqref{expansion_scalar_prop_pure_A_minus_5}, encoded by the $z^-$ dependence of the Wilson lines.
If one neglects this dependence of the Wilson lines on $z^-$ in Eq.~\eqref{expansion_scalar_prop_pure_A_minus_5}, one recovers the standard Eikonal expression for the scalar propagator upon integration over $z^-$.
Since the expression in Eq.~\eqref{expansion_scalar_prop_pure_A_minus_5} has the form of the standard Eikonal one, but with $z^-$ dependent Wilson lines encoding non-Eikonal effects from the dynamics of the target, we refer to it as the Generalized Eikonal approximation for the scalar propagator.

As a remark, the explicit NEik corrections associated with $z^-$ dependence have cancelled out in Eq.~\eqref{p_n_plus_fraction_expansion}, provided the denominator is written as $(q^++k^+)$. Indeed, we could instead take $(2q^+)$ or $(2k^+)$ as the denominator in Eq.~\eqref{expansion_scalar_prop_pure_A_minus_5} but in such cases we would obtain a non-zero explicit NEik correction on top of the generalized Eikonal term.

For completeness, we now have to understand more systematically the difference between the Generalized Eikonal expression \eqref{expansion_scalar_prop_pure_A_minus_5} and the strict Eikonal expression usually considered in the literature. For that purpose, let us remember that the $z^-$ dependence of the Wilson lines is parametrically slow due to Lorentz time dilation of the target in the case of a large Lorentz boost of the target. It is thus meaningful to perform a Taylor expansion of the Wilson lines around $z^-=0$. Under the $z^-$ integral, this expansion gives
\begin{align}
\int d z^-\: e^{i z^-   (q^+ -k^+)}\, {\cal U}_F(x^+,y^+;\z,z^-)
=& \, \int d z^-\: e^{i z^-   (q^+ -k^+)}\,
\Big\{
{\cal U}_F(x^+,y^+;\z,0) + z^-\; \partial_-{\cal U}_F(x^+,y^+;\z,0) + \textrm{NNEik}
\Big\}
\nonumber\\
=& \, 2\pi\delta(q^+\!-\!k^+)\; {\cal U}_F(x^+,y^+;\z,0)\; 
\nonumber\\
&  - 2\pi\, i\, \delta'(q^+\!-\!k^+)\; \big[\partial_-{\cal U}_F(x^+,y^+;\z,0)\big]\;
+ \textrm{NNEik}
\label{expansion_Wilson_line_zminus_equal_0}
\, .
\end{align}
Inserting the expansion \eqref{expansion_Wilson_line_zminus_equal_0} into the Generalized Eikonal expression \eqref{expansion_scalar_prop_pure_A_minus_5}, one can integrate by parts with respect to  ${q^+}$ and/or ${k^+}$, before using $\delta(q^+\!-\!k^+)$ to remove one of the integrations (in $q^+$ or in $k^+$). In such a way, one obtains, in the case $x^+>y^+$,
\begin{align}
G_F(x,y)\bigg|_{\textrm{Eik }+\, z^{-}\textrm{ dep.}}
=& \,
\int \frac{d^{D-2} \q}{(2\pi)^{D-2}}\:  e^{i \q\cdot \x}\,
\int \frac{d^{D-2} \k}{(2\pi)^{D-2}}\:  e^{-i \k\cdot \y}\,
\int \frac{d p^+}{2\pi}\:
\frac{\theta(p^+)}{2p^+}\:
e^{-i p^+(x^- -y^-)}\:
e^{-i \frac{(\q^2+m^2)}{2p^+}\, x^+}\:
e^{i \frac{(\k^2+m^2)}{2p^+}\, y^+}\:
\nonumber\\
&\, 
\times\:
\int d^{D-2} \z\: e^{-i \z \cdot (\q -\k)}\;
\bigg\{
{\cal U}_F(x^+,y^+;\z,0)
\nonumber\\
&\, 
\hspace{1cm}
+ \left[
   \frac{x^-\!+\!y^-}{2} -\frac{(\q^2+m^2)}{(2p^+)^2}\, x^+ -\frac{(\k^2+m^2)}{(2p^+)^2}\, y^+
 \right]
 \Big[\partial_-{\cal U}_F(x^+,y^+;\z,0)\Big]
\bigg\}
+\textrm{NNEik}
\label{expansion_scalar_prop_pure_A_minus_6}
\, .
\end{align}
In this expansion of the Generalized Eikonal scalar propagator \eqref{expansion_scalar_prop_pure_A_minus_5}, the first term corresponds to the usual strict Eikonal term, whereas the last line corresponds to an extra NEik correction. The $\partial_-$ derivative of the Wilson line indeed provides a suppression as $1/\gamma_T$ in the limit of large Lorentz boost of the target.

The presence of the external coordinates $x$ and $y$ in the numerator in the coefficient of the NEik term in Eq.~\eqref{expansion_scalar_prop_pure_A_minus_6} makes that expression cumbersome to use in the calculation of scattering observables beyond Eikonal accuracy. Our general recommendation is thus to use in such calculations an expansion for the propagators around the Generalized Eikonal approximation \eqref{expansion_scalar_prop_pure_A_minus_5}. Then, it would still be possible to expand the Wilson lines around $z^-=0$ in the result at the cross section level using Eq.~\eqref{expansion_Wilson_line_zminus_equal_0}, in order to isolate the strictly Eikonal result from any non-Eikonal effect. For that reason, in the rest of the present paper, we will always use the Generalized Eikonal approximation as the leading power contribution at high energy, and calculate NEik corrections on top of it. But one should remember that part of the non-Eikonal contributions associated with the $z^-$ dependence of the target are resummed (or more precisely non-expanded) into the Generalized Eikonal contribution.


\subsection{Scalar propagator in a pure $A^{-}(z)$ background : finite $z^+$ width effects}

After having discussed the non-Eikonal effects associated with the $z^-$ dependence for the scalar propagator in a pure $A^{-}(z)$ background, we have to study the non-Eikonal effects associated with the finite width of the target in $z^+$. For simplicity we will focus on the case $x^+>y^+$, starting from the exact expression~\eqref{expansion_scalar_prop_pure_A_minus_1} for the propagator in a pure $A^{-}(z)$ background. Keeping this time the $z_n^+$ dependent non-Eikonal phase, we can expand the expression~\eqref{expansion_scalar_prop_pure_A_minus_1} in the regime defined in Eq.~\eqref{hyp_p_plus_power_counting}, and integrate over $\Delta  p_n^+$. Alternatively, one can use Eq.~\eqref{expansion_scalar_prop_pure_A_minus_4_app}, derived in appendix~\ref{app:NEik_from_zmin_dep_scalar_prop}. 
 In either way, one obtains
\begin{align}
\delta G_F(x,y)\bigg|_{\textrm{pure }A^{-}}
= &\, \int \frac{d^{D-1} \underline{q}}{(2\pi)^{D-1}}\:
\int \frac{d^{D-1} \underline{k}}{(2\pi)^{D-1}}\:
\frac{\theta(q^+)\theta(k^+)}{(q^++k^+)}\: 
\int dz^- e^{-i q^+ (x^--z^-)}\, e^{-i k^+ (z^--y^-)}\,
\nonumber\\
&\, \times \; 
\sum_{N=1}^{+\infty} \int 
\bigg[\prod_{n=1}^{N} d^{D-1} \underline{z_n}\bigg]\,
\bigg[{\cal P}_n\, \prod_{n=1}^{N}\Big(-i g\, A^{-}(\underline{z_n},z^-)\Big)\bigg]\;
\bigg[\prod_{n=0}^{N}
\theta(z_{n+1}^+\!-\!z_n^+)
\bigg]
\nonumber\\
&\, 
\times\;
e^{i \q\cdot (\x-\z_N)}\; e^{i \k\cdot (\z_1-\y)}\;
e^{-i\, \frac{(\q^2+m^2)}{2q^+}\, (x^+-z_N^+)}\; 
e^{-i\, \frac{(\k^2+m^2)}{2k^+}\, (z_{1}^+-y^+)}\,
\nonumber\\
&\, 
\,\times\;
\int\bigg[\prod_{n=1}^{N-1}
\frac{d^{D-2} \p_n}{(2\pi)^{D-2}}\:
e^{i \p_{n} \cdot (\z_{n+1}-\z_n)}\,
e^{-i\, \frac{(\p_n^2+m^2)}{2p^+}\, (z_{n+1}^+-z_n^+)}
\bigg]
+\textrm{NNEik}
\label{expansion_scalar_prop_pure_A_minus_brownian_1}
\\
= &\, \int \frac{d^{D-1} \underline{q}}{(2\pi)^{D-1}}\:
\int \frac{d^{D-1} \underline{k}}{(2\pi)^{D-1}}\:
\frac{\theta(q^+)\theta(k^+)}{(q^+\!+\!k^+)}\: e^{-i \check{q}\cdot x}\, e^{i \check{k}\cdot y}\,
\int dz^-\, e^{i z^-   (q^+ -k^+)}\,
\nonumber\\
&\, \times \; 
\sum_{N=1}^{+\infty} \int \bigg[\prod_{n=1}^{N} d^{D-1} \underline{z_n}\bigg]\,
\bigg[{\cal P}_n\, \prod_{n=1}^{N}\Big(-i g\, A^{-}(\underline{z_n},z^-)\Big)\bigg]\;
\bigg[\prod_{n=0}^{N}
\theta(z_{n+1}^+\!-\!z_n^+)
\bigg]
\nonumber\\
&\, 
\times\;
e^{-i \q\cdot \z_N}\, e^{i \check{q}^- z_N^+}\,
e^{i \k\cdot \z_1}\, e^{-i \check{k}^- z_1^+}\,
\nonumber\\
&\, 
\,\times\;
e^{-i\, \frac{m^2}{2p^+}\, (z_{N}^+-z_1^+)}
\Bigg[\prod_{n=1}^{N-1}
\bigg(\frac{p^+}{2\pi i (z_{n+1}^+\!-\!z_n^+)}\bigg)^{\frac{D}{2}-1}
\:
e^{\frac{i  p^+ (\z_{n+1}-\z_n)^2}{2(z_{n+1}^+-z_n^+)}}
\Bigg]
+\textrm{NNEik}
\label{expansion_scalar_prop_pure_A_minus_brownian_2}
\, .
\end{align}
In this expression, the last three lines are identical to the last three lines
of Eq.~(A.1) from Ref.~\cite{Altinoluk:2020oyd} in the quark propagator case (once restricted to $x^+>y^+$). The derivation performed in the Appendix A of Ref.~\cite{Altinoluk:2020oyd} is thus valid for Eq.~\eqref{expansion_scalar_prop_pure_A_minus_brownian_2}, and by comparison with Eq.~(A.14) from Ref.~\cite{Altinoluk:2020oyd}, one obtains
\begin{align}
&\delta G_F(x,y)\bigg|_{\textrm{pure }A^{-}}
=  \int \frac{d^{D-1} \underline{q}}{(2\pi)^{D-1}}\:
\int \frac{d^{D-1} \underline{k}}{(2\pi)^{D-1}}\:
\frac{\theta(q^+)\theta(k^+)}{(q^+\!+\!k^+)}\: e^{-i \check{q}\cdot x}\, e^{i \check{k}\cdot y}\,
\int dz^-\, e^{i z^-   (q^+ -k^+)}\,
\int d^{D-2}\z\,\, e^{-i \z \cdot (\q -\k)}\,
\nonumber\\
&\, \times \; 
\Bigg\{\Big[{\cal U}_F(x^+,y^+;\z,z^-)-\mathbf{1}\Big]
+\frac{(\q^j+\k^j)}{(q^+\!+\!k^+)}
\int_{y^+}^{x^+}\! dv^+\, v^+\,
{\cal U}_F(x^+,v^+;\z,z^-)  \big[-ig\, \d_j A^{-}(v^+,z^-,\z)\big]
{\cal U}_F(v^+,y^+;\z,z^-)
\nonumber\\
&\, \hspace{0.8cm}
-\frac{i}{(q^+\!+\!k^+)}
\int_{y^+}^{x^+}\! dv^+\int_{v^+}^{x^+}\! dw^+\, (w^+\!-\!v^+)\,
{\cal U}_F(x^+,w^+;\z,z^-)  \big[-ig\, \d_j A^{-}(w^+,z^-,\z)\big]
{\cal U}_F(w^+,v^+;\z,z^-)
\nonumber\\
&\, \hspace{2cm} \times \;
 \big[-ig\, \d_j A^{-}(v^+,z^-,\z)\big]
{\cal U}_F(v^+,y^+;\z,z^-)
\Bigg\}
+\textrm{NNEik}
\label{expansion_scalar_prop_pure_A_minus_brownian_3}
\, .
\end{align}

Restricting ourselves to the case of $x^+>L^+/2$ and $y^+<-L^+/2$ (corresponding to propagation from before to after the target), we can use Eqs.~(A.18) and (A.20) from Ref.~\cite{Altinoluk:2020oyd} to write (including the vacuum contribution)
\begin{align}
&G_F(x,y)\bigg|_{\textrm{pure }A^{-}}
=  \int \frac{d^{D-1} \underline{q}}{(2\pi)^{D-1}}\:
\int \frac{d^{D-1} \underline{k}}{(2\pi)^{D-1}}\:
\frac{\theta(q^+)\theta(k^+)}{(q^+\!+\!k^+)}\: e^{-i \check{q}\cdot x}\, e^{i \check{k}\cdot y}\,
\int dz^-\, e^{i z^-   (q^+ -k^+)}\,
\int d^{D-2}\z\,\, e^{-i \z \cdot (\q -\k)}\,
\nonumber\\
&\, \times \; 
\Bigg\{{\cal U}_F\left(\frac{L^+}{2},-\frac{L^+}{2};\z,z^-\right)
-\frac{(\q^j+\k^j)}{2(q^+\!+\!k^+)}
\int_{-\frac{L^+}{2}}^{\frac{L^+}{2}}\! dz^+\,
\left[{\cal U}_F\left(\frac{L^+}{2},z^+;\z,z^-\right)   \overleftrightarrow{\d}_{\z^j}
{\cal U}_F\left(z^+,-\frac{L^+}{2};\z,z^-\right)\right]
\nonumber\\
&\, \hspace{0.8cm}
-\frac{i}{(q^+\!+\!k^+)}
\int_{-\frac{L^+}{2}}^{\frac{L^+}{2}}\! dz^+\,
\left[{\cal U}_F\left(\frac{L^+}{2},z^+;\z,z^-\right)   \overleftarrow{\d}_{\z^j}\overrightarrow{\d}_{\z^j}
{\cal U}_F\left(z^+,-\frac{L^+}{2};\z,z^-\right)\right]
\Bigg\}
+\textrm{NNEik}
\label{expansion_scalar_prop_pure_A_minus_brownian_4}
\, .
\end{align}
Note that in Eq.~\eqref{expansion_scalar_prop_pure_A_minus_brownian_4}, the transverse derivatives acts only within the square brackets, and not on the phase factors.


\subsection{NEik corrections from transverse $A^{j}(z)$ field components\label{sec:A_j_insert_scalar}}

With the scalar propagator in a pure $A^{-}(z)$ background field~\eqref{expansion_scalar_prop_pure_A_minus_brownian_4} known at NEik accuracy, it is time to consider the NEik contributions from the other components of the background field, starting with the transverse components $A^{j}(z)$. At NEik accuracy, the relevant contribution from $A^{j}(z)$ corresponds to a
single insertion of $A^{j}(z)$ or an instantaneous double $A^{j}(z)$ insertion, into the scalar propagator in pure $A^{-}(z)$ field. This contribution stems from the second bracket in Eqs.~\eqref{sol_scalar_prop_L_3} or \eqref{sol_scalar_prop_L_4}, and is defined as
\begin{align}
\delta G_F(x,y)\bigg|_{A^{j}}\, =  &\, \int d^Dz\,\, G_{F}(x,z)\bigg|_{\textrm{Pure }A^{-}}\,
\Big[-g\, A_{j}(z)\, \overrightarrow{\d}_{z^{j}}
+ \overleftarrow{\d}_{z^{j}}\,  g\, A_{j}(z)
-ig^2\, A_{j}(z)\, A_{j}(z)
\Big]
\, G_F(z,y)\bigg|_{\textrm{Pure }A^{-}}
\nonumber\\
=  &\, \int d^Dz\,\, G_{F}(x,z)\bigg|_{\textrm{Pure }A^{-}}\,
\Big[-i\, \overleftarrow{D}_{z^{j}}\, \overrightarrow{D}_{z^{j}}
+i\, \overleftarrow{\d}_{z^{j}}\, \overrightarrow{\d}_{z^{j}}
\Big]
\, G_F(z,y)\bigg|_{\textrm{Pure }A^{-}}
\label{single_A_perp_scalar_prop_1}
\, .
\end{align}
Due to the integration over $z^+$, this contribution starts only at NEik accuracy. It is thus sufficient to insert in Eq.~\eqref{single_A_perp_scalar_prop_1} the Generalized Eikonal expression~\eqref{expansion_scalar_prop_pure_A_minus_5} for the scalar propagators in pure $A^{-}(z)$ field. Focusing again on the case $x^+>y^+$ for simplicity, one thus finds
\begin{align}
&\, \delta G_F(x,y)\bigg|_{A^{j}}\,
=   \int d^Dz\,\,
\int \frac{d^{D-1} \underline{q}}{(2\pi)^{D-1}}\: e^{-i \check{q}\cdot x}\, \theta(q^+)
\int \frac{d^{D-1} \underline{k}}{(2\pi)^{D-1}}\: e^{i \check{k}\cdot y}\, \theta(k^+)
\int \frac{d^{D-1} \underline{p_1}}{(2\pi)^{D-1}}\:  \theta(p_1^+)
\int \frac{d^{D-1} \underline{p_2}}{(2\pi)^{D-1}}\: \theta(p_2^+)
\nonumber\\
&\, \times\;
\frac{1}{(q^+ + p_2^+)}\: \frac{1}{(p_1^+ + k^+)}\:
\int d^{D-2} \v\: e^{-i \v \cdot (\p_1 -\k)}\,
\int d v^-\: e^{i v^-   (p_1^+ -k^+)}\,
\int d^{D-2} \w\: e^{-i \w \cdot (\q -\p_2)}\,
\int d w^-\: e^{i w^-   (q^+ -p_2^+)}\,
\nonumber\\
&\, \times\;
\bigg[\theta(x^+\!-\!z^+)\,  e^{i \check{p}_2\cdot z}\,   {\cal U}_F(x^+,z^+;\w,w^-)\bigg]
\Big[-g\, A_{j}(z)\, \overrightarrow{\d}_{z^{j}}
+ \overleftarrow{\d}_{z^{j}}\,  g\, A_{j}(z)
-ig^2\, A_{j}(z)\, A_{j}(z)
\Big]
\nonumber\\
&\, \times\;
\bigg[\theta(z^+\!-\!y^+)\,  e^{-i \check{p}_1\cdot z}\,   {\cal U}_F(z^+,y^+;\v,v^-)\bigg]
+\textrm{NNEik}
\label{single_A_perp_scalar_prop_2}
\, .
\end{align}
On the one hand, the $z^+$ dependent phases in Eq.~\eqref{single_A_perp_scalar_prop_2} would contribute only at NNEik accuracy, and can thus be dropped. Then, the $\p_1$ and $\p_2$ transverse integrations become trivial, giving delta functions forcing $\w=\v=\z$.
On the other hand, we can Taylor expand the Wilson lines in Eq.~\eqref{single_A_perp_scalar_prop_2} around $w^-=z^-$ and $v^-=z^-$ respectively. Higher order terms in these Taylor expansions, suppressed by higher derivatives in $z^-$, would only contribute to the scalar propagators at NNEik accuracy and beyond. Hence, it is sufficient to keep only the leading terms in these Taylor expansions, which amounts to replace  $w^-$ and $v^-$
by $z^-$ as arguments of the Wilson lines. The integrations over $w^-$ and $v^-$ then become trivial, giving delta functions forcing $p_1^+=k^+$ and $p_2^+=q^+$.
All in all, one finds
\begin{align}
 \delta G_F(x,y)\bigg|_{A^{j}}\,
= &\,
\int \frac{d^{D-1} \underline{q}}{(2\pi)^{D-1}}\: e^{-i \check{q}\cdot x}\, \frac{\theta(q^+)}{(2q^+)}
\int \frac{d^{D-1} \underline{k}}{(2\pi)^{D-1}}\: e^{i \check{k}\cdot y}\, \frac{\theta(k^+)}{(2k^+)}
\int_{y^+}^{x^+} dz^+\,
\int dz^-\,\, e^{i z^-(q^+ -k^+)}\,
\int d^{D-2}\z\,\,
\nonumber\\
&\, \times\;
\Big[e^{-i \z \cdot \q }\, {\cal U}_F(x^+,z^+;\z,z^-)\Big]
\Big[-g\, A_{j}(z)\, \overrightarrow{\d}_{z^{j}}
+ \overleftarrow{\d}_{z^{j}}\,  g\, A_{j}(z)
-ig^2\, A_{j}(z)\, A_{j}(z)
\Big]
\nonumber\\
&\, \times\;
\Big[e^{i \z \cdot\k}\,{\cal U}_F(z^+,y^+;\z,z^-)\Big]
+\textrm{NNEik}
\label{single_A_perp_scalar_prop_5}
\end{align}
In Eq.~\eqref{single_A_perp_scalar_prop_5}, each transverse derivatives acts both on a Wilson line and on a phase factor. In
order to combine Eq.~\eqref{single_A_perp_scalar_prop_5} with Eq.~\eqref{expansion_scalar_prop_pure_A_minus_brownian_4}, it is convenient to work out the action of the derivatives on the phase factors as
\begin{align}
 \delta G_F(x,y)\bigg|_{A^{j}}\,
= &\,
\int \frac{d^{D-1} \underline{q}}{(2\pi)^{D-1}}\: e^{-i \check{q}\cdot x}\, \frac{\theta(q^+)}{(2q^+)}
\int \frac{d^{D-1} \underline{k}}{(2\pi)^{D-1}}\: e^{i \check{k}\cdot y}\, \frac{\theta(k^+)}{(2k^+)}
\int_{y^+}^{x^+} dz^+\,
\int dz^-\,\, e^{i z^-(q^+ -k^+)}\,
\int d^{D-2}\z\,\, e^{-i \z \cdot (\q-\k) }\,
\nonumber\\
&\, \times\;
\bigg[ {\cal U}_F(x^+,z^+;\z,z^-)
\Big(-g\, A_{j}(z)\, (\overrightarrow{\d}_{z^{j}}+i\k^j)
+ (\overleftarrow{\d}_{z^{j}}-i\q^j)\,  g\, A_{j}(z)
-ig^2\, A_{j}(z)\, A_{j}(z)
\Big)
{\cal U}_F(z^+,y^+;\z,z^-)\bigg]
\nonumber\\
&\,
+\textrm{NNEik}
\label{single_A_perp_scalar_prop_6}
\\
= &\,
\int \frac{d^{D-1} \underline{q}}{(2\pi)^{D-1}}\: e^{-i \check{q}\cdot x}\, \frac{\theta(q^+)}{(2q^+)}
\int \frac{d^{D-1} \underline{k}}{(2\pi)^{D-1}}\: e^{i \check{k}\cdot y}\, \frac{\theta(k^+)}{(2k^+)}
\int_{y^+}^{x^+} dz^+\,
\int dz^-\,\, e^{i z^-(q^+ -k^+)}\,
\int d^{D-2}\z\,\, e^{-i \z \cdot (\q-\k) }\,
\nonumber\\
&\, \times\;
\bigg[ {\cal U}_F(x^+,z^+;\z,z^-)
\Big(-i\, \overleftarrow{D}_{z^{j}}\, \overrightarrow{D}_{z^{j}}
+i\, \overleftarrow{\d}_{z^{j}}\, \overrightarrow{\d}_{z^{j}}
-\frac{(\q^j\!+\!\k^j)}{2}\, (\overleftrightarrow{D}_{z^{j}}-\overleftrightarrow{\d}_{z^{j}})
\Big)
{\cal U}_F(z^+,y^+;\z,z^-)\bigg]
+\textrm{NNEik}
\label{single_A_perp_scalar_prop_7}
\, ,
\end{align}
where the derivatives (covariant or not) act only on the Wilson lines, within the square bracket. To the accuracy of interest, in the regime defined in Eq.~\eqref{hyp_p_plus_power_counting}, the denominators $(2q^+)$ or $(2k^+)$ are equivalent to $(q^++k^+)$. Eq.~\eqref{single_A_perp_scalar_prop_7} is thus equivalent to
\begin{align}
 \delta & G_F(x,y)\bigg|_{A^{j}}\,
= \,
\int \frac{d^{D-1} \underline{q}}{(2\pi)^{D-1}}\: e^{-i \check{q}\cdot x}\, \theta(q^+)
\int \frac{d^{D-1} \underline{k}}{(2\pi)^{D-1}}\: e^{i \check{k}\cdot y}\, \theta(k^+)
\int_{y^+}^{x^+} dz^+\,
\int dz^-\,\, e^{i z^-(q^+ -k^+)}\,
\int d^{D-2}\z\,\, e^{-i \z \cdot (\q-\k) }\,
\nonumber\\
&\, \times\; \frac{1}{(q^++k^+)^2}
\bigg[ {\cal U}_F(x^+,z^+;\z,z^-)
\Big(-i\, \overleftarrow{D}_{z^{j}}\, \overrightarrow{D}_{z^{j}}
+i\, \overleftarrow{\d}_{z^{j}}\, \overrightarrow{\d}_{z^{j}}
-\frac{(\q^j\!+\!\k^j)}{2}\, (\overleftrightarrow{D}_{z^{j}}-\overleftrightarrow{\d}_{z^{j}})
\Big)
{\cal U}_F(z^+,y^+;\z,z^-)\bigg]
+\textrm{NNEik}
\label{single_A_perp_scalar_prop_8}
\, .
\end{align}


\subsection{Contribution from the $A^{+}(x)$ field component\label{sec:A_plus_insert_scalar}}

Finally, it remains to study the NEik contribution to the scalar propagator from single $A^{+}(x)$ insertion, corresponding to the third bracket in Eq.~\eqref{sol_scalar_prop_L_4}. It is defined as
\begin{align}
\delta G_F(x,y)\bigg|_{\textrm{Single }A^{+}}\, =  &\, \int d^Dz\,\, G_{F}(x,z)\bigg|_{\textrm{Pure }A^{-}}\,
\Big[
g\, A^{+}(z)\, \overrightarrow{D}_{z^{+}}
- \overleftarrow{D}_{z^{+}}\,  g\, A^{+}(z)
\Big]
\, G_F(z,y)\bigg|_{\textrm{Pure }A^{-}}
\label{single_A_plus_scalar_prop_1}
\, .
\end{align}
Again, it is sufficient to use the Generalized Eikonal expression~\eqref{expansion_scalar_prop_pure_A_minus_5} for the scalar propagators in pure $A^{-}(z)$ background in Eq.~\eqref{single_A_plus_scalar_prop_1}, in order to reach NEik precision. Focusing on the case $x^+>y^+$, one finds
\begin{align}
&\, \delta G_F(x,y)\bigg|_{\textrm{Single }A^{+}}\,
=   \int d^Dz\,\,
\int \frac{d^{D-1} \underline{q}}{(2\pi)^{D-1}}\: e^{-i \check{q}\cdot x}\, \theta(q^+)
\int \frac{d^{D-1} \underline{k}}{(2\pi)^{D-1}}\: e^{i \check{k}\cdot y}\, \theta(k^+)
\int \frac{d^{D-1} \underline{p_1}}{(2\pi)^{D-1}}\:  \theta(p_1^+)
\int \frac{d^{D-1} \underline{p_2}}{(2\pi)^{D-1}}\: \theta(p_2^+)
\nonumber\\
&\, \times\;
\frac{1}{(q^+ + p_2^+)}\: \frac{1}{(p_1^+ + k^+)}\:
\int d^{D-2} \v\: e^{-i \v \cdot (\p_1 -\k)}\,
\int d v^-\: e^{i v^-   (p_1^+ -k^+)}\,
\int d^{D-2} \w\: e^{-i \w \cdot (\q -\p_2)}\,
\int d w^-\: e^{i w^-   (q^+ -p_2^+)}\,
\nonumber\\
&\, \times\;
\bigg[\theta(x^+\!-\!z^+)\,  e^{i \check{p}_2\cdot z}\,   {\cal U}_F(x^+,z^+;\w,w^-)\bigg]
\Big[
g\, A^{+}(z)\, \overrightarrow{D}_{z^{+}}
- \overleftarrow{D}_{z^{+}}\,  g\, A^{+}(z)
\Big]
\nonumber\\
&\, \times\;
\bigg[\theta(z^+\!-\!y^+)\,  e^{-i \check{p}_1\cdot z}\,   {\cal U}_F(z^+,y^+;\v,v^-)\bigg]
+\textrm{NNEik}
\label{single_A_plus_scalar_prop_2}
\, .
\end{align}
In Eq.~\eqref{single_A_plus_scalar_prop_2}, each covariant derivative acts on the product of a Wilson line, a phase factor and a theta function. By definition a covariant derivative acting in such a way on a Wilson line gives zero. Moreover, the action of $\partial_{z^+}$ on the $z^+$ dependent phase factor will produce a contribution to Eq.~\eqref{single_A_plus_scalar_prop_2} only at NNEik accuracy.
We are thus left with the action of the derivative on the theta functions, as
\begin{align}
&\, \delta G_F(x,y)\bigg|_{\textrm{Single }A^{+}}\,
=   \int d^Dz\,\,
\int \frac{d^{D-1} \underline{q}}{(2\pi)^{D-1}}\: e^{-i \check{q}\cdot x}\, \theta(q^+)
\int \frac{d^{D-1} \underline{k}}{(2\pi)^{D-1}}\: e^{i \check{k}\cdot y}\, \theta(k^+)
\int \frac{d^{D-1} \underline{p_1}}{(2\pi)^{D-1}}\:  \theta(p_1^+)
\int \frac{d^{D-1} \underline{p_2}}{(2\pi)^{D-1}}\: \theta(p_2^+)
\nonumber\\
&\, \times\;
\frac{1}{(q^+ + p_2^+)}\: \frac{1}{(p_1^+ + k^+)}\:
\int d^{D-2} \v\: e^{-i \v \cdot (\p_1 -\k)}\,
\int d v^-\: e^{i v^-   (p_1^+ -k^+)}\,
\int d^{D-2} \w\: e^{-i \w \cdot (\q -\p_2)}\,
\int d w^-\: e^{i w^-   (q^+ -p_2^+)}\,
e^{i (\check{p}_2-\check{p}_1)\cdot z}
\nonumber\\
&\, \times\;
\bigg[\theta(x^+\!-\!z^+)\, \delta(z^+\!-\!y^+) + \delta(x^+\!-\!z^+)\, \theta(z^+\!-\!y^+)    \bigg]
{\cal U}_F(x^+,z^+;\w,w^-)\, g\, A^{+}(z) {\cal U}_F(z^+,y^+;\v,v^-)
+\textrm{NNEik}
\label{single_A_plus_scalar_prop_3}
\, .
\end{align}
Since we are assuming that $A^{+}(z)=0$ outside of the target, the expression \eqref{single_A_plus_scalar_prop_3} is non-zero only if $x^+$ or $y^+$ lies inside the target. 
In such case, it is still justified to drop the $z^+$ dependent phases at the considered accuracy, in order to simplify the $\p_1$ and $\p_2$ integrations and obtain the constraints $\w=\v=\z$. To the accuracy of interested, it is again sufficient to keep only the leading term in the Taylor expansion of the Wilson lines around $w^-=z^-$ and $v^-=z^-$ respectively, which amounts to make the replacements $w^-\mapsto z^-$ and $v^-\mapsto z^-$ in the arguments of the Wilson lines. The integrations over $w^-$ and $v^-$ then become trivial and provide the constraints $p_1^+=k^+$ and $p_2^+=q^+$.
Performing these steps, one arrives at
\begin{align}
 \delta G_F(x,y)\bigg|_{\textrm{Single }A^{+}}\,
= &\,
\int \frac{d^{D-1} \underline{q}}{(2\pi)^{D-1}}\: e^{-i \check{q}\cdot x}\, \frac{\theta(q^+)}{(2q^+)}
\int \frac{d^{D-1} \underline{k}}{(2\pi)^{D-1}}\: e^{i \check{k}\cdot y}\, \frac{\theta(k^+)}{(2k^+)}
\int d^{D-2}\z\,\, e^{-i \z \cdot (\q -\k)}\,
\int dz^-\,\, e^{i z^-(q^+ -k^+)}\,
\nonumber\\
&\, \times\;
\bigg[g\, A^{+}(x^+,z^-,\z)\, {\cal U}_F(x^+,y^+;\z,z^-)
+{\cal U}_F(x^+,y^+;\z,z^-)\, g\, A^{+}(y^+,z^-,\z)
\bigg]
+\textrm{NNEik}
\label{single_A_plus_scalar_prop_5}
\, .
\end{align}
The expression \eqref{single_A_plus_scalar_prop_5} is indeed a NEik contribution to $G_F(x,y)$, and it vanishes if both endpoints $x$ or $y$ lies outside of the target, due to our assumption that $A^{\mu}(z)=0$ outside of the target.


\subsection{Final result for the scalar propagator through a shockwave at NEik accuracy}

We have calculated the various contributions to $G_F(x,y)$ at NEik accuracy in the previous subsections. In the case of propagation from before to after the target, meaning $x^+>L^+/2$ and $y^+<-L^+/2$, the contribution \eqref{single_A_plus_scalar_prop_5} vanishes, whereas the other ones, \eqref{expansion_scalar_prop_pure_A_minus_brownian_4} and \eqref{single_A_perp_scalar_prop_8}, can be combined into
\begin{align}
&G_F(x,y)
=  \int \frac{d^{D-1} \underline{q}}{(2\pi)^{D-1}}\:
\int \frac{d^{D-1} \underline{k}}{(2\pi)^{D-1}}\:
\frac{\theta(q^+)\theta(k^+)}{(q^+\!+\!k^+)}\: e^{-i \check{q}\cdot x}\, e^{i \check{k}\cdot y}\,
\int dz^-\, e^{i z^-   (q^+ -k^+)}\,
\int d^{D-2}\z\,\, e^{-i \z \cdot (\q -\k)}\,
\nonumber\\
&\, \times \; 
\Bigg\{{\cal U}_F\left(\frac{L^+}{2},-\frac{L^+}{2};\z,z^-\right)
-\frac{(\q^j+\k^j)}{2(q^+\!+\!k^+)}
\int_{-\frac{L^+}{2}}^{\frac{L^+}{2}}\! dz^+\,
\left[{\cal U}_F\left(\frac{L^+}{2},z^+;\z,z^-\right)   \overleftrightarrow{D}_{\z^j}
{\cal U}_F\left(z^+,-\frac{L^+}{2};\z,z^-\right)\right]
\nonumber\\
&\, \hspace{0.8cm}
-\frac{i}{(q^+\!+\!k^+)}
\int_{-\frac{L^+}{2}}^{\frac{L^+}{2}}\! dz^+\,
\left[{\cal U}_F\left(\frac{L^+}{2},z^+;\z,z^-\right)   \overleftarrow{D}_{\z^j}\overrightarrow{D}_{\z^j}
{\cal U}_F\left(z^+,-\frac{L^+}{2};\z,z^-\right)\right]
\Bigg\}
+\textrm{NNEik}
\label{expansion_scalar_prop_total_NEik_before_after}
\, .
\end{align}
As a reminder, the first term in Eq.~\eqref{expansion_scalar_prop_total_NEik_before_after} corresponds to the Generalized Eikonal approximation introduced in Sec.~\ref{sec:zmin_dep_scalar_prop}. It goes beyond the strict Eikonal approximation due to the $z^-$ dependence of the Wilson line, but it can be further expanded as in Eq.~\eqref{expansion_scalar_prop_pure_A_minus_6} in order to recover the usual Eikonal expression plus a new NEik correction.

In Appendix \ref{app:scalar_comparison_Giovanni}, we further study the scalar propagator in a more generic situation, without assuming that the endpoints $x$ and $y$ are outside of the target and without assuming that the gauge field is vanishing outside of the target. We compare these results for the NEik scalar propagator with the earlier ones from Ref.~\cite{Chirilli:2018kkw}, and show that they are equivalent, up to the $z^-$ dependence effects which are included for the first time in the present paper.

\section{Quark propagator through a shockwave at NEik accuracy\label{sec:quark_prop}}


The aim of this section is to combine the results from the sections~\ref{sec:quark_vs_scal_prop} and \ref{sec:scal_prop}
in order to obtain an expression for the quark propagator through a gluon shockwave at NEik accuracy, and compare it to
earlier results in the literature.

\subsection{Derivation\label{sec:quark_prop_deriv}}


The general result~\eqref{quark_prop_NEik_from_scalar_result} from Sec.~\ref{sec:quark_vs_scal_prop} involves $\Delta(x,y)$, defined in Eq.~\eqref{def_Delta}. Since $\Delta(x,y)$ contains a factor $\delta(x^+-y^+)$, the enhanced contribution to $S_F(x,y)$ is instantaneous, and thus vanishes in the case of propagation from before to after the target, meaning $x^+>L^+/2$ and $y^+<-L^+/2$. For the same reason, the ${F^+}_{j}(w)$ insertion occurs at $w^+=y^+$ (resp. $w^+=x^+$) in the fourth (resp. fifth) term in Eq.~\eqref{quark_prop_NEik_from_scalar_result}. Since $F_{\mu\nu}(w)=0$ outside of the target, the fourth and fifth term in Eq.~\eqref{quark_prop_NEik_from_scalar_result} vanish if $x^+>L^+/2$ and $y^+<-L^+/2$.
In this case, we are thus left with
\begin{align}
  S_F(x,y)
= &\,
\big( i\,\slashed{D}_{x} +m \big)\frac{\gamma^+}{2}\int d^{D}w\,
\bigg\{G_F(x,w)\,
\nonumber\\
&\hspace{1cm}
+\int d^{D}z\, G_F(x,z)\bigg[
\frac{1}{4}\, [\gamma^{i},\gamma^{j}]\, g F_{ij}(z)
+g F_{+-}(z)
\bigg]G_F(z,w)
\bigg\}
\Delta(w,y)\big( -i\,\overleftarrow{\slashed{D}}_{y} +m \big)
\nonumber\\
&
+\big( i\,\slashed{D}_{x} +m \big)\frac{\gamma^+}{2}\int d^{D}w\, \Delta(x,w)\,
\bigg\{
G_F(w,y)
\nonumber\\
&\hspace{1cm}
+\int d^{D}z\, G_F(w,z)\bigg[
\frac{1}{4}\, [\gamma^{i},\gamma^{j}]\, g F_{ij}(z)
-g F_{+-}(z)
\bigg]G_F(z,y)
\bigg\}\big( -i\,\overleftarrow{\slashed{D}}_{y} +m \big)
+\textrm{NNEik}
\label{quark_prop_NEik_from_scalar_result_BA_1}
\, .
\end{align}
The covariant derivatives and the $\Delta$ are outside of the target in this case. Hence, assuming in addition that $A^{\mu}=0$ outside of the target, the covariant derivatives reduce to ordinary ones and the Wilson line in the $\Delta$ becomes trivial. Hence, for $x^+>L^+/2$ and $y^+<-L^+/2$, one obtains
\begin{align}
  S_F(x,y)
= &\,
\big( i\,\slashed{\d}_{x} +m \big)\frac{\gamma^+}{2}\int d^{D}w\,
\bigg\{G_F(x,w)\,
+\int d^{D}z\, G_F(x,z)\bigg[
\frac{1}{4}\, [\gamma^{i},\gamma^{j}]\, g F_{ij}(z)
+g F_{+-}(z)
\bigg]G_F(z,w)
\bigg\}
\nonumber\\
&\hspace{2cm}\times\;
\frac{(-i)}{4}\, \delta^{(D-1)}(\uw\!-\!\uy)\, \textrm{sgn}(w^-\!-\!y^-)\,
\big( -i\,\overleftarrow{\slashed{\d}}_{y} +m \big)
\nonumber\\
&
+\big( i\,\slashed{\d}_{x} +m \big)\frac{\gamma^+}{2}\int d^{D}w\,
\frac{(-i)}{4}\, \delta^{(D-1)}(\ux\!-\!\uw)\, \textrm{sgn}(x^-\!-\!w^-)\,
\bigg\{
G_F(w,y)
\nonumber\\
&\hspace{1cm}
+\int d^{D}z\, G_F(w,z)\bigg[
\frac{1}{4}\, [\gamma^{i},\gamma^{j}]\, g F_{ij}(z)
-g F_{+-}(z)
\bigg]G_F(z,y)
\bigg\}\big( -i\,\overleftarrow{\slashed{\d}}_{y} +m \big)
+\textrm{NNEik}
\label{quark_prop_NEik_from_scalar_result_BA_2}
\end{align}

Thanks to the relation
\begin{align}
\int_{0}^{+\infty} \!\!\! dp^+\, \frac{\sin\big(p^+ (x^-\!-\!y^-)\big)}{p^+}
= &\, \frac{\pi}{2}\; \textrm{sgn}(x^-\!-\!y^-)
\label{sign_func_rep_1}
\, ,
\end{align}
we have the Fourier representation
\begin{align}
\frac{(-i)}{4}\, \delta^{(D-1)}(\ux\!-\!\uy)\, \textrm{sgn}(x^-\!-\!y^-)\,
= &\,
\int \frac{d^{D} {p}}{(2\pi)^{D}}\: \frac{e^{-i {p}\cdot (x\!-\!y)}}{2p^+}
\label{sign_func_rep_2}
\end{align}
for $\Delta(x,y)$ in vacuum. Introducing this Fourier representation into Eq.~\eqref{quark_prop_NEik_from_scalar_result_BA_2} gives
\begin{align}
  S_F(x,y)
= &\,
\big( i\,\slashed{\d}_{x} +m \big)\frac{\gamma^+}{2}\int d^{D}w\,
\bigg\{G_F(x,w)\,
+\int d^{D}z\, G_F(x,z)\bigg[
\frac{1}{4}\, [\gamma^{i},\gamma^{j}]\, g F_{ij}(z)
+g F_{+-}(z)
\bigg]G_F(z,w)
\bigg\}
\nonumber\\
&\hspace{2cm}\times\;
\int \frac{d^{D} {k}}{(2\pi)^{D}}\: e^{-i {k}\cdot (w\!-\!y)}\;
\frac{\big( {\slashed{k}} +m \big)}{2k^+}
\nonumber\\
&
+
\int d^{D}w\,
\int \frac{d^{D} {q}}{(2\pi)^{D}}\: e^{-i {q}\cdot (x\!-\!w)}\:
\frac{\big(\slashed{q} +m \big)}{2q^+}\; \frac{\gamma^+}{2}
\bigg\{
G_F(w,y)
\nonumber\\
&\hspace{1cm}
+\int d^{D}z\, G_F(w,z)\bigg[
\frac{1}{4}\, [\gamma^{i},\gamma^{j}]\, g F_{ij}(z)
-g F_{+-}(z)
\bigg]G_F(z,y)
\bigg\}\big( -i\,\overleftarrow{\slashed{\d}}_{y} +m \big)
+\textrm{NNEik}
\label{quark_prop_NEik_from_scalar_result_BA_3}
\, .
\end{align}

In order to go further, let us study the insertion of $g F_{+-}(z)$ on the scalar propagator. We can split $g F_{+-}(z)$ into two terms as
\begin{align}
g F_{+-}(z) = &\, -i\big[D_{z^+},D_{z^-}\big]
=-i\big[\d_{z^+} +ig A^-(z),\d_{z^-}\big]
-i\big[D_{z^+},ig A^+(z)\big]
=-g\big( \d_{z^-}A^-(z)\big)
+\big[D_{z^+},g A^+(z)\big]
\label{F_plus_minus}
\, ,
\end{align}
so that
\begin{align}
\int d^Dz\, G_F(x,z)\, g F_{+-}(z)\, G_F(z,y)
=&\,
-\int d^Dz\, G_F(x,z)\, g\big( \d_{z^-}A^-(z)\big)\, G_F(z,y)
\nonumber\\
&\,
+\int d^Dz\, G_F(x,z)\, \big[D_{z^+},g A^+(z)\big]\, G_F(z,y)
\label{F_plus_minus_insertion}
\, .
\end{align}
In the expression \eqref{F_plus_minus_insertion}, both terms are NEik due to the $z^+$ integration. Hence, it is sufficient to insert the Generalized Eikonal expression for the scalar propagators. The first term in Eq.~\eqref{F_plus_minus_insertion} then becomes
\begin{align}
-\int d^Dz\, G_F(x,z)\, & g\big( \d_{z^-}A^-(z)\big)
  G_F(z,y)
=\,
\int d^Dz\,
 \int \frac{d^{D-1} \underline{q}}{(2\pi)^{D-1}}\:
\int \frac{d^{D-1} \underline{p_2}}{(2\pi)^{D-1}}\:
\frac{\theta(q^+)\theta(p_2^+)}{(q^+\!+\!p_2^+)}\: e^{-i \check{q}\cdot x}\, e^{i \check{p_2}\cdot z}\,
\nonumber\\
&\times\;
\int dw^-\, e^{i w^-   (q^+ -p_2^+)}\,
\int d^{D-2}\w\,\, e^{-i \w \cdot (\q -\p_2)}\,
\theta(x^+\!-\!z^+)\, {\cal U}_F\left(x^+,z^+;\w,w^-\right)
\nonumber\\
&\times\;
\Big[-g\big( \d_{z^-}A^-(z)\big)\Big]\,\int \frac{d^{D-1} \underline{p_1}}{(2\pi)^{D-1}}\:
\int \frac{d^{D-1} \underline{k}}{(2\pi)^{D-1}}\:
\frac{\theta(p_1^+)\theta(k^+)}{(p_1^+\!+\!k^+)}\: e^{-i \check{p}_1\cdot z}\, e^{i \check{k}\cdot y}\,
\nonumber\\
&\times\;
\int dv^-\, e^{i v^-   (p_1^+ -k^+)}\,
\int d^{D-2}\v\,\, e^{-i \v \cdot (\p_1 -\k)}\,
\theta(z^+\!-\!y^+)\, {\cal U}_F\left(z^+,y^+;\v,v^-\right)
+\textrm{NNEik}
\label{F_plus_minus_insertion_zminus_dep}
\, .
\end{align}
Keeping only the leading term in the regime defined in Eq.~\eqref{hyp_p_plus_power_counting} (using the definitions \eqref{def_p_plus} and \eqref{def_Delta_p_n_plus}), and dropping the $z^+$ dependent phases, one arrives at
\begin{align}
-&\int d^Dz\, G_F(x,z)\, g\big( \d_{z^-}A^-(z)\big)\, G_F(z,y)
=
\int \frac{d^{D-1} \underline{q}}{(2\pi)^{D-1}}\: e^{-i \check{q}\cdot x}\, \theta(q^+)\,
\int \frac{d^{D-1} \underline{k}}{(2\pi)^{D-1}}\: e^{i \check{k}\cdot y}\, \frac{\theta(k^+)}{(q^+\!+\!k^+)^2}\,
\int d^{D-2}\z\,\, e^{-i \z \cdot (\q -\k)}\,
\nonumber\\
&\times\;
\int dz^-\, e^{i z^-   (q^+ -k^+)}
\int_{y^+}^{x^+}\!\!\!dz^+\,
{\cal U}_F\left(x^+,z^+;\z,z^-\right)\,
\Big[-g\big( \d_{z^-}A^-(z)\big)\Big]\,
{\cal U}_F\left(z^+,y^+;\z,z^-\right)
+\textrm{NNEik}
\label{F_plus_minus_insertion_zminus_dep_2}
\\
=&
\int \frac{d^{D-1} \underline{q}}{(2\pi)^{D-1}}\: e^{-i \check{q}\cdot x}\, \theta(q^+)\,
\int \frac{d^{D-1} \underline{k}}{(2\pi)^{D-1}}\: e^{i \check{k}\cdot y}\, \frac{\theta(k^+)}{(q^+\!+\!k^+)^2}\,
\int d^{D-2}\z\,\, e^{-i \z \cdot (\q -\k)}\,
\nonumber\\
&\times\;
\int dz^-\, e^{i z^-   (q^+ -k^+)}
(-i)\d_{z^-}
{\cal U}_F\left(x^+,y^+;\z,z^-\right)
+\textrm{NNEik}
\label{F_plus_minus_insertion_zminus_dep_3}
\\
=&
\int \frac{d^{D-1} \underline{q}}{(2\pi)^{D-1}}\: e^{-i \check{q}\cdot x}\, \theta(q^+)\,
\int \frac{d^{D-1} \underline{k}}{(2\pi)^{D-1}}\: e^{i \check{k}\cdot y}\, \theta(k^+)\,
(-1)\frac{(q^+ \!-\!k^+)}{(q^+\!+\!k^+)^2}
\nonumber\\
&\times\;
\int d^{D-2}\z\,\, e^{-i \z \cdot (\q -\k)}\,
\int dz^-\, e^{i z^-   (q^+ -k^+)}
{\cal U}_F\left(x^+,y^+;\z,z^-\right)
+\textrm{NNEik}
\label{F_plus_minus_insertion_zminus_dep_4}
\end{align}

By contrast, for the second term in Eq.~\eqref{F_plus_minus_insertion}, one finds
\begin{align}
&\int d^Dz\, G_F(x,z)\, \big[D_{z^+},g A^+(z)\big]\, G_F(z,y)
=
\int d^Dz\, G_F(x,z)\, \Big(- \overleftarrow{D}_{z^+}g A^+(z)-g A^+(z)\overrightarrow{D}_{z^+}   \Big)\, G_F(z,y)
\label{F_plus_minus_insertion_Aplus_Dzplus_1}
\\
=&\,
\int d^Dz\,
 \int \frac{d^{D-1} \underline{q}}{(2\pi)^{D-1}}\:
\int \frac{d^{D-1} \underline{p_2}}{(2\pi)^{D-1}}\:
\frac{\theta(q^+)\theta(p_2^+)}{(q^+\!+\!p_2^+)}\: e^{-i \check{q}\cdot x}\, e^{i \check{p_2}\cdot z}\,
\nonumber\\
&\times\;
\int dw^-\, e^{i w^-   (q^+ -p_2^+)}\,
\int d^{D-2}\w\,\, e^{-i \w \cdot (\q -\p_2)}\,
\theta(x^+\!-\!z^+)\, {\cal U}_F\left(x^+,z^+;\w,w^-\right)
\nonumber\\
&\times\;
\Big[- \overleftarrow{D}_{z^+}g A^+(z)-g A^+(z)\overrightarrow{D}_{z^+}   \big)\Big]\,\int \frac{d^{D-1} \underline{p_1}}{(2\pi)^{D-1}}\:
\int \frac{d^{D-1} \underline{k}}{(2\pi)^{D-1}}\:
\frac{\theta(p_1^+)\theta(k^+)}{(p_1^+\!+\!k^+)}\: e^{-i \check{p}_1\cdot z}\, e^{i \check{k}\cdot y}\,
\nonumber\\
&\times\;
\int dv^-\, e^{i v^-   (p_1^+ -k^+)}\,
\int d^{D-2}\v\,\, e^{-i \v \cdot (\p_1 -\k)}\,
\theta(z^+\!-\!y^+)\, {\cal U}_F\left(z^+,y^+;\v,v^-\right)
+\textrm{NNEik}
\label{F_plus_minus_insertion_Aplus_Dzplus_2}
\, .
\end{align}
The situation is the same as in Sec.~\eqref{sec:A_plus_insert_scalar}: the covariant derivatives give zero when acting on the Wilson lines, and only the action of the $\partial_{z^+}$ derivatives on the theta function can provide a contribution to Eq.~\eqref{F_plus_minus_insertion_Aplus_Dzplus_2} at NEik accuracy. Then, keeping only the leading terms in the regime \eqref{hyp_p_plus_power_counting} and neglecting $z^+$ dependent phases leads to
\begin{align}
\int d^Dz\, G_F(x,z)\, \big[D_{z^+},g A^+(z)\big]\, G_F(z,y)
=&\,
 \int \frac{d^{D-1} \underline{q}}{(2\pi)^{D-1}}\: e^{-i \check{q}\cdot x}\, \theta(q^+)
 \int \frac{d^{D-1} \underline{k}}{(2\pi)^{D-1}}\: e^{i \check{k}\cdot y}\, \frac{\theta(k^+)}{(q^+\!+\!k^+)^2}\:
\int d^{D-2}\z\,\, e^{-i \z \cdot (\q -\k)}\,
\nonumber\\
&\times\;
\int dz^+ \Big[\delta(x^+\!-\!z^+)\theta(z^+\!-\!y^+) -\theta(x^+\!-\!z^+)\delta(z^+\!-\!y^+)\Big]\;
\nonumber\\
&\times\;
\int dz^-\, e^{i z^-   (q^+ -k^+)}
{\cal U}_F\left(x^+,z^+;\z,z^-\right)\, g A^+(z)\, {\cal U}_F\left(z^+,y^+;\z,z^-\right)
+\textrm{NNEik}
\label{F_plus_minus_insertion_Aplus_Dzplus_3}
\\
=&\,
 \int \frac{d^{D-1} \underline{q}}{(2\pi)^{D-1}}\: e^{-i \check{q}\cdot x}\, \theta(q^+)
 \int \frac{d^{D-1} \underline{k}}{(2\pi)^{D-1}}\: e^{i \check{k}\cdot y}\, \theta(k^+)\:
\int d^{D-2}\z\,\, e^{-i \z \cdot (\q -\k)}\,
\nonumber\\
&\times\; \frac{\theta(x^+\!-\!y^+)}{(q^+\!+\!k^+)^2}
\int dz^-\, e^{i z^-   (q^+ -k^+)}\,
\bigg\{
g A^+(x^+,z^-,\z)\, {\cal U}_F\left(x^+,y^+;\z,z^-\right)
\nonumber\\
&\hspace{2cm}\;
-{\cal U}_F\left(x^+,y^+;\z,z^-\right)\, g A^+(y^+,z^-,\z)
\bigg\}
+\textrm{NNEik}
\label{F_plus_minus_insertion_Aplus_Dzplus_4}
\, .
\end{align}
Assuming that $A_{\mu}(z)=0$ outside of the target, the expression \eqref{F_plus_minus_insertion_Aplus_Dzplus_4} vanishes for
$x^+>L^+/2$ and $y^+<-L^+/2$.

Using the relations \eqref{expansion_scalar_prop_total_NEik_before_after} and \eqref{F_plus_minus_insertion_zminus_dep_4} into Eq.~\eqref{quark_prop_NEik_from_scalar_result_BA_3}, we obtain
\begin{align}
& S_F(x,y) =  \int \frac{d^{D-1} \underline{q}}{(2\pi)^{D-1}}\:
\int \frac{d^{D-1} \underline{k}}{(2\pi)^{D-1}}\:
\theta(q^+)\theta(k^+)\: e^{-i \check{q}\cdot x}\, e^{i \check{k}\cdot y}\,
\int dz^-\, e^{i z^-   (q^+ -k^+)}\,
\int d^{D-2}\z\,\, e^{-i \z \cdot (\q -\k)}\,
\nonumber\\
&\, \times \; 
 \big(\check{\slashed{q}} +m \big)\, \frac{\gamma^+}{2}
 \Bigg\{\frac{1}{(q^+\!+\!k^+)}\, {\cal U}_F\left(\frac{L^+}{2},-\frac{L^+}{2};\z,z^-\right)
-\frac{(q^+\!-\!k^+)}{(q^+\!+\!k^+)^2}\, {\cal U}_F\left(\frac{L^+}{2},-\frac{L^+}{2};\z,z^-\right)
\nonumber\\
&\, \hspace{2cm}
-\frac{(\q^j+\k^j)}{2(q^+\!+\!k^+)^2}
\int_{-\frac{L^+}{2}}^{\frac{L^+}{2}}\! dz^+\,
\left[{\cal U}_F\left(\frac{L^+}{2},z^+;\z,z^-\right)   \overleftrightarrow{D}_{\z^j}
{\cal U}_F\left(z^+,-\frac{L^+}{2};\z,z^-\right)\right]
\nonumber\\
&\, \hspace{2cm}
-\frac{i}{(q^+\!+\!k^+)^2}
\int_{-\frac{L^+}{2}}^{\frac{L^+}{2}}\! dz^+\,
\left[{\cal U}_F\left(\frac{L^+}{2},z^+;\z,z^-\right)   \overleftarrow{D}_{\z^j}\overrightarrow{D}_{\z^j}
{\cal U}_F\left(z^+,-\frac{L^+}{2};\z,z^-\right)\right]
\nonumber\\
&\, \hspace{2cm}
+\frac{1}{(q^+\!+\!k^+)^2}
\frac{1}{4}\, [\gamma^{i},\gamma^{j}]\,
\int_{-\frac{L^+}{2}}^{\frac{L^+}{2}}\! dz^+\,
\left[{\cal U}_F\left(\frac{L^+}{2},z^+;\z,z^-\right)\,  g F_{ij}(z)\,
{\cal U}_F\left(z^+,-\frac{L^+}{2};\z,z^-\right)\right]
 \Bigg\}\frac{\big( {\slashed{k}} +m \big)}{2k^+}
 \nonumber\\
 &\, +
 \int \frac{d^{D-1} \underline{q}}{(2\pi)^{D-1}}\:
\int \frac{d^{D-1} \underline{k}}{(2\pi)^{D-1}}\:
\theta(q^+)\theta(k^+)\: e^{-i \check{q}\cdot x}\, e^{i \check{k}\cdot y}\,
\int dz^-\, e^{i z^-   (q^+ -k^+)}\,
\int d^{D-2}\z\,\, e^{-i \z \cdot (\q -\k)}\,
\nonumber\\
&\, \times \; 
 \frac{\big(\check{\slashed{q}} +m \big)}{2q^+}\, \frac{\gamma^+}{2}
 \Bigg\{\frac{1}{(q^+\!+\!k^+)}\, {\cal U}_F\left(\frac{L^+}{2},-\frac{L^+}{2};\z,z^-\right)
+\frac{(q^+\!-\!k^+)}{(q^+\!+\!k^+)^2}\, {\cal U}_F\left(\frac{L^+}{2},-\frac{L^+}{2};\z,z^-\right)
\nonumber\\
&\, \hspace{2cm}
-\frac{(\q^j+\k^j)}{2(q^+\!+\!k^+)^2}
\int_{-\frac{L^+}{2}}^{\frac{L^+}{2}}\! dz^+\,
\left[{\cal U}_F\left(\frac{L^+}{2},z^+;\z,z^-\right)   \overleftrightarrow{D}_{\z^j}
{\cal U}_F\left(z^+,-\frac{L^+}{2};\z,z^-\right)\right]
\nonumber\\
&\, \hspace{2cm}
-\frac{i}{(q^+\!+\!k^+)^2}
\int_{-\frac{L^+}{2}}^{\frac{L^+}{2}}\! dz^+\,
\left[{\cal U}_F\left(\frac{L^+}{2},z^+;\z,z^-\right)   \overleftarrow{D}_{\z^j}\overrightarrow{D}_{\z^j}
{\cal U}_F\left(z^+,-\frac{L^+}{2};\z,z^-\right)\right]
\nonumber\\
&\, \hspace{2cm}
+\frac{1}{(q^+\!+\!k^+)^2}
\frac{1}{4}\, [\gamma^{i},\gamma^{j}]\,
\int_{-\frac{L^+}{2}}^{\frac{L^+}{2}}\! dz^+\,
\left[{\cal U}_F\left(\frac{L^+}{2},z^+;\z,z^-\right)\,  g F_{ij}(z)\,
{\cal U}_F\left(z^+,-\frac{L^+}{2};\z,z^-\right)\right]
 \Bigg\} \big({\slashed{k}} +m \big)
 +\textrm{NNEik}
\label{quark_prop_NEik_BA_2}
\, .
\end{align}

In the first part of Eq.~\eqref{quark_prop_NEik_BA_2}, the first two terms can be combined thanks to the relation
\begin{align}
\frac{1}{(q^+\!+\!k^+)}-\frac{(q^+\!-\!k^+)}{(q^+\!+\!k^+)^2}
= &\,
\frac{2k^+}{(q^+\!+\!k^+)^2}
=
\frac{1}{2q^+} - \frac{(q^+\!-\!k^+)^2}{(2q^+)(q^+\!+\!k^+)^2}
=
\frac{1}{2q^+}\, \Big[1+ \textrm{NNEik} \Big]
\label{qplus_kplus_in_first_part_quark_prop_2}
\, .
\end{align}
By contrast, in second part Eq.~\eqref{quark_prop_NEik_BA_2}, the first two terms can be combined thanks to
\begin{align}
\frac{1}{(q^+\!+\!k^+)}+\frac{(q^+\!-\!k^+)}{(q^+\!+\!k^+)^2}
= &\,
\frac{2q^+}{(q^+\!+\!k^+)^2}
=
\frac{1}{2k^+} - \frac{(q^+\!-\!k^+)^2}{(2k^+)(q^+\!+\!k^+)^2}
=
\frac{1}{2k^+}\, \Big[1+ \textrm{NNEik} \Big]
\label{qplus_kplus_in_second_part_quark_prop_2}
\, .
\end{align}
The two parts in Eq.~\eqref{quark_prop_NEik_BA_2}, then give the same result (up to NNEik corrections), so that
\begin{align}
& S_F(x,y) =  \int \frac{d^{D-1} \underline{q}}{(2\pi)^{D-1}}\:
\int \frac{d^{D-1} \underline{k}}{(2\pi)^{D-1}}\:
\theta(q^+)\theta(k^+)\: e^{-i \check{q}\cdot x}\, e^{i \check{k}\cdot y}\,
\int dz^-\, e^{i z^-   (q^+ -k^+)}\,
\int d^{D-2}\z\,\, e^{-i \z \cdot (\q -\k)}\,
\nonumber\\
&\, \times \; 
 \frac{\big(\check{\slashed{q}} +m \big)}{2q^+}\,\, \gamma^+
 \Bigg\{ {\cal U}_F\left(\frac{L^+}{2},-\frac{L^+}{2};\z,z^-\right)
\nonumber\\
&\, \hspace{2cm}
-\frac{(\q^j+\k^j)}{2(q^+\!+\!k^+)}
\int_{-\frac{L^+}{2}}^{\frac{L^+}{2}}\! dz^+\,
\left[{\cal U}_F\left(\frac{L^+}{2},z^+;\z,z^-\right)   \overleftrightarrow{D}_{\z^j}
{\cal U}_F\left(z^+,-\frac{L^+}{2};\z,z^-\right)\right]
\nonumber\\
&\, \hspace{2cm}
-\frac{i}{(q^+\!+\!k^+)}
\int_{-\frac{L^+}{2}}^{\frac{L^+}{2}}\! dz^+\,
\left[{\cal U}_F\left(\frac{L^+}{2},z^+;\z,z^-\right)   \overleftarrow{D}_{\z^j}\overrightarrow{D}_{\z^j}
{\cal U}_F\left(z^+,-\frac{L^+}{2};\z,z^-\right)\right]
\nonumber\\
&\, \hspace{2cm}
+\frac{1}{(q^+\!+\!k^+)}
\frac{1}{4}\, [\gamma^{i},\gamma^{j}]\,
\int_{-\frac{L^+}{2}}^{\frac{L^+}{2}}\! dz^+\,
\left[{\cal U}_F\left(\frac{L^+}{2},z^+;\z,z^-\right)\,  g F_{ij}(z)\,
{\cal U}_F\left(z^+,-\frac{L^+}{2};\z,z^-\right)\right]
 \Bigg\}\frac{\big( \check{\slashed{k}} +m \big)}{2k^+}
 +\textrm{NNEik}
\label{quark_prop_NEik_BA_3}
\, .
\end{align}
The first term in Eq.~\eqref{quark_prop_NEik_BA_3} corresponds to the Generalized Eikonal approximation for the quark propagator through a gluon shockwave, involving a $z^-$ dependent Wilson line. It could be further expanded using Eq.~\eqref{expansion_Wilson_line_zminus_equal_0} into the strict Eikonal approximation, plus extra NEik corrections. However, these extra NEik terms obtained after integration by parts in $q^+$ or $k^+$ would be quite numerous and cumbersome, as is clear upon examination of Eq.~\eqref{quark_prop_NEik_BA_3}. Hence, the expression \eqref{quark_prop_NEik_BA_3} starting with the Generalized Eikonal contribution is a more convenient starting point for future applications.


\subsection{Comparison with earlier results in the literature}

\subsubsection{Comparison with Ref.~\cite{Altinoluk:2020oyd}}

In Ref.~\cite{Altinoluk:2020oyd}, the NEik quark propagator through a gluon shockwave was calculated, using a Feynman diagram approach analog to the one used in Sec.~\ref{sec:scal_prop} of the present paper for the scalar propagator. A major difference is however that non-Eikonal effects associated with the $z^-$ dependence of the target were thoroughly neglected in Ref.~\cite{Altinoluk:2020oyd}.

Dropping the $z^-$ dependence of all Wilson lines in Eq.~\eqref{quark_prop_NEik_BA_3}, one obtains a trivial $z^-$ integral giving $2\pi \delta(q^+\!-\!k^+)$, and the results of Ref.~\cite{Altinoluk:2020oyd} (given in Eqs.~(2.29), (2.30) and (2.31) there) are exactly recovered. Eq.~\eqref{quark_prop_NEik_BA_3} is thus fully consistent with the result of Ref.~\cite{Altinoluk:2020oyd}, and extends it to include the effects of $z^-$ dependence in the target.

Moreover, the new techniques developed here in Sec.~\ref{sec:scal_prop} to address the $z^-$ dependence effects in the scalar case can be applied straightforwardly to extend the derivation of the NEik quark propagator from Ref.~\cite{Altinoluk:2020oyd}. This corresponds to a faster and easier route to obtain the result given in Eq.~\eqref{quark_prop_NEik_BA_3} than the one followed in the present paper. These two different derivations are thus providing a non trivial cross-check of each other, which adds confidence in their
 common result, Eq.~\eqref{quark_prop_NEik_BA_3}.


\subsubsection{Comparison with Refs.~\cite{Chirilli:2018kkw} and \cite{Chirilli:2021lif}}

In Ref.~\cite{Chirilli:2018kkw}, NEik corrections to the scalar, quark and gluon propagators were calculated both in gluon and in quark background fields. These results were then applied in Ref.~\cite{Chirilli:2021lif} to calculate the high-energy operator product expansion of two electromagnetic currents at NEik accuracy, and derive the evolution equations for the operators involved.
Since we have not yet presented a full derivation of the NEik gluon propagator in our formalism, or of NEik propagators in quark background field, the only meaningful quantities to compare between the results of Refs.~\cite{Chirilli:2018kkw} and \cite{Chirilli:2021lif} and ours are the scalar and quark propagators at NEik accuracy in a gluon background field. The comparison of the results for the scalar propagator is performed in Appendix \ref{app:scalar_comparison_Giovanni}.
The result for the NEik quark propagator in a gluon background is given in Eq.~(4.19) in Ref.~\cite{Chirilli:2018kkw} or in Eqs.~(3.1) and (B.12) in Ref.~\cite{Chirilli:2021lif}, and should be compared with our final result given in Eq.~\eqref{quark_prop_NEik_BA_3} here, or with our relation \eqref{quark_prop_NEik_from_scalar_result} between the quark and scalar propagators.
In Refs.~\cite{Chirilli:2018kkw,Chirilli:2021lif}, the NEik corrections to the quark propagator are split into two contributions $\hat{\cal O}_1$ and $\hat{\cal O}_2$, which themselves can be split further: $\hat{\cal O}_1$ contains an unpolarized piece (sharing the same Dirac structure as the Eikonal contribution) and a helicity polarized piece, whereas $\hat{\cal O}_2$ contains the contributions $\hat{\cal O}_{\bullet\ast}$ and $\hat{\cal O}_j$, defined in Eqs.~(B.17) and (B.15) of Ref.~\cite{Chirilli:2021lif}. Let us discuss each piece separately.
\begin{description}
  \item[Helicity polarized piece of $\hat{\cal O}_1$] This contribution is the term including a $[\gamma^i,\gamma^j]\, g\, F_{ij}$ insertion. It fully agrees between Refs.~\cite{Chirilli:2018kkw,Chirilli:2021lif} and Eq.~\eqref{quark_prop_NEik_BA_3}. It is also present in Eq.~\eqref{quark_prop_NEik_from_scalar_result}.

    \item[Unpolarized piece of $\hat{\cal O}_1$] Both in Refs.~\cite{Chirilli:2018kkw,Chirilli:2021lif} and in the present paper, it is found that the unpolarized piece of the NEik quark propagator is determined by the NEik scalar propagator.
        As stated earlier, we have shown explicitly in Appendix \ref{app:scalar_comparison_Giovanni} the equivalence between the NEik scalar propagators derived in Refs.~\cite{Chirilli:2018kkw,Chirilli:2021lif} and in the present paper, in the case of a $z^-$ independent background field.
        On the other hand, in the present paper, we demonstrated that the $z^-$ dependence of the background field (or equivalently the non-conservation of $p^+$ along the propagation) is a new source of NEik corrections, and discussed how to include
        such effect systematically.
        It would be interesting to see whether the approach of Refs.~\cite{Chirilli:2018kkw,Chirilli:2021lif} can be extended in order to include this missing NEik effect. In the formalism of Refs.~\cite{Chirilli:2018kkw,Chirilli:2021lif}, this would correspond to treat the Sudakov variable $\alpha$ associated with $p^+$ as an operator instead of a number, and address the effects of the non-trivial commutation relation of this operator $\hat{\alpha}$ with the gauge field, in particular with the component ${A}_{\bullet}$ (corresponding to ${A}^-$ in our notations).


  \item[$\hat{\cal O}_{\bullet\ast}$ contribution] The operator $\hat{\cal O}_{\bullet\ast}$ present in the results of Refs.~\cite{Chirilli:2018kkw,Chirilli:2021lif} corresponds, in our notations, to the insertion of $g\, F_{+-}(z)$ into a Wilson line. We indeed find corrections with a $g\, F_{+-}(z)$ insertion in the expression \eqref{quark_prop_NEik_from_scalar_result} for quark propagator in terms of the scalar propagator, and these corrections should be related to $\hat{\cal O}_{\bullet\ast}$. However, note that there is a slight mismatch in the ordering of operators. In our case, the $g\, F_{+-}(z)$ insertion is happening in between the two Dirac differential operators (see Eq.~\eqref{quark_prop_NEik_from_scalar_result}). By contrast, in each of the terms involving $\hat{\cal O}_{\bullet\ast}$ in Ref.~\cite{Chirilli:2021lif}, all of the momentum operators are on the same side of $\hat{\cal O}_{\bullet\ast}$ (see the left hand side of Eq.~(B.18) there). In principle, the commutation relations are under control, so that it should be possible to write the $\hat{\cal O}_{\bullet\ast}$-type contributions with the same operator ordering in the formalism of Ref.~\cite{Chirilli:2021lif} as in our Eq.~\eqref{quark_prop_NEik_from_scalar_result}, up to new corrections which might behave as extra terms in $\hat{\cal O}_j$.

      Thanks to our Eq.~\eqref{F_plus_minus}, the field strength component $g\, F_{+-}(z)$ can be split into two terms. The first term probes the $z^-$ dependence of the background field. It is thus vanishing in the approximations of Refs.~\cite{Chirilli:2018kkw,Chirilli:2021lif} as well as of Ref.~\cite{Altinoluk:2020oyd}, and is relevant only in the present paper. The second term in $g\, F_{+-}(z)$ in Eq.~\eqref{F_plus_minus} involves $A^+(z)$ and $D_{z^+}$. Eventually, such contributions reduce to insertions of $A^+(z)$ outside of the target, which are usually made to vanish by gauge choice. The insertion of $g\, F_{+-}(z)$ is thus primarily associated with the $z^-$ dependence of the target, absent from previous studies. For that reason, it might be premature to try to fully compare such contributions between Refs.~\cite{Chirilli:2018kkw,Chirilli:2021lif} and the present work, until non-Eikonal effects of the $z^-$ dependence of the background field are studied in the approach of Refs.~\cite{Chirilli:2018kkw,Chirilli:2021lif}.

      Nevertheless, the results of the present paper allow us to gain a physical interpretation for the $g\, F_{+-}(z)$ insertions (and thus possibly for the $\hat{\cal O}_{\bullet\ast}$ contribution as well). In the Generalized Eikonal approximation introduced here, one keeps track of the overall difference between the incoming $k^+$ and the outgoing $q^+$ due to scattering on the target. By contrast, any other effects of the $z^-$ dependence of the background field are expanded as NEik corrections and beyond. Both in the case of the scalar propagator and in the case of the quark propagator, it is possible to remove such extra NEik corrections by an appropriate choice of the light-cone momentum scale in the denominators of the Generalized Eikonal contribution, $1/(q^++k^+)$ in the case of the scalar propagator~\eqref{expansion_scalar_prop_pure_A_minus_5} and $1/(4q^+k^+)$ in the case of the quark propagator~\eqref{quark_prop_NEik_BA_3}. The role of the $g\, F_{+-}(z)$ terms in Eq.~\eqref{quark_prop_NEik_from_scalar_result} is then to make the translation between these two different prescriptions for the denominators, as is clear from the derivation in Sec.~\ref{sec:quark_prop_deriv}. This explains why terms with $g\, F_{+-}(z)$ insertions are present in the relation \eqref{quark_prop_NEik_from_scalar_result} between the quark and scalar propagators, but not in our results for the NEik expansion of the scalar~\eqref{expansion_scalar_prop_total_NEik_before_after} and quark~\eqref{quark_prop_NEik_BA_3} propagators around the Generalized Eikonal approximation.

  \item[$\hat{\cal O}_j$ contribution] In Eq.~\eqref{quark_prop_NEik_from_scalar_result}, the last two terms have the same Dirac structure as the terms involving $\hat{\cal O}_j$ in Ref.~\cite{Chirilli:2021lif} (see the left hand side of Eq.~(B.16) there). However, the operators accompanying these Dirac structures look very different from $\hat{\cal O}_j$. In particular, in the last two terms in Eq.~\eqref{quark_prop_NEik_from_scalar_result}, the $g {F^+}_{j}$ is  always inserted at the initial or final times $y^+$ or $x^+$ of the propagator. For that reason, these $\hat{\cal O}_j$-type contributions vanish when considering the propagator from before to after the target, which explains their absence in the result~\eqref{quark_prop_NEik_BA_3} for the quark propagator, valid only in that kinematics.

      Apart from the possible mixing with the $\hat{\cal O}_{\bullet\ast}$ contribution mentioned earlier, the main cause for the apparent difference between the $\hat{\cal O}_j$-type contributions found in Refs.~\cite{Chirilli:2018kkw,Chirilli:2021lif} or in the present paper is most likely the different approach to calculate the quark propagator in a symmetric way.
      As discussed in Sec.~\ref{sec:quark_vs_scal_prop}, the quark propagator obeys an evolution equation on the left hand side and another one on the right hand side. Each of the two is sufficient to determine the quark propagator, and both should lead to equivalent results. In the formalism of Refs.~\cite{Chirilli:2018kkw,Chirilli:2021lif}, the counterpart of the two evolution equations is possibility to put the $\slashed{P}$ in the numerator of the propagator either on the left or on the right of the scalar-like factor involving the denominator. In Refs.~\cite{Chirilli:2018kkw,Chirilli:2021lif}, the quark propagator is calculated separately from the evolution equations on the left and on the right. Since an asymmetric result is obtained in both cases at NEik accuracy, the average of the two results is taken (see Eq.~(4.18) from Ref.~\cite{Chirilli:2018kkw}), in order to obtain a symmetric result. By contrast, in Sec.~\ref{sec:quark_vs_scal_prop}, we have first projected the quark propagator and its evolution equations as Dirac matrices into four blocks. After checking the equivalence of the results obtained from the left and right evolutions for each block separately, we have been able to construct a more refined symmetrization procedure. Since the blocks ${\cal P}_G S_F(x,y) {\cal P}_G$ and ${\cal P}_B S_F(x,y) {\cal P}_B$ are symmetric to each other, and that for both of them the evolution equation on one side gives a compact result and the evolution on the other side gives a cumbersome result, we have chosen take the result \eqref{GG_block_R_NEik_2} for ${\cal P}_G S_F(x,y) {\cal P}_G$ obtained from the evolution on the right, and the result \eqref{BB_block_L_NEik_2} for ${\cal P}_B S_F(x,y) {\cal P}_B$ obtained from the evolution on the left. By contrast, for each of the two other blocks ${\cal P}_G S_F(x,y) {\cal P}_B$ and ${\cal P}_B S_F(x,y) {\cal P}_G$, the results from evolution on the left or right are averaged in order to obtain a symmetric result at block level. Combining these expressions, we have obtained an expression \eqref{quark_prop_NEik_from_scalar_result} for the quark propagator which is both symmetric and quite compact, with minimal contributions of the type $\hat{\cal O}_j$. Note that by using an overall symmetrization procedure like in
      Refs.~\cite{Chirilli:2018kkw,Chirilli:2021lif} we would have obtained extra contributions coming from Eqs.~\eqref{GG_block_L_NEik} and \eqref{BB_block_R_NEik}, some of which are reminiscent of terms present in $\hat{\cal O}_j$ in Ref.~\cite{Chirilli:2021lif} (see Eq.~(B.15) there).
      Hence, it would be interesting to check to what extent one can simplify $\hat{\cal O}_j$  by using the same projections into blocks and the same refined symmetrization procedure in the formalism of Refs.~\cite{Chirilli:2018kkw,Chirilli:2021lif} as in the present study.

\end{description}


\section{Summary and outlook}
\label{sec:Discussions}

In this article, we have revisited the calculation of the NEik corrections to the quark propagator in a highly boosted gluon background field.
First, we have derived in Sec.~\ref{sec:quark_vs_scal_prop} a very general expression~\eqref{quark_prop_NEik_from_scalar_result} at NEik accuracy of the quark propagator written in terms of the full scalar propagator in the same background field. That expression~\eqref{quark_prop_NEik_from_scalar_result} is valid no matter if the endpoints of the propagator are inside or outside of the target. It is also fully gauge-covariant, since the whole derivation of Sec.~\ref{sec:quark_vs_scal_prop} involves the covariant derivative and the field strength, but never the gauge field itself.  Eq.~\eqref{quark_prop_NEik_from_scalar_result} can be understood as a partial expansion beyond the Eikonal approximation, limited to the non-Eikonal effects which are specific to the quark case. Indeed, universal non-Eikonal effects stay fully included in the scalar propagator in the leading term of Eq.~\eqref{quark_prop_NEik_from_scalar_result}.

Second, we have calculated the NEik corrections to the scalar propagator in a gluon background field, using the same approach as used in Ref.~\cite{Altinoluk:2020oyd} for the quark propagator. However, we have significantly extended that approach, by deriving the NEik corrections associated to the $z^-$ dependence of the background field or equivalently to the non-conservation of $p^+$ along the propagation in the background field. To our knowledge, this is the first time that these particular effects are addressed in a systematic way beyond the Eikonal approximation at high energy. It is now known how to include each of the possibe non-Eikonal corrections to CGC-type calculations at NEik precision.

%

However, the NEik corrections associated to the $z^-$ dependence of the background field are found to have a cumbersome expression in view of further applications. For convenience, we have introduced a Generalized Eikonal approximation, which retains the $z^-$ dependence in an approximate way in the leading power contribution. In particular, this corresponds to a finite $p^+$ momentum transfer from the target cumulated over the course of the propagation.
By contrast, further details about the $z^-$ dependence of the background field (or equivalently about the $p^+$ transfer process) are explicitly treated as subleading power corrections. It is actually found that these further subleading power corrections are of NNEik order. Hence, the Generalized Eikonal approximation accounts for $z^-$ dependence effects at NEik accuracy in a compact and convenient way. It is still possible to further expand the Generalized Eikonal expression in order to isolate the $z^-$ dependence effects from the strict Eikonal contribution, leading to new NEik corrections, see Eq.~\eqref{expansion_scalar_prop_pure_A_minus_6}. However, we recommend to perform that final expansion only at cross section level, in order to avoid cumbersome intermediate expressions.

Finally, combining the results of the various sections, we have obtained a complete NEik expression for the quark propagator in a gluon background field, and checked that it is equivalent to our earlier result from Ref.~\cite{Altinoluk:2020oyd} plus the NEik corrections due to the $z^-$ dependence of the background field.

As a future study, we plan to calculate the full NEik corrections to the gluon propagator in a gluon background field, including the ones associated with its $z^-$ dependence,  using the techniques developed in Ref.~\cite{Altinoluk:2020oyd} and in the present paper. We will also calculate the NEik corrections to the propagators in the presence of a quark background field. Then, we will have all of the building blocks required in order to perform the calculation of NEik corrections to high-energy scattering observables, beyond the Eikonal CGC result. We plan to perform such NEik calculations for various observables as further studies, in particular for observables of interest at the future EIC.


\acknowledgments

We thank Giovanni Chirilli and Cyrille Marquet for useful discussions.
TA is supported in part by the National Science Centre (Poland) under the research Grant No. 2018/31/D/ST2/00666 (SONA\-TA 14).
GB is supported in part by the National Science Centre (Poland) under the research Grant No. 2020/38/E/ST2/00122 (SONA\-TA BIS 10).
This work has been performed in the framework of COST Action CA 15213 ``Theory of hot matter and relativistic heavy-ion collisions" (THOR), MSCA RISE 823947 ``Heavy ion collisions: collectivity and precision in saturation physics''  (HI\-EIC) and has received funding from the European Un\-ion's Horizon 2020 research and innovation programme under grant agreement No. 824093.


\appendix
\section{Left versus right covariant derivatives acting on the scalar propagator\label{app:L_vs_R_cov_deriv_on_GF}}

The purpose of this appendix is to derive the relation \eqref{deriv_scalar_prop_flip_generic} between
$D_{x^{\rho}} G_F(x,y)$ and $G_F(x,y) \overleftarrow{D}_{y^{\rho}}$ from the scalar Green's equation, Eq.~\eqref{scalar_prop_eq_L}. As a preliminary, note that the relation
\begin{align}
D_{x^{\rho}} \delta^{(D)}(x\!-\!y) = &\, - \delta^{(D)}(x\!-\!y) \overleftarrow{D}_{y^{\rho}}
\label{passing_cov_deriv_through_delta}
\end{align}
follows from the definition of the left and right covariant derivatives, Eqs.~\eqref{Cov_deriv} and \eqref{Cov_deriv_back}.
Moreover, one finds
\begin{align}
D_{x^{\rho}}D_{x^{\mu}}D_{x^{\nu}} G_F(x,y)
= &\, \Big([D_{x^{\rho}},D_{x^{\mu}}]D_{x^{\nu}}+D_{x^{\mu}}[D_{x^{\rho}},D_{x^{\nu}}]+D_{x^{\mu}}D_{x^{\nu}}D_{x^{\rho}}
\Big)G_F(x,y)
\nonumber\\
= &\, \Big(ig F_{\rho\mu}(x) D_{x^{\nu}}+D_{x^{\mu}}ig F_{\rho\nu}(x)+D_{x^{\mu}}D_{x^{\nu}}D_{x^{\rho}}
\Big)G_F(x,y)
\label{triple_cov_deriv}
\, ,
\end{align}
by commuting the $D_{x^{\rho}}$ covariant derivative to the right, step by step.
Hence, when acting with $D_{x^{\rho}}$ on the left on the scalar Green's equation~\eqref{scalar_prop_eq_L}, and moving $D_{x^{\rho}}$ towards the right on both sides of the equation thanks to Eqs.~\eqref{passing_cov_deriv_through_delta} and \eqref{triple_cov_deriv}, one obtains
\begin{align}
\left[-g^{\mu\nu}\, D_{x^{\mu}}D_{x^{\nu}}-m^2
\right]
D_{x^{\rho}} G_F(x,y)
-ig {F_{\rho}}^{\mu}(x)\, D_{x^{\mu}} G_F(x,y)
-D_{x^{\mu}} \Big(ig {F_{\rho}}^{\mu}(x)\,  G_F(x,y)\Big)
 = &\, -i \delta^{(D)}(x\!-\!y)\overleftarrow{D}_{y^{\rho}}
\label{deriv_scalar_prop_eq_L_1_app}
\, .
\end{align}
The same operator appears in the square bracket in Eq.~\eqref{deriv_scalar_prop_eq_L_1_app} as in \eqref{scalar_prop_eq_L}. One can thus rewrite Eq.~\eqref{deriv_scalar_prop_eq_L_1_app} as an integral equation as
\begin{align}
D_{x^{\rho}} G_F(x,y)
 = &\, -i\int d^Dz\, G_F(x,z)\, \left\{-i \delta^{(D)}(z\!-\!y)\overleftarrow{D}_{y^{\rho}}
 +ig {F_{\rho}}^{\mu}(z)\, D_{z^{\mu}} G_F(z,y)
+D_{z^{\mu}} \Big(ig {F_{\rho}}^{\mu}(z)\,  G_F(z,y)\Big)
\right\}
\label{deriv_scalar_prop_eq_L_2_app}
\, ,
\end{align}
using $G_F(x,y)$ as the Green's function for that operator.
Eq.~\eqref{deriv_scalar_prop_eq_L_2_app} is indeed equivalent to Eq.~\eqref{deriv_scalar_prop_flip_generic}, through integration by parts.


\section{Equivalence of left and right evolution for the ${\cal P}_G S_F(x,y) {\cal P}_G$ block\label{app:LR_equiv_other_blocks}}

The aim of this appendix is to show the equivalence between the expressions \eqref{GG_block_L_NEik} and
\eqref{GG_block_R_NEik_2} for the ${\cal P}_G S_F(x,y) {\cal P}_G$ block, obtained from the evolution on the left or right side respectively.
For that purpose, it is necessary to understand the relation between $D_{x^{j}} G_F(x,y)$ and
$G_F(x,y) \overleftarrow{D}_{y^{j}}$. The $\rho=j$ component of the identity~\eqref{deriv_scalar_prop_flip_generic} is
\begin{align}
D_{x^{j}} G_F(x,y)
 = &\, -G_F(x,y) \overleftarrow{D}_{y^{j}}
 +\int d^Dz\, G_F(x,z)\, \Big[
 -g {F^{-}}_j(z)\, \overrightarrow{D}_{z^{-}} +\overleftarrow{D}_{z^{-}} g {F^{-}}_j(z)
 +g {F_j}^{i}(z)\, \overrightarrow{D}_{z^{i}} -\overleftarrow{D}_{z^{i}} g {F_j}^{i}(z)
\nonumber\\
&\hspace{6cm}
 -g {F^{+}}_j(z)\, \overrightarrow{D}_{z^{+}} +\overleftarrow{D}_{z^{+}} g {F^{+}}_j(z)
 \Big]G_F(z,y)
\label{deriv_scalar_prop_flip_perp}
\, ,
\end{align}
which is equivalent to
\begin{align}
D_{x^{j}} G_F(x,y)
 = &\, -G_F(x,y) \overleftarrow{D}_{y^{j}}
 \nonumber\\
 &\,
 +\int d^Dz\, G_F(x,z)\, \Big\{
 2\overleftarrow{D}_{z^{-}} g {F^{-}}_j(z)
 -\big[g {F^{-}}_j(z),\, \overrightarrow{D}_{z^{-}}\big]
 +2g {F_j}^{i}(z)\, \overrightarrow{D}_{z^{i}} +\big[\overrightarrow{D}_{z^{i}}, g {F_j}^{i}(z)\big]
\nonumber\\
&\, \hspace{4cm}
-2g {F^{+}}_j(z)\, \overrightarrow{D}_{z^{+}} -\big[\overrightarrow{D}_{z^{+}}, g {F^{+}}_j(z)\big]
\Big\}G_F(z,y)
\label{deriv_scalar_prop_flip_perp_exact_2}
\, .
\end{align}
Using Jacobi's identity, one finds
\begin{align}
-\big[g {F^{-}}_j(z),\, {D}_{z^{-}}\big] -\big[{D}_{z^{+}}, g {F^{+}}_j(z)\big]
=&\, i \big[[{D}_{z^{+}}, {D}_{z^{j}}],\, {D}_{z^{-}}\big]
+i \big[{D}_{z^{+}},\, [{D}_{z^{-}}, {D}_{z^{j}}]\big]
\nonumber\\
=&\,
i \big[[{D}_{z^{+}}, {D}_{z^{j}}],\, {D}_{z^{-}}\big]
+i \big[[{D}_{z^{j}}, {D}_{z^{-}}],\, {D}_{z^{+}}\big]
\nonumber\\
=&\, -i \big[[{D}_{z^{-}}, {D}_{z^{+}}],\, {D}_{z^{j}}\big]
\nonumber\\
=&\, -\big[g F_{+-}(z),\, {D}_{z^{j}}\big]
\label{plus_minus_perp_double_commutator}
\, .
\end{align}
Then, from Eqs.~\eqref{Dxmin_scalar_prop_eq_L_sol_final_3}, \eqref{deriv_scalar_prop_flip_minus_NEik_1}, \eqref{deriv_scalar_prop_flip_perp_exact_2} and \eqref{plus_minus_perp_double_commutator}, one obtains
\begin{align}
& \left(i\, \gamma^j D_{x^j}+m\right)  G_F(x,y)+2i\gamma^j\, D_{x^-}\int d^Dz\, G_F(x,z)\, g {F^{-}}_{j}(z)\, G_F(z,y)
= 
G_F(x,y)\, \left(-i\, \gamma^j  \overleftarrow{D}_{y^j}+m\right)
 \nonumber\\
 &\, \hspace{2cm}
+4i\, \gamma^j\int d^Dz\, \int d^Dw\, G_F(x,w)\, g F_{+-}(w)\, G_F(w,z) \overleftarrow{D}_{z^{-}}\,
 g {F^{-}}_{j}(z)\, G_F(z,y)
  \nonumber\\
 &\, \hspace{2cm}
 +i\, \gamma^j\int d^Dz\, G_F(x,z)\, \Big\{
 2g {F_j}^{i}(z)\, \overrightarrow{D}_{z^{i}} +\big[{D}_{z^{i}}, g {F_j}^{i}(z)\big]
+\big[{D}_{z^{j}},\, g F_{+-}(z)\big]
\Big\}G_F(z,y)
\nonumber\\
&\, \hspace{2cm}
-2i\, \gamma^j\int d^Dz\, G_F(x,z)\, g {F^{+}}_j(z)\,
\Delta(z,y)
 +\textrm{NNEik}
\label{GG_perp_deriv_flip_NEik_1}
\, .
\end{align}

Using Eq.~\eqref{GG_perp_deriv_flip_NEik_1} (and~\eqref{deriv_scalar_prop_flip_minus_NEik_1}) to simplify Eq.~\eqref{GG_block_L_NEik} then leads to
\begin{align}
[{\cal P}_G S_F(x,y) {\cal P}_G] = &\,  {\cal P}_G\, G_F(x,y)\, \left(-i\, \gamma^j  \overleftarrow{D}_{y^j}+m\right)
-2i{\cal P}_G\, \gamma^j\int d^Dz\, G_F(x,z)\, g {F^{+}}_j(z)\, \Delta(z,y)
\nonumber\\
&
+4i{\cal P}_G\,\, \gamma^j\int d^Dz\, \int d^Dw\, G_F(x,w)\, g F_{+-}(w)\, G_F(w,z) \overleftarrow{D}_{z^{-}}\,
 g {F^{-}}_{j}(z)\, G_F(z,y)
  \nonumber\\
 &
 +i{\cal P}_G\,\, \gamma^j\int d^Dz\, G_F(x,z)\, \Big\{
 2g {F_j}^{i}(z)\, \overrightarrow{D}_{z^{i}} +\big[{D}_{z^{i}}, g {F_j}^{i}(z)\big]
+\big[{D}_{z^{j}},\, g F_{+-}(z)\big]
\Big\}G_F(z,y)
  \nonumber\\
 &\hspace{-1cm}
 -2i\, {\cal P}_G \int d^Dz\,
\int d^Dw\, G_F(x,w) \overleftarrow{D}_{w^-}
\left(\frac{1}{4}\, [\gamma^{i},\gamma^{j}]\, g F_{ij}(w) + g F_{+-}(w)\right)\,
G_F(w,z)\gamma^{l}\, g {F^{-}}_{l}(z)\, G_F(z,y)
\nonumber\\
&
+ {\cal P}_G \int d^Dz\,
G_F(x,z)\, \left(-i\, \gamma^l  \overleftarrow{D}_{z^l}+m\right)\,
\left(\frac{1}{4}\, [\gamma^{i},\gamma^{j}]\, g F_{ij}(z) - g F_{+-}(z)\right)\,
G_F(z,y)
+\textrm{NNEik}
\\
= &\,
  {\cal P}_G\, G_F(x,y)\, \left(-i\, \gamma^j  \overleftarrow{D}_{y^j}+m\right)
-2i{\cal P}_G\, \gamma^j\int d^Dz\, G_F(x,z)\, g {F^{+}}_j(z)\, \Delta(z,y)
  \nonumber\\
 &\hspace{-1cm}
 +{\cal P}_G\,\int d^Dz\, G_F(x,z)\, \Big\{
 2ig {F_j}^{i}(z)\, \gamma^j\, \overrightarrow{D}_{z^{i}} +i\gamma^j\, \big[{D}_{z^{i}}, g {F_j}^{i}(z)\big]
- g F_{+-}(z)\, \left(i\, \gamma^l  \overrightarrow{D}_{z^l}+m\right)\,
\Big\}G_F(z,y)
  \nonumber\\
 &\hspace{-1cm}
 +2i\, {\cal P}_G \int d^Dz\,
\int d^Dw\, G_F(x,w)
\left(\frac{1}{4}\, [\gamma^{i},\gamma^{j}]\, g F_{ij}(w) - g F_{+-}(w)\right)\, \overrightarrow{D}_{w^-}
G_F(w,z)\gamma^{l}\, g {F^{-}}_{l}(z)\, G_F(z,y)
\nonumber\\
&
+ {\cal P}_G \int d^Dz\,
G_F(x,z)\,
\bigg\{
 \Big[i\, \gamma^l  \overrightarrow{D}_{z^l},\,
\frac{1}{4}\, [\gamma^{i},\gamma^{j}]\, g F_{ij}(z)\Big] \,
\nonumber\\
&\, \hspace{4cm}
+\frac{1}{4}\, [\gamma^{i},\gamma^{j}]\, g F_{ij}(z) \,
\left(i\, \gamma^l  \overrightarrow{D}_{z^l}+m\right)
\bigg\}
G_F(z,y)
+\textrm{NNEik}
\label{GG_block_L_NEik_app_2}
\, .
\end{align}
On the other hand,
\begin{align}
\Big[i\, \gamma^l  {D}_{z^l},\,
\frac{1}{4}\, [\gamma^{i},\gamma^{j}]\, g F_{ij}(z)\Big]
=&\Big[\, \gamma^l  {D}_{z^l},\,
\frac{1}{4}\, [\gamma^{i},\gamma^{j}]\, [{D}_{z^i},\, {D}_{z^j}]\Big]
\nonumber\\
=&\Big[\, \gamma^l  {D}_{z^l},\,
\frac{1}{2}\, [\gamma^{i},\gamma^{j}]\, {D}_{z^i}\, {D}_{z^j}\Big]
\nonumber\\
=&\Big[\, \gamma^l  {D}_{z^l},\,
\left(\gamma^{i}\gamma^{j} -\frac{1}{2}\, \{\gamma^{i},\gamma^{j}\}\right)\, {D}_{z^i}\, {D}_{z^j}\Big]
\nonumber\\
=&\Big[\gamma^l  {D}_{z^l}\, ,\: \gamma^i  {D}_{z^i}\, \gamma^j  {D}_{z^j}\Big]
-g^{ij}\, \gamma^l
\Big[ {D}_{z^l}\, ,\, {D}_{z^i}\, {D}_{z^j}\Big]
\nonumber\\
=& -g^{ij}\, \gamma^l
\bigg(
\Big[ {D}_{z^l}\, ,\, {D}_{z^i}\Big]\, {D}_{z^j}
+{D}_{z^i}\,\Big[ {D}_{z^l}\, ,\,  {D}_{z^j}\Big]
\bigg)
\nonumber\\
=& -\, \gamma^l
\bigg(ig {F_{l}}^{i}(z)\, {D}_{z^i}
+{D}_{z^i}\,ig {F_{l}}^{i}(z)
\bigg)
\nonumber\\
=& -2ig {F_{j}}^{i}(z)\,\gamma^j\, {D}_{z^i}
-\, \gamma^j
\Big[{D}_{z^i}\, ,\,ig {F_{j}}^{i}(z)
\Big]
\label{perp_perp_perp_double_commutator}
\, ,
\end{align}
leading to some cancelation in Eq.~\eqref{GG_block_L_NEik_app_2}, which then becomes
\begin{align}
[{\cal P}_G S_F(x,y) {\cal P}_G]
= &\,
  {\cal P}_G\, G_F(x,y)\, \left(-i\, \gamma^j  \overleftarrow{D}_{y^j}+m\right)
-2i{\cal P}_G\, \gamma^j\int d^Dz\, G_F(x,z)\, g {F^{+}}_j(z)\, \Delta(z,y)
  \nonumber\\
 &
 +{\cal P}_G\,\int d^Dz\, G_F(x,z)\, \left(
\frac{1}{4}\, [\gamma^{i},\gamma^{j}]\, g F_{ij}(z) \,- g F_{+-}(z)\right)
  \nonumber\\
 &
 \,\times\,
\bigg\{\left(i\, \gamma^l  \overrightarrow{D}_{z^l}+m\right)\,
G_F(z,y)
 + 2i\, \overrightarrow{D}_{z^-}
\int d^Dw\,
G_F(z,w)\gamma^{l}\, g {F^{-}}_{l}(w)\, G_F(w,y)\bigg\}
+\textrm{NNEik}
\label{GG_block_L_NEik_app_3}
\, .
\end{align}
After using Eq.~\eqref{GG_perp_deriv_flip_NEik_1} once again, Eq.~\eqref{GG_block_L_NEik_app_3} is found to be equivalent
to Eq.~\eqref{GG_block_R_NEik_2}. This completes our explicit proof of the equivalence of the expressions
~\eqref{GG_block_L_NEik}
and~\eqref{GG_block_R_NEik_2} for the ${\cal P}_G S_F(x,y) {\cal P}_G$ block.


\section{Details on the NEik expansion for the BG block}
\label{app:BG_block_exp}

Inserting the NEik expansion~\eqref{BB_block_L_NEik_2} for ${\cal P}_B S_F(x,y) {\cal P}_B$ into Eq.~\eqref{BG_block_L_exact_subtr}, one can write
\begin{align}
& \Big[{\cal P}_B S_F(x,y) {\cal P}_G - i\, \gamma^+ \Delta(x,y)\Big]
= \, {\cal I}_1 + {\cal I}_2 + {\cal I}_3 +{\cal I}_4 +{\cal I}_5
\nonumber\\
&\, \hspace{2cm}
+\int d^Dz\, \Big[{\cal P}_B S_F(x,z) {\cal P}_G- i\, \gamma^+ \Delta(x,z)\Big]\,
\left\{\frac{1}{4}\, [\gamma^{i},\gamma^{j}]\, g F_{ij}(z) - g F_{+-}(z)
\right\}\, G_F(z,y)
+\textrm{NNEik}
\label{BG_block_L_subtr_1}
\end{align}
with the inhomogeneous terms
\begin{align}
{\cal I}_1 \equiv &\,
i\, \gamma^+ \Big[D_{x^+} G_F(x,y) -\Delta(x,y)\Big]
\label{BG_block_L_I1}
\\
{\cal I}_2 \equiv &\,
i\, \gamma^+\int d^Dz\,  \Delta(x,z)\,
\left\{\frac{1}{4}\, [\gamma^{i},\gamma^{j}]\, g F_{ij}(z) - g F_{+-}(z)
\right\}\, G_F(z,y)
\label{BG_block_L_I2}
\\
{\cal I}_3 \equiv &\,
\int d^Dz\, {\cal P}_B\left(i\, \gamma^l D_{x^l}+m\right) G_F(x,z)\, \gamma^{+}\gamma^{j}\, g {F^{-}}_{j}(z)\, G_F(z,y)
\label{BG_block_L_I3}
\\
{\cal I}_4 \equiv &\,
\int d^Dz\, {\cal P}_B\left(i\, \gamma^l D_{x^l}+m\right) \int d^Dw\, G_F(x,w)\,
\left(\frac{1}{4}\, [\gamma^{i},\gamma^{j}]\, g F_{ij}(w) + g F_{+-}(w)
\right)\, G_F(w,z)\, \gamma^{+}\gamma^{m}\, g {F^{-}}_{m}(z)\, G_F(z,y)
\label{BG_block_L_I4}
\\
{\cal I}_5 \equiv &\,
\int d^Dz\, 2i\,{\cal P}_B\,  \int d^Dw\,  \Delta(x,w)\, \gamma^{i}\, g {F^+}_{i}(w)\, G_F(w,z)\, \gamma^{+}\gamma^{j}\, g {F^{-}}_{j}(z)\, G_F(z,y)
\label{BG_block_L_I5}
\end{align}
Note that the terms ${\cal I}_1$, ${\cal I}_2$ and ${\cal I}_3$ contribute at Eik accuracy (meaning at order $(\gamma_T)^0$), whereas ${\cal I}_4$ and ${\cal I}_5$ are NEik contributions (meaning of order $1/\gamma_T$). Hence, only the terms ${\cal I}_1$, ${\cal I}_2$ and ${\cal I}_3$ need to be inserted in the homogeneous term on the right-hand side of Eq.~\eqref{BG_block_L_subtr_1} in order to calculate ${\cal P}_B S_F(x,y) {\cal P}_G$ at NEik accuracy.

From Eq.\eqref{Dxmin_scalar_prop_eq_L_sol_final_3}, one has
\begin{align}
{\cal I}_1 = &\,
 \gamma^+
\int d^{D}z\, \Delta(x,z)\;
 \Big[ -\delta^{ij}\, D_{z^{i}}D_{z^{j}} +m^2 +ig F_{+-}(z)
\Big]\, G_F(z,y)+\cdots
\label{BG_block_L_I1_1}
\, ,
\end{align}
where the dots stand for the zero mode contribution which should not appear in the high-energy power series, as discussed in Sec.~\eqref{sec:taming_enhancement}.
Then, combining Eqs.~\eqref{BG_block_L_I2} and \eqref{BG_block_L_I1_1} leads to
\begin{align}
{\cal I}_1+{\cal I}_2 =&\,
 \gamma^+
\int d^{D}z\, \Delta(x,z)\;
 \left[ \left(\frac{\{\gamma^i,\gamma^j\}}{2}\, D_{z^{i}}D_{z^{j}} +m^2 +ig F_{+-}(z)\right)
 +\left(\frac{[\gamma^{i},\gamma^{j}]}{2}\, D_{z^{i}}D_{z^{j}} - ig F_{+-}(z)
\right)
\right]\, G_F(z,y)+\cdots
\nonumber\\
=&\,
 \gamma^+
\int d^{D}z\, \Delta(x,z)\;
 \Big[ \gamma^i D_{z^{i}} \gamma^jD_{z^{j}} +m^2
\Big]\, G_F(z,y)+\cdots
\nonumber\\
=&\,
\int d^{D}z\, \Delta(x,z)\; \big( -i\gamma^i\,\overleftarrow{D}_{z^{i}} +m \big)\gamma^+
\big( i\gamma^j\,\overrightarrow{D}_{z^{j}} +m \big)
\, G_F(z,y)+\cdots
\label{BG_block_L_I12_1}
\end{align}
Noting that
\begin{align}
-2\int d^Dz\,   \Delta(x,z)\, g {F^+}_{j}(z)\,  \Delta(z,y)
=&\,
2i\int d^Dz\,   \Delta(x,z)\, \big( -\overleftarrow{D}_{z^-}  \overrightarrow{D}_{z^j}   +\overleftarrow{D}_{z^j}  \overrightarrow{D}_{z^-}\big)\,  \Delta(z,y)
\nonumber\\
=&\,  \overrightarrow{D}_{x^j}\Delta(x,y) +\Delta(x,y)\overleftarrow{D}_{y^j}
\, ,
\label{Fplusperp_insertion_in_Delta}
\end{align}
thanks to Eq.~\eqref{eq_Delta}, one obtains
\begin{align}
{\cal I}_1+{\cal I}_2 =&\,
\big( i\gamma^i\,{D}_{x^{i}} +m \big)\gamma^+\int d^{D}z\, \Delta(x,z)\;
\big( i\gamma^j\,\overrightarrow{D}_{z^{j}} +m \big)
\, G_F(z,y)
\\
&\, +2i\gamma^i\gamma^+\int d^{D}z\, \int d^{D}w\, \Delta(x,w)\, g {F^+}_{i}(w)\,  \Delta(w,z)
\big( i\gamma^j\,\overrightarrow{D}_{z^{j}} +m \big)
\, G_F(z,y)
+\cdots
\label{BG_block_L_I12_2}
\end{align}

On the other hand,
${\cal I}_3$, ${\cal I}_4$ and ${\cal I}_5$ contain a ${F^{-}}_{l}(z)$ insertion into the scalar propagator $G_F$, which can be expressed as
\begin{align}
\int d^Dz\,  G_F(x,z)\, g {F^{-}}_{l}(z)\, G_F(z,y)
= &\,
-i\int d^Dz\,  G_F(x,z)\, \big( -\overleftarrow{D}_{z^+}  \overrightarrow{D}_{z^l}   -\overrightarrow{D}_{z^l}  \overrightarrow{D}_{z^+}\big)\, G_F(z,y)
\nonumber\\
= &\, -i\int d^Dz\, \bigg[\Delta(x,z)\, \overrightarrow{D}_{z^l}\, G_F(z,y)
    -G_F(x,z)\, \overrightarrow{D}_{z^l}\Delta(z,y) \bigg]
\nonumber\\
&\hspace{-2cm}
+\int d^Dz\,
\int d^{D}w\, G_F(x,w)\;
 \Big[ \delta^{ij}\, \overrightarrow{D}_{w^{i}}\overrightarrow{D}_{w^{j}} -m^2 +ig F_{+-}(w)
\Big]\, \Delta(w,z)
\overrightarrow{D}_{z^l}\, G_F(z,y)
\nonumber\\
&\hspace{-2cm}
-\int d^Dz\,
\int d^{D}w\, G_F(x,w)\;
\overrightarrow{D}_{w^l}\,
\Delta(w,z)
 \Big[ \delta^{ij}\, \overrightarrow{D}_{z^{i}}\overrightarrow{D}_{z^{j}} -m^2 -ig F_{+-}(z)
\Big]\,
G_F(z,y)
+\cdots
\label{F_minus_perp_insertion}
\end{align}
using Eqs.~\eqref{Dxmin_scalar_prop_eq_L_sol_final_3} and \eqref{Dxmin_scalar_prop_eq_R_sol_final}.
Hence, one has
\begin{align}
{\cal I}_3 = &\,
\left(i\, \gamma^l D_{x^l}+m\right)\gamma^{+}\int d^Dz\, \bigg[\Delta(x,z)\, (-i)\gamma^j\overrightarrow{D}_{z^j}\, G_F(z,y)
    + G_F(x,z)\, i\gamma^{j}\,\overrightarrow{D}_{z^j}\Delta(z,y) \bigg]
\nonumber\\
&\hspace{-2cm}
+\left(i\, \gamma^l D_{x^l}+m\right)\gamma^{+}\int d^Dz\,
\int d^{D}w\, G_F(x,w)\;
 \Big[ \delta^{ij}\, \overrightarrow{D}_{w^{i}}\overrightarrow{D}_{w^{j}} -m^2 +ig F_{+-}(w)
\Big]\, \Delta(w,z)
\gamma^{m}\overrightarrow{D}_{z^m}\, G_F(z,y)
\nonumber\\
&\hspace{-2cm}
-\left(i\, \gamma^l D_{x^l}+m\right)\gamma^{+}\int d^Dz\,
\int d^{D}w\, G_F(x,w)\;
\gamma^{m}\overrightarrow{D}_{w^m}\,
\Delta(w,z)
 \Big[ \delta^{ij}\, \overrightarrow{D}_{z^{i}}\overrightarrow{D}_{z^{j}} -m^2 -ig F_{+-}(z)
\Big]\,
G_F(z,y)
+\textrm{NNEik}
\label{BG_block_L_I3_1}
\, ,
\end{align}
\begin{align}
{\cal I}_4 \equiv &\,
\left(i\, \gamma^l D_{x^l}+m\right) \gamma^{+}
\int d^Dz\,  \int d^Dw\, G_F(x,w)\,
\left(\frac{1}{4}\, [\gamma^{i},\gamma^{j}]\, g F_{ij}(w) + g F_{+-}(w)
\right)\,
\nonumber\\
&\hspace{2cm}\times\;
\bigg[\Delta(w,z)\, (-i)\gamma^m\overrightarrow{D}_{z^m}\, G_F(z,y)
    + G_F(w,z)\, i\gamma^{m}\,\overrightarrow{D}_{z^m}\Delta(z,y) \bigg]
\label{BG_block_L_I4_1}
\end{align}
and
\begin{align}
{\cal I}_5 \equiv &\,
2i\,\int d^Dz\,   \int d^Dw\,  \Delta(x,w)\,  g {F^+}_{i}(w)\, \gamma^{i}\gamma^{+}\, \bigg[\Delta(w,z)\, (-i)\gamma^j\overrightarrow{D}_{z^j}\, G_F(z,y)
    + G_F(w,z)\, i\gamma^{j}\,\overrightarrow{D}_{z^j}\Delta(z,y) \bigg]
\label{BG_block_L_I5_1}
\, .
\end{align}

Moreover, consecutive $G_F$ and $\Delta$ factors convoluted with each other can be flipped up to corrections as
\begin{align}
\int d^Dz\,  G_F(x,z) \Delta(z,y)
=&\,
\int d^Dz\,  G_F(x,z)\, D_{z^+}G_F(z,y)
\nonumber\\
&\,
-i\int d^Dz\,  G_F(x,z)\,
\int d^{D}w\, \Delta(z,w)\;
 \Big[ \delta^{ij}\, D_{w^{i}}D_{w^{j}} -m^2 -ig F_{+-}(w)
\Big]\, G_F(w,y)
+\cdots
\nonumber\\
=&\,
\int d^Dz\,  \Delta(x,z)\, G_F(z,y)
\nonumber\\
&\,
+i\int d^Dz\,\int d^{D}w\, G_F(x,w)\;
 \Big[ \delta^{ij}\, D_{w^{i}}D_{w^{j}} -m^2 +ig F_{+-}(w)
\Big]\, \Delta(w,z)\, G_F(z,y)
\nonumber\\
&\,
-i\int d^Dz\,  G_F(x,z)\,
\int d^{D}w\, \Delta(z,w)\;
 \Big[ \delta^{ij}\, D_{w^{i}}D_{w^{j}} -m^2 -ig F_{+-}(w)
\Big]\, G_F(w,y)
+\cdots
\nonumber\\
=&\,
\int d^Dz\,  \Delta(x,z)\, G_F(z,y)
\nonumber\\
&\,
-\int d^Dz\,\int d^{D}w\, G_F(x,w)\;
 \Big[  g F_{+-}(w)\, \Delta(w,z)
 +\Delta(w,z)\, g F_{+-}(z)
\Big]\, G_F(z,y)
+\textrm{NNEik}
\label{Delta_GF_flip}
\end{align}
using Eq.~\eqref{Fplusperp_insertion_in_Delta} in the last step.
The first term is an Eikonal contribution, whereas the other two are NEik corrections.

Thanks to this relation, Eqs.~\eqref{BG_block_L_I12_2} and \eqref{BG_block_L_I5_1} can be combined as
\begin{align}
{\cal I}_1+{\cal I}_2 +{\cal I}_5
=&\,
\big( i\gamma^i\,{D}_{x^{i}} +m \big)\gamma^+\int d^{D}z\, \Delta(x,z)\;
\big( i\gamma^j\,\overrightarrow{D}_{z^{j}} +m \big)
\, G_F(z,y)
\nonumber\\
&\, +2i\gamma^i\gamma^+\int d^{D}z\, \int d^{D}w\, \Delta(x,w)\, g {F^+}_{i}(w)\,  G_F(w,z)
\big( i\gamma^j\,\overrightarrow{D}_{z^{j}} +m \big)
\, \Delta(z,y)
+\textrm{NNEik}
\nonumber\\
=&\,
\big( i\gamma^i\,{D}_{x^{i}} +m \big)\gamma^+\int d^{D}z\, \Delta(x,z)\;
\big( i\gamma^j\,\overrightarrow{D}_{z^{j}} +m \big)
\, G_F(z,y)
\nonumber\\
&\, +2i\gamma^i\gamma^+\int d^{D}z\, \int d^{D}w\, \Delta(x,w)\, g {F^+}_{i}(w)\,  G_F(w,z)
\, \Delta(z,y)\, \big( -i\gamma^j\,\overleftarrow{D}_{y^{j}} +m \big)
+\textrm{NNEik}
\label{BG_block_L_I125_1}
\, .
\end{align}

When summing ${\cal I}_3$ and ${\cal I}_4$, the term in the second line of Eq.~\eqref{BG_block_L_I3_1} and the first term in Eq.~\eqref{BG_block_L_I4_1} can be combined using the same steps as in Eq.~\eqref{BG_block_L_I12_1}. Then:
\begin{align}
{\cal I}_3+{\cal I}_4
= &\,
\left(i\, \gamma^l D_{x^l}+m\right)\gamma^{+}\int d^Dz\, \bigg[\Delta(x,z)\, (-i)\gamma^j\overrightarrow{D}_{z^j}\, G_F(z,y)
    + G_F(x,z)\, i\gamma^{j}\,\overrightarrow{D}_{z^j}\Delta(z,y) \bigg]
\nonumber\\
&\hspace{-2cm}
-\left(i\, \gamma^l D_{x^l}+m\right)\gamma^{+}\int d^Dz\,
\int d^{D}w\, G_F(x,w)\;
 \Big[  \gamma^i\overrightarrow{D}_{w^{i}}\gamma^j\overrightarrow{D}_{w^{j}} +m^2
\Big]\, \Delta(w,z)
\gamma^{m}\overrightarrow{D}_{z^m}\, G_F(z,y)
\nonumber\\
&\hspace{-2cm}
-\left(i\, \gamma^l D_{x^l}+m\right)\gamma^{+}\int d^Dz\,
\int d^{D}w\, G_F(x,w)\;
\gamma^{m}\overrightarrow{D}_{w^m}\,
\Delta(w,z)
 \Big[ \delta^{ij}\, \overrightarrow{D}_{z^{i}}\overrightarrow{D}_{z^{j}} -m^2 -ig F_{+-}(z)
\Big]\,
G_F(z,y)
\nonumber\\
&\hspace{-2cm}
+\left(i\, \gamma^l D_{x^l}+m\right) \gamma^{+}
\int d^Dz\,  \int d^Dw\, G_F(x,w)\,
\left(\frac{1}{4}\, [\gamma^{i},\gamma^{j}]\, g F_{ij}(w) + g F_{+-}(w)
\right)\,
 G_F(w,z)\, i\gamma^{m}\,\overrightarrow{D}_{z^m}\Delta(z,y)
+\textrm{NNEik}
\nonumber\\
= &\,
\left(i\, \gamma^l D_{x^l}+m\right)\gamma^{+}\int d^Dz\, \bigg[\Delta(x,z)\, (-i)\gamma^j\overrightarrow{D}_{z^j}\, G_F(z,y)
    + G_F(x,z)\, i\gamma^{j}\,\overrightarrow{D}_{z^j}\Delta(z,y) \bigg]
\nonumber\\
&\hspace{-2cm}
-\left(i\, \gamma^l D_{x^l}+m\right)\gamma^{+}\int d^Dz\,
\int d^{D}w\, G_F(x,w)\, \Delta(w,z)\;
 i\gamma^{m}\overrightarrow{D}_{z^m}\,\left( \frac{1}{4}\, [\gamma^{i},\gamma^{j}]\, g F_{ij}(z) -g F_{+-}(z)
\right)
\, G_F(z,y)
\nonumber\\
&\hspace{-2cm}
+\left(i\, \gamma^l D_{x^l}+m\right) \gamma^{+}
\int d^Dz\,  \int d^Dw\, G_F(x,w)\,
\left(\frac{1}{4}\, [\gamma^{i},\gamma^{j}]\, g F_{ij}(w) + g F_{+-}(w)
\right)\,
 G_F(w,z)\, i\gamma^{m}\,\overrightarrow{D}_{z^m}\Delta(z,y)
+\textrm{NNEik}
\label{BG_block_L_I34_1}
\end{align}
using Eq.~\eqref{Fplusperp_insertion_in_Delta} to perform simplifications.

In both Eqs.~\eqref{BG_block_L_I125_1} and \eqref{BG_block_L_I34_1}, the terms in the first line are Eikonal, whereas the following ones are NEik corrections.
There is an obvious cancellation between the first term of Eq.~\eqref{BG_block_L_I125_1} and the first term of Eq.~\eqref{BG_block_L_I34_1}. The remaining Eikonal terms can be rewritten as
\begin{align}
&\,
\big( i\gamma^i\,{D}_{x^{i}} +m \big)\gamma^+\int d^{D}z\,
\bigg\{m\, \Delta(x,z)\, G_F(z,y)
+G_F(x,z)\, i\gamma^{j}\,\overrightarrow{D}_{z^j}\Delta(z,y)
\bigg\}
\nonumber\\
=&\, \big( i\gamma^i\,{D}_{x^{i}} +m \big)\gamma^+\int d^{D}z\,
\bigg\{m\, \Delta(x,z)\, G_F(z,y)
+G_F(x,z)\, \Delta(z,y)\, (-i)\gamma^{j}\,\overleftarrow{D}_{y^j}
\bigg\}
\nonumber\\
&\, -2i\big( i\gamma^i\,{D}_{x^{i}} +m \big)\gamma^+\gamma^{j}\int d^{D}z\, G_F(x,z)\,
\int d^{D}w\,  \Delta(z,w)\, g {F^+}_{j}(w)\, \Delta(w,y)
+\textrm{NNEik}
\nonumber\\
=&\, \big( i\gamma^i\,{D}_{x^{i}} +m \big)\frac{\gamma^+}{2}\int d^{D}z\,
\Big\{\Delta(x,z)\, G_F(z,y)
+G_F(x,z)\, \Delta(z,y)\,
\Big\}\big( -i\gamma^{j}\,\overleftarrow{D}_{y^j} +m \big)
\nonumber\\
&\,   +\big( i\gamma^i\,{D}_{x^{i}} +m \big)\, \frac{\gamma^+}{2}\, m
 \int d^Dz\,\int d^{D}w\, G_F(x,w)\;
 \Big[  g F_{+-}(w)\, \Delta(w,z)+\Delta(w,z)\, g F_{+-}(z)
\Big]\, G_F(z,y)
\nonumber\\
&\,
-\big( i\gamma^i\,{D}_{x^{i}} +m \big)\, \frac{\gamma^+}{2}\,
\int d^Dz\,\int d^{D}w\, G_F(x,w)\;
 \Big[  g F_{+-}(w)\, \Delta(w,z)+\Delta(w,z)\, g F_{+-}(z)
\Big]\, G_F(z,y)
(-i)\gamma^{j}\,\overleftarrow{D}_{y^j}
\nonumber\\
&\,
-2i\big( i\gamma^i\,{D}_{x^{i}} +m \big)\gamma^+\gamma^{j}\int d^{D}z\, \int d^{D}w\, G_F(x,z)\,
  \Delta(z,w)\, g {F^+}_{j}(w)\, \Delta(w,y)
+\textrm{NNEik}
\label{BG_block_L_I12345_Eik_1}
\end{align}
using first the relation \eqref{Fplusperp_insertion_in_Delta}, and then the relation \eqref{Delta_GF_flip} in order to symmetrize the Eikonal term.
Using the result \eqref{BG_block_L_I12345_Eik_1} together with Eqs.~\eqref{BG_block_L_I125_1} and \eqref{BG_block_L_I34_1}
one finds
\begin{align}
&\, {\cal I}_1+{\cal I}_2 +{\cal I}_3+{\cal I}_4 +{\cal I}_5
=
\, \big( i\gamma^i\,{D}_{x^{i}} +m \big)\frac{\gamma^+}{2}\int d^{D}z\,
\Big\{\Delta(x,z)\, G_F(z,y)
+G_F(x,z)\, \Delta(z,y)\,
\Big\}\big( -i\gamma^{j}\,\overleftarrow{D}_{y^j} +m \big)
\nonumber\\
&\,
-\left(i\, \gamma^l D_{x^l}+m\right)\gamma^{+}\int d^Dz\,
\int d^{D}w\, G_F(x,w)\, \Delta(w,z)\;
 i\gamma^{m}\overrightarrow{D}_{z^m}\,\left( \frac{1}{4}\, [\gamma^{i},\gamma^{j}]\, g F_{ij}(z) -g F_{+-}(z)
\right)
\, G_F(z,y)
\nonumber\\
&\, 
+\left(i\, \gamma^l D_{x^l}+m\right) \gamma^{+}
\int d^Dz\,  \int d^Dw\, G_F(x,w)\,
\left(\frac{1}{4}\, [\gamma^{i},\gamma^{j}]\, g F_{ij}(w) + g F_{+-}(w)
\right)\,
 G_F(w,z)\, i\gamma^{m}\,\overrightarrow{D}_{z^m}\Delta(z,y)
\nonumber\\
&\,
+\big( i\gamma^i\,{D}_{x^{i}} +m \big)\, \frac{\gamma^+}{2}\,
\int d^Dz\,\int d^{D}w\, G_F(x,w)\;
 \Big[  g F_{+-}(w)\, \Delta(w,z)+\Delta(w,z)\, g F_{+-}(z)
\Big]\, G_F(z,y)
\big( i\gamma^{j}\,\overleftarrow{D}_{y^j} +m \big)
\nonumber\\
&\,
-2i\big( i\gamma^i\,{D}_{x^{i}} +m \big)\gamma^+\gamma^{j}\int d^{D}z\, \int d^{D}w\, G_F(x,z)\,
  \Delta(z,w)\, g {F^+}_{j}(w)\, \Delta(w,y)
\nonumber\\
&\, +2i\gamma^i\gamma^+\int d^{D}z\, \int d^{D}w\, \Delta(x,w)\, g {F^+}_{i}(w)\,  G_F(w,z)
\, \Delta(z,y)\, \big( -i\gamma^j\,\overleftarrow{D}_{y^{j}} +m \big)
+\textrm{NNEik}
\label{BG_block_L_I12345_NEik_1}
\,
\end{align}
where the first line is the Eikonal contribution, and the rest is the NEik contribution.

In order to find the NEik expansion of $[{\cal P}_B S_F(x,y) {\cal P}_G ]$, one still has to insert the expression \eqref{BG_block_L_I12345_NEik_1} for the inhomogeneous term into Eq.~\eqref{BG_block_L_subtr_1}. At strict NEik accuracy, in addition to the NEik expression \eqref{BG_block_L_I12345_NEik_1} for the inhomogeneous term, one has to include one step of iteration applied to the Eikonal contribution to the inhomogeneous term, which leads to
\begin{align}
& \Big[{\cal P}_B S_F(x,y) {\cal P}_G - i\, \gamma^+ \Delta(x,y)\Big]
= \,
\big( i\gamma^i\,{D}_{x^{i}} +m \big)\frac{\gamma^+}{2}\int d^{D}z\,
\Big\{\Delta(x,z)\, G_F(z,y)
+G_F(x,z)\, \Delta(z,y)\,
\Big\}\big( -i\gamma^{j}\,\overleftarrow{D}_{y^j} +m \big)
\nonumber\\
&\,
+\big( i\gamma^i\,{D}_{x^{i}} +m \big)\frac{\gamma^+}{2}\int d^{D}z\,\int d^{D}w\,
\Big\{\Delta(x,w)\, G_F(w,z)
+G_F(x,w)\, \Delta(w,z)\,
\Big\}\big( -i\gamma^{j}\,\overleftarrow{D}_{z^j} +m \big)
\nonumber\\
&\, \hspace{5cm}\times \;
\left\{\frac{1}{4}\, [\gamma^{i},\gamma^{j}]\, g F_{ij}(z) - g F_{+-}(z)
\right\}\, G_F(z,y)
\nonumber\\
&\,
-\left(i\, \gamma^l D_{x^l}+m\right)\gamma^{+}\int d^Dz\,
\int d^{D}w\, G_F(x,w)\, \Delta(w,z)\;
 i\gamma^{m}\overrightarrow{D}_{z^m}\,\left( \frac{1}{4}\, [\gamma^{i},\gamma^{j}]\, g F_{ij}(z) -g F_{+-}(z)
\right)
\, G_F(z,y)
\nonumber\\
&\, 
+\left(i\, \gamma^l D_{x^l}+m\right) \gamma^{+}
\int d^Dz\,  \int d^Dw\, G_F(x,w)\,
\left(\frac{1}{4}\, [\gamma^{i},\gamma^{j}]\, g F_{ij}(w) + g F_{+-}(w)
\right)\,
 G_F(w,z)\, i\gamma^{m}\,\overrightarrow{D}_{z^m}\Delta(z,y)
\nonumber\\
&\,
+\big( i\gamma^i\,{D}_{x^{i}} +m \big)\, \frac{\gamma^+}{2}\,
\int d^Dz\,\int d^{D}w\, G_F(x,w)\;
 \Big[  g F_{+-}(w)\, \Delta(w,z)+\Delta(w,z)\, g F_{+-}(z)
\Big]\, G_F(z,y)
\big( i\gamma^{j}\,\overleftarrow{D}_{y^j} +m \big)
\nonumber\\
&\,
-2i\big( i\gamma^i\,{D}_{x^{i}} +m \big)\gamma^+\gamma^{j}\int d^{D}z\, \int d^{D}w\, G_F(x,z)\,
  \Delta(z,w)\, g {F^+}_{j}(w)\, \Delta(w,y)
\nonumber\\
&\, +2i\gamma^i\gamma^+\int d^{D}z\, \int d^{D}w\, \Delta(x,w)\, g {F^+}_{i}(w)\,  G_F(w,z)
\, \Delta(z,y)\, \big( -i\gamma^j\,\overleftarrow{D}_{y^{j}} +m \big)
+\textrm{NNEik}
\label{BG_block_L_subtr_NEik}
\, .
\end{align}
Inside the NEik contributions, one can make simple manipulations based on Eqs.~\eqref{Fplusperp_insertion_in_Delta} and \eqref{Delta_GF_flip}, keeping only the leading terms. In particular, one can freely perform the flips $\Delta(x,z)\,  G_F(z,y)\leftrightarrow G_F(x,z)\,  \Delta(z,y)$  under the $z$ integral, and ${D}_{x^{j}}\Delta(x,y)\leftrightarrow -\Delta(x,y)\overleftarrow{D}_{y^{j}}$, within the NEik contributions, since the subleading corrections would matter only at NNEik accuracy at the propagator level. Moreover, remembering that $ig F_{ij}(z)\equiv [{D}_{z^{i}},{D}_{z^{j}}]$, one can also flip $g F_{ij}(x)\, \Delta(x,y)\leftrightarrow \Delta(x,y)\, g F_{ij}(y)$ within the NEik contribution. In such a way, the expression \eqref{BG_block_L_subtr_NEik} can be rewritten as in 
Eq.~\eqref{BG_block_L_subtr_NEik_result_2}.


\section{Study of NEik corrections for the scalar propagator from the $z^-$ dependence of $A^{-}(z)$\label{app:NEik_from_zmin_dep_scalar_prop}}

In this appendix, we perform a study of the possible NEik corrections to the scalar propagator arising from the  $z^-$ dependence of the gluon background field $A^{\mu}(z)$. Since the $A^{j}(z)$ and  $A^{+}(z)$ components are  power suppressed compared to the $A^{-}(z)$ component, it is sufficient to restrict ourselves to a pure $A^{-}(z)$ background, and start from Eq.~\eqref{expansion_scalar_prop_pure_A_minus_1}. For convenience, let us write Eq.~\eqref{expansion_scalar_prop_pure_A_minus_1} as 
\begin{align}
&\delta G_F(x,y)\bigg|_{\textrm{pure }A^{-}}
= 
\sum_{N=1}^{+\infty} \int \bigg[\prod_{n=1}^{N} d^{D-1} \underline{z_n}\bigg]\,
\int \bigg[\prod_{n=0}^{N} \frac{d^{D-2} \p_n}{(2\pi)^{D-2}}\;
e^{i \p_{n} \cdot (\z_{n+1}-\z_n)}
\bigg]\,
\nonumber\\
&\, 
\times\;
\int \frac{d q^+}{2\pi}\:
\frac{1}{2q^+} \,
e^{-i\, \frac{(\q^2+m^2)}{2q^+}\, (x^+-z_N^+)}\,
\Big[
\theta(x^+\!-\!z_N^+)\, \theta(q^+) - \theta(z_N^+\!-\!x^+)\, \theta(-q^+)
\Big]
\nonumber\\
&\, 
\times\;
\int  \frac{d k^+}{2\pi}\:
\frac{1}{2k^+} \,
e^{-i\, \frac{(\k^2+m^2)}{2k^+}\, (z_{1}^+-y^+)}\,
\Big[
\theta(z_{1}^+\!-\!y^+)\, \theta(k^+) - \theta(y^+\!-\!z_{1}^+)\, \theta(-k^+)
\Big]
\bigg]
\, I_{p^+}
\label{expansion_scalar_prop_pure_A_minus_1_app}
\, ,
\end{align}
introducing the notations
\begin{align}
I_{p^+}\equiv &\,
\int \bigg[\prod_{n=1}^{N-1} \frac{d p_n^+}{2\pi}\:
\frac{1}{2p_n^+} \,
e^{-i\, \frac{(\p_n^2+m^2)}{2p_n^+}\, (z_{n+1}^+-z_n^+)}\,
\Big[
\theta(z_{n+1}^+\!-\!z_n^+)\, \theta(p_n^+) - \theta(z_n^+\!-\!z_{n+1}^+)\, \theta(-p_n^+)
\Big]
\bigg]\,
\nonumber\\
&\, 
\,\times\;
\bigg[\prod_{n=1}^{N} \big(p_n^+ + p_{n-1}^+  \big)\bigg]\;
I_{z^-}
\label{Ipplus_def_app}
\end{align}
and
\begin{align}
I_{z^-}\equiv &\,
 \int \bigg[\prod_{n=1}^{N} d z_n^-\bigg]\,
\bigg[{\cal P}_n\, \prod_{n=1}^{N}\Big(-i g\, A^{-}(z_n)\Big)\bigg]\,
 \bigg[\prod_{n=0}^{N}
e^{-i p_{n}^+  (z_{n+1}^- -z_n^-)}\,
\bigg]
\nonumber\\
= &\,
 \int \bigg[\prod_{n=1}^{N} d z_n^-\bigg]\,
\bigg[{\cal P}_n\, \prod_{n=1}^{N}\Big(-i g\, A^{-}(z_n)\Big)\bigg]\,
e^{-i q^+  (x^- -z_N^-)}\,
e^{-i k^+  (z_{1}^- -y^-)}\,
 \bigg[\prod_{n=1}^{N-1}
e^{-i p_{n}^+  (z_{n+1}^- -z_n^-)}\,
\bigg]
\label{Izminus_def_app}
\, .
\end{align}
In order to derive systematically the non-Eikonal corrections associated with the non-trivial dependence of  $A^{-}(z)$ on $z^-$, we can Taylor expand each factor of $A^{-}(z_n)$ with respect to $z_n^-$ around a common typical value $z^-$. For that purpose, let us first perform in Eq.~\eqref{Izminus_def_app} the change of variables from $z_n^-$ (for $n$ from $1$ to $N$) to $w_n^-$ (for $n$ from $1$ to $N-1$) and $z^-$, defined as 
\begin{align}
w_n^- \equiv &\, z_{n+1}^- -  z_n^-
\nonumber\\
z^-  \equiv &\, \frac{(z_{N}^- +  z_1^-)}{2}
\label{ch_var_znminus}
\, .
\end{align}
The change of variables \eqref{ch_var_znminus} has a unit Jacobian, and can be inverted as
\begin{align}
z_n^- = &\, z^- +\frac{1}{2}\, \sum_{l=1}^{n-1}w_{l}^- -\frac{1}{2}\, \sum_{l=n}^{N-1}w_{l}^-
\label{ch_var_znminus_inverse}
\, , 
\end{align}
so that Eq.~\eqref{Izminus_def_app} becomes
\begin{align}
I_{z^-}= &\, \int  dz^-\,   e^{-i q^+  (x^- -z^-)}\,
e^{-i k^+  (z^- -y^-)}\,
 \int \bigg[\prod_{n=1}^{N-1} d w_n^-\: 
e^{-i \Delta p_{n}^+\,  w_n^-}\,
\bigg]\,
\nonumber\\
 &\, \times \,\, 
\left[{\cal P}_n\, \prod_{n=1}^{N}\left(-i g\, A^{-}\left(\underline{z_n}, z^- +\frac{1}{2} \sum_{l=1}^{n-1}w_{l}^- -\frac{1}{2} \sum_{l=n}^{N-1}w_{l}^-\right)\right)\right]\,
\label{Izminus_exp_1}
\, ,
\end{align}
using the notation $\Delta p_{n}^+$ defined in Eq.~\eqref{def_Delta_p_n_plus}. 
Derivatives of $A^{-}$ with respect to $z^-$ are suppressed as 
\begin{align}
\partial_-^m A^{-} \propto \frac{1}{(\gamma_T)^m}\,  A^{-}
\label{powercount_minus_deriv_Aminus}
\end{align}
in the limit of large Lorentz boosts of the target. Hence, to NEik accuracy, it is enough to keep only the first two terms in the Taylor expansion of each $A^{-}(z_n)$ around $z_n^-=z^-$. At the level of Eq.~\eqref{Izminus_exp_1}, the Eikonal contribution is obtained by keeping only the leading term in the Taylor expansion of each $A^{-}(z_n)$ factor. By contrast, the NEik contributions are obtained by taking the first order correction in the Taylor expansion of one $A^{-}(z_n)$ factor and the leading term in each of the other $A^{-}(z_n)$ factors. Thus, we obtain
\begin{align}
I_{z^-}= &\, \int  dz^-\,   e^{-i q^+  (x^- -z^-)}\,
e^{-i k^+  (z^- -y^-)}\,
 \int \bigg[\prod_{n=1}^{N-1} d w_n^-\: 
e^{-i \Delta p_{n}^+\,  w_n^-}\,
\bigg]\,
\Bigg\{
\left[{\cal P}_n\, \prod_{n=1}^{N}\left(-i g\, A^{-}\left(\underline{z_n}, z^- \right)\right)\right]\,
\nonumber\\
 &\, \quad 
+ 
\sum_{n_1=1}^{N} \frac{1}{2} \left[\sum_{l=1}^{{n_1}-1}w_{l}^- -\sum_{l=n_1}^{N-1}w_{l}^-\right]
\Bigg[
{\cal P}_n\, \left(-ig \partial_- A^-(\underline{z_{n_1}},z^-)\right) 
\prod_{\scriptsize{\begin{array}{c}
           n=1 \\
           n\neq n_1 \\
         \end{array}}}^{N}
\left(-ig  A^-(\underline{z_{n}},z^-)\right)
\Bigg]
+\textrm{NNEik}
\Bigg\}
\label{Izminus_exp_2}
\, .
\end{align}
The integrals over the $w_n^-$ variables can be performed using the relations
\begin{align}
\int dw^-\, e^{-i \Delta p^+\, w^-} = &\, 2\pi\, \delta(\Delta p^+)
\nonumber\\
\int dw^-\, w^-\,e^{-i \Delta p^+\, w^-} = &\, 2\pi\, i\, \delta'(\Delta p^+)
\, .
\end{align}
One finds
\begin{align}
I_{z^-}= &\, \int  dz^-\,   e^{-i q^+  (x^- -z^-)}\,
e^{-i k^+  (z^- -y^-)}\,
\Bigg\{
\left[\prod_{n=1}^{N-1} 
2\pi\, \delta(\Delta p_{n}^+)
\right]
\left[{\cal P}_n\, \prod_{n=1}^{N}\left(-i g\, A^{-}\left(\underline{z_n}, z^- \right)\right)\right]\,
\nonumber\\
 &\, \; 
+ 
\sum_{n_1=1}^{N} \frac{1}{2} 
\Bigg[
{\cal P}_n\, \left(-ig \partial_- A^-(\underline{z_{n_1}},z^-)\right) 
\prod_{\scriptsize{\begin{array}{c}
           n=1 \\
           n\neq n_1 \\
         \end{array}}}^{N}
\left(-ig  A^-(\underline{z_{n}},z^-)\right)
\Bigg]
\nonumber\\
 &\, \;\;\; \times\;
\Bigg[\sum_{l=1}^{{n_1}-1} 2\pi\, i\, \delta'(\Delta p_l^+)
\Bigg[
\prod_{\scriptsize{\begin{array}{c}
           n=1 \\
           n\neq l \\
         \end{array}}}^{N-1} 
2\pi\, \delta(\Delta p_n^+)
\Bigg]
- \sum_{l=n_1}^{N-1}
2\pi\, i\, \delta'(\Delta p_l^+)
\Bigg[
\prod_{\scriptsize{\begin{array}{c}
           n=1 \\
           n\neq l \\
         \end{array}}}^{N-1} 
2\pi\, \delta(\Delta p_n^+)
\Bigg]
\Bigg]
+\textrm{NNEik}
\Bigg\}
\label{Izminus_exp_3}
\, .
\end{align}
In Eq.~\eqref{Izminus_exp_3}, the sum over $n_1$ in the NEik contribution can be performed explicitly, leading to
\begin{align}
I_{z^-}= &\, \int  dz^-\,   e^{-i q^+  (x^- -z^-)}\,
e^{-i k^+  (z^- -y^-)}\,
\Bigg\{
\left[\prod_{n=1}^{N-1} 
2\pi\, \delta(\Delta p_{n}^+)
\right]
\left[{\cal P}_n\, \prod_{n=1}^{N}\left(-i g\, A^{-}\left(\underline{z_n}, z^- \right)\right)\right]\,
\nonumber\\
 &\, \; 
-\frac{1}{2} 
\sum_{l=1}^{N-1} 
\Bigg[ 
\left({\cal P}_n \prod_{ n=l+1}^{N}
\left(-ig A^-(\underline{z_{n}},z^-)\right) 
\right)
\overleftrightarrow{\partial_{z^-}}
\left({\cal P}_n\prod_{ n=1}^{l}
\left(-ig A^-(\underline{z_{n}},z^-)\right) 
\right)
\Bigg]
\nonumber\\
 &\, \;\hspace{2cm} \times \;
2\pi\, i\, \delta'(\Delta p_l^+)
\Bigg[
\prod_{\scriptsize{\begin{array}{c}
           n=1 \\
           n\neq l \\
         \end{array}}}^{N-1} 
2\pi\, \delta(\Delta p_n^+)
\Bigg]
+\textrm{NNEik}
\Bigg\}
\label{Izminus_exp_4}
\, ,
\end{align}
where the derivative with respect to $z^-$ acts within the square bracket, but not on the phase factors. The next step is to insert this result for $I_{z^-}$ in Eq.~\eqref{Ipplus_def_app}, in order to evaluate $I_{p^+}$. For simplicity, we decompose  $I_{p^+}$ as $I_{p^+}=I_{p^+}^{(0)}+I_{p^+}^{(1)}+\textrm{NNEik}$, where $I_{p^+}^{(0)}$ is the contribution arising from the term in the first line of Eq.~\eqref{Izminus_exp_4} in $I_{z^-}$, and $I_{p^+}^{(1)}$ is the contribution arising from the subleading term in $I_{z^-}$, corresponding to the second and third lines of Eq.~\eqref{Izminus_exp_4}.
For $I_{p^+}^{(0)}$, the delta functions present in the first term of Eq.~\eqref{Izminus_exp_4} allow us to perform all of the integrations present in Eq.~\eqref{Ipplus_def_app} (using the definitions \eqref{def_p_plus} and \eqref{def_Delta_p_n_plus}), which leads to
\begin{align}
I_{p^+}^{(0)} = &\,
 \int  dz^-\,   e^{-i q^+  (x^- -z^-)}\, e^{-i k^+  (z^- -y^-)}\,
\left[{\cal P}_n\, \prod_{n=1}^{N}\left(-i g\, A^{-}\left(\underline{z_n}, z^- \right)\right)\right]\,
 \big(q^+ + p^+  \big)\, \bigg[\prod_{n=2}^{N-1} \big(2p^+ \big)\bigg]\,  \big(p^+ + k^+  \big)
\nonumber\\
&\, 
\,\times\;
\bigg[\prod_{n=1}^{N-1}
\frac{1}{2p^+} \,
e^{-i\, \frac{(\p_n^2+m^2)}{2p^+}\, (z_{n+1}^+-z_n^+)}\,
\Big[
\theta(z_{n+1}^+\!-\!z_n^+)\, \theta(p^+) - \theta(z_n^+\!-\!z_{n+1}^+)\, \theta(-p^+)
\Big]
\bigg]\,
\nonumber\\
= &\,
\frac{1}{2p^+} \, \left[(2p^+)^2-\frac{(q^+\!-\!k^+)^2}{4}\right]
\bigg[\prod_{n=1}^{N-1}
e^{-i\, \frac{(\p_n^2+m^2)}{2p^+}\, (z_{n+1}^+-z_n^+)}\,
\Big[
\theta(z_{n+1}^+\!-\!z_n^+)\, \theta(p^+) - \theta(z_n^+\!-\!z_{n+1}^+)\, \theta(-p^+)
\Big]
\bigg]\,
\nonumber\\
&\, 
\,\times\;
 \int  dz^-\,   e^{-i q^+  (x^- -z^-)}\, e^{-i k^+  (z^- -y^-)}\,
\left[{\cal P}_n\, \prod_{n=1}^{N}\left(-i g\, A^{-}\left(\underline{z_n}, z^- \right)\right)\right]\,
\label{Ipplus_exp_0_1}
\, .
\end{align}
In Eq.~\eqref{Ipplus_exp_0_1}, each power of $(q^+\!-\!k^+)$ can be represented as a derivative with respect to $z^-$, following
\begin{align}
&\, (q^+\!-\!k^+) \int  dz^-\,   e^{-i q^+  (x^- -z^-)}\, e^{-i k^+  (z^- -y^-)}\,
\left[{\cal P}_n\, \prod_{n=1}^{N}\left(-i g\, A^{-}\left(\underline{z_n}, z^- \right)\right)\right]\,
\nonumber\\
= &\,
\int  dz^-\,   e^{-i q^+  (x^- -z^-)}\, e^{-i k^+  (z^- -y^-)}\;   (-i)  \overleftarrow{\partial_{z^-}}
\left[{\cal P}_n\, \prod_{n=1}^{N}\left(-i g\, A^{-}\left(\underline{z_n}, z^- \right)\right)\right]\,
\nonumber\\
= &\,
\int  dz^-\,   e^{-i q^+  (x^- -z^-)}\, e^{-i k^+  (z^- -y^-)}\;   (+i)  \overrightarrow{\partial_{z^-}}
\left[{\cal P}_n\, \prod_{n=1}^{N}\left(-i g\, A^{-}\left(\underline{z_n}, z^- \right)\right)\right]\,
\, .
\label{qplus_min_kplus_as_power_suppressed}
\end{align}
Hence, due to the power counting relation \eqref{powercount_minus_deriv_Aminus}, each factor of $(q^+\!-\!k^+)$ amounts to a suppression by a factor $1/\gamma_T$ in the limit of large Lorentz boost of the target. The second term in the first bracket in Eq.~\eqref{Ipplus_exp_0_1} is thus suppressed as $1/(\gamma_T)^2$ compared to the first term, so that it is a NNEik correction. Then, Eq.~\eqref{Ipplus_exp_0_1} becomes
\begin{align}
I_{p^+}^{(0)} = &\,
(2p^+)
\bigg[\prod_{n=1}^{N-1}
e^{-i\, \frac{(\p_n^2+m^2)}{2p^+}\, (z_{n+1}^+-z_n^+)}\,
\Big[
\theta(z_{n+1}^+\!-\!z_n^+)\, \theta(p^+) - \theta(z_n^+\!-\!z_{n+1}^+)\, \theta(-p^+)
\Big]
\bigg]\,
\nonumber\\
&\, 
\,\times\;
 \int  dz^-\,   e^{-i q^+  (x^- -z^-)}\, e^{-i k^+  (z^- -y^-)}\,
\left[{\cal P}_n\, \prod_{n=1}^{N}\left(-i g\, A^{-}\left(\underline{z_n}, z^- \right)\right)\right]\,
+\textrm{NNEik}
\label{Ipplus_exp_0_2}
\, .
\end{align}

Inserting the second contribution to $I_{z^-}$ from Eq.~\eqref{Izminus_exp_4}  into Eq.~\eqref{Ipplus_def_app}, one finds
\begin{align}
 I_{p^+}^{(1)} = &\,
 \int  dz^-\,   e^{-i q^+  (x^- -z^-)}\, e^{-i k^+  (z^- -y^-)}\,
\frac{(-1)}{2} 
\sum_{l=1}^{N-1} 
\Bigg[ 
\left({\cal P}_n \prod_{ n=l+1}^{N}
\left(-ig A^-(\underline{z_{n}},z^-)\right) 
\right)
\overleftrightarrow{\partial_{z^-}}
\left({\cal P}_n\prod_{ n=1}^{l}
\left(-ig A^-(\underline{z_{n}},z^-)\right) 
\right)
\Bigg]
\nonumber\\
&\, \hspace{-1cm}
\,\times\;
\int \bigg[\prod_{n=1}^{N-1} \frac{d \Delta p_n^+}{2\pi}\:
\frac{1}{2(p^++\Delta p_n^+)} \,
e^{-i\, \frac{(\p_n^2+m^2)}{2(p^++\Delta p_n^+)}\, (z_{n+1}^+-z_n^+)}\,
\Big[
\theta(z_{n+1}^+\!-\!z_n^+)\, \theta(p^++\Delta p_n^+) - \theta(z_n^+\!-\!z_{n+1}^+)\, \theta(-p^+-\Delta p_n^+)
\Big]
\bigg]\,
\nonumber\\
&\, \hspace{-1cm}
\,\times\;
\bigg[\prod_{n=1}^{N} \big(p_n^+ + p_{n-1}^+  \big)\bigg]\;  
2\pi\, i\, \delta'(\Delta p_l^+)
\Bigg[
\prod_{\scriptsize{\begin{array}{c}
           n=1 \\
           n\neq l \\
         \end{array}}}^{N-1} 
2\pi\, \delta(\Delta p_n^+)
\Bigg]
\nonumber\\
 = &\,
\int  dz^-\,   e^{-i q^+  (x^- -z^-)}\, e^{-i k^+  (z^- -y^-)}\,
\frac{(-1)}{2} 
\sum_{l=1}^{N-1} 
\Bigg[ 
\left({\cal P}_n \prod_{ n=l+1}^{N}
\left(-ig A^-(\underline{z_{n}},z^-)\right) 
\right)
\overleftrightarrow{\partial_{z^-}}
\left({\cal P}_n\prod_{ n=1}^{l}
\left(-ig A^-(\underline{z_{n}},z^-)\right) 
\right)
\Bigg]
\nonumber\\
&\, \hspace{-1cm}
\,\times\;
(-i)\partial_{(\Delta p_l^+)} \Bigg\{ {\cal R}(\{\Delta p_n^+\})\;
\prod_{n=1}^{N-1}  
e^{-i\, \frac{(\p_n^2+m^2)}{2(p^++\Delta p_n^+)}\, (z_{n+1}^+-z_n^+)}\,
\Big[
\theta(z_{n+1}^+\!-\!z_n^+)\, \theta(p^++\Delta p_n^+) - \theta(z_n^+\!-\!z_{n+1}^+)\, \theta(-p^+-\Delta p_n^+)
\Big]
\Bigg\}\Bigg|_{\Delta p_n^+\rightarrow 0}
\label{Ipplus_exp_1_1}
\, ,
\end{align}
where
\begin{align}
{\cal R}(\{\Delta p_n^+\}) 
\equiv &\,
\left(2p^+ \!+\!\frac{(q^+\!-\!k^+)}{2}\!+\!\Delta p_{N-1}^+\right)
\left(2p^+\!-\!\frac{(q^+\!-\!k^+)}{2}\!+\!\Delta p_{1}^+\right)
\left[\prod_{ n=2}^{N-1}  \left(2p^+ \!+\!\Delta p_{n}^+\!+\!\Delta p_{n-1}^+\right)
\right]
\left[\prod_{ n=1}^{N-1}
 \frac{1}{ 
\left(2p^+ \!+\!2\Delta p_{n}^+\right)
}
\right]
\label{def_rat_funct}
\, .
\end{align}

There are three possible types of contributions to Eq.~\eqref{Ipplus_exp_1_1}, with $\partial_{(\Delta p_l^+)}$ acting either on ${\cal R}(\{\Delta p_n^+\})$, or on a phase factor, or on a $\theta$ function. Let us discuss each of them separately.
When this derivative acts on ${\cal R}(\{\Delta p_n^+\})$, we have to distinguish between the cases $l=1$, $l=N-1$, or $2 \leq l \leq N-2$. For $2 \leq l \leq N-2$, one finds
\begin{align}
\partial_{(\Delta p_l^+)}  {\cal R}(\{\Delta p_n^+\}) \bigg|_{\Delta p_n^+\rightarrow 0}
= &\,
\left(2p^+ \!+\!\frac{(q^+\!-\!k^+)}{2}\right)
\left(2p^+\!-\!\frac{(q^+\!-\!k^+)}{2}\right)\left\{
\frac{(2p^+)^{N-3}}{(2p^+)^{N-1}} +\frac{(2p^+)^{N-3}}{(2p^+)^{N-1}}
-2\,\frac{(2p^+)^{N-2}}{(2p^+)^{N}}
\right\}
=0
\label{deriv_rat_funct_gen_case}
\, .
\end{align}
By contrast, for $l=1$,
\begin{align}
\partial_{(\Delta p_1^+)}  {\cal R}(\{\Delta p_n^+\}) \bigg|_{\Delta p_n^+\rightarrow 0}
= &\,
\left(2p^+ \!+\!\frac{(q^+\!-\!k^+)}{2}\right)
\bigg\{
\frac{(2p^+)^{N-2}}{(2p^+)^{N-1}}
+\left(2p^+\!-\!\frac{(q^+\!-\!k^+)}{2}\right)\, \frac{(2p^+)^{N-3}}{(2p^+)^{N-1}}
\nonumber\\
& \hspace{2cm}
-2 \left(2p^+\!-\!\frac{(q^+\!-\!k^+)}{2}\right)\, \frac{(2p^+)^{N-2}}{(2p^+)^{N}}
\bigg\}
\nonumber\\
= &\,
\left(2p^+ \!+\!\frac{(q^+\!-\!k^+)}{2}\right)\, \frac{1}{(2p^+)^2}\, \frac{(q^+\!-\!k^+)}{2}
\label{deriv_rat_funct_l_eq_1}
\end{align}
and for $l=N-1$,
\begin{align}
\partial_{(\Delta p_{N-1}^+)}  {\cal R}(\{\Delta p_n^+\}) \bigg|_{\Delta p_n^+\rightarrow 0}
= &\,
\left(2p^+ \!-\!\frac{(q^+\!-\!k^+)}{2}\right)
\bigg\{
\frac{(2p^+)^{N-2}}{(2p^+)^{N-1}}
+\left(2p^+\!+\!\frac{(q^+\!-\!k^+)}{2}\right)\, \frac{(2p^+)^{N-3}}{(2p^+)^{N-1}}
\nonumber\\
& \hspace{2cm}
-2 \left(2p^+\!+\!\frac{(q^+\!-\!k^+)}{2}\right)\, \frac{(2p^+)^{N-2}}{(2p^+)^{N}}
\bigg\}
\nonumber\\
= &\,
\left(2p^+ \!-\!\frac{(q^+\!-\!k^+)}{2}\right)\, \frac{(-1)}{(2p^+)^2}\, \frac{(q^+\!-\!k^+)}{2}
\label{deriv_rat_funct_l_eq_Nmin1}
\, .
\end{align}
The expressions  \eqref{deriv_rat_funct_l_eq_1} and \eqref{deriv_rat_funct_l_eq_Nmin1} are proportional to $(q^+\!-\!k^+)$, and thus bring a suppression factor of $1/\gamma_T$, as shown in Eq.~\eqref{qplus_min_kplus_as_power_suppressed}. However, in Eq.~\eqref{Ipplus_exp_1_1}, the derivative with respect to $z^-$ is already providing a suppression factor of $1/\gamma_T$. Hence, the contributions to Eq.~\eqref{Ipplus_exp_1_1} in which $\partial_{(\Delta p_l^+)} $ acts on ${\cal R}(\{\Delta p_n^+\})$ are only NNEik corrections to the scalar propagator, which are beyond our scope.

In Eq.~\eqref{Ipplus_exp_1_1}, when $\partial_{(\Delta p_l^+)} $ acts on a phase factor, one has
\begin{align}
\partial_{(\Delta p_l^+)}  \bigg[e^{-i\, \frac{(\p_l^2+m^2)}{2(p^++\Delta p_l^+)}\, (z_{l+1}^+-z_l^+)}
 \bigg]\bigg|_{\Delta p_l^+\rightarrow 0}
= &\,
\frac{i\, (\p_l^2\!+\!m^2)\, (z_{l+1}^+\!-\!z_l^+)}{2(p^+)^2}\;
e^{-i\, \frac{(\p_l^2+m^2)}{2p^+}\, (z_{l+1}^+-z_l^+)}
\, .
\label{deriv_delta_pplus_phase}
\end{align}
At the level of Eq.~\eqref{expansion_scalar_prop_pure_A_minus_1_app},
upon integration over $z_{l}^+$ and $z_{l+1}^+$ (at which background field insertions occur), the  factor $(z_{l+1}^+\!-\!z_l^+)$ in Eq.~\eqref{deriv_delta_pplus_phase} will provide an extra suppression factor $L^+\propto 1/\gamma_T$, on top of the suppression as $1/\gamma_T$ provided by the $\partial_{z^-}$ in  Eq.~\eqref{Ipplus_exp_1_1}. Hence, acting with $\partial_{(\Delta p_l^+)}$ on phase factors in  Eq.~\eqref{Ipplus_exp_1_1} provides contributions to the scalar propagator which are simultaneously NEik corrections with respect to the $z^-$ dependence of the background field and NEik corrections with respect to the width of the target, so that they are of order 
 NNEik overall, and beyond our scope.

Finally, when $\partial_{(\Delta p_l^+)} $ acts on a $\theta$ function in Eq.~\eqref{Ipplus_exp_1_1}, one gets
\begin{align}
&
\partial_{(\Delta p_l^+)} \Big[
\theta(z_{l+1}^+\!-\!z_l^+)\, \theta(p^+\!+\!\Delta p_l^+) - \theta(z_l^+\!-\!z_{l+1}^+)\, \theta(-p^+\!-\!\Delta p_l^+)
\Big]\bigg|_{\Delta p_l^+\rightarrow 0}
\nonumber\\
= &\,
\theta(z_{l+1}^+\!-\!z_l^+)\, \delta(p^+) - \theta(z_l^+\!-\!z_{l+1}^+)\, (-1)\, \delta(p^+)
=  \delta(p^+)
\, .
\label{deriv_delta_pplus_theta}
\end{align}
Such contribution is a zero mode $p^+=0$ of the projectile. As argued in Sec.~\ref{sec:taming_enhancement}, such zero mode cannot contribute to scattering processes and should be discarded in the expansion of propagators beyond the Eikonal approximation.

Hence, $ I_{p^+}^{(1)} $ only contains contributions which are $\textrm{NNEik}$ or zero modes, and no contribution of order $\textrm{NEik}$ after all.
All in all, at Eikonal accuracy, only $ I_{p^+}^{(0)} $ as given in Eq.~\eqref{Ipplus_exp_0_2} contributes to the scalar 
propagator in a pure $A^-$ background field, and Eq.~\eqref{expansion_scalar_prop_pure_A_minus_1_app} becomes
\begin{align}
&\delta G_F(x,y)\bigg|_{\textrm{pure }A^{-}}
= 
\sum_{N=1}^{+\infty} \int \bigg[\prod_{n=1}^{N} d^{D-1} \underline{z_n}\bigg]\,
\int \bigg[\prod_{n=0}^{N} \frac{d^{D-2} \p_n}{(2\pi)^{D-2}}\;
e^{i \p_{n} \cdot (\z_{n+1}-\z_n)}
\bigg]\,
\nonumber\\
&\, 
\times\;
\int \frac{d q^+}{2\pi}\:
\frac{1}{2q^+} \,
e^{-i\, \frac{(\q^2+m^2)}{2q^+}\, (x^+-z_N^+)}\,
\Big[
\theta(x^+\!-\!z_N^+)\, \theta(q^+) - \theta(z_N^+\!-\!x^+)\, \theta(-q^+)
\Big]
\nonumber\\
&\, 
\times\;
\int  \frac{d k^+}{2\pi}\:
\frac{1}{2k^+} \,
e^{-i\, \frac{(\k^2+m^2)}{2k^+}\, (z_{1}^+-y^+)}\,
\Big[
\theta(z_{1}^+\!-\!y^+)\, \theta(k^+) - \theta(y^+\!-\!z_{1}^+)\, \theta(-k^+)
\Big]
\nonumber\\
&\, 
\times\;
(2p^+)
\bigg[\prod_{n=1}^{N-1}
e^{-i\, \frac{(\p_n^2+m^2)}{2p^+}\, (z_{n+1}^+-z_n^+)}\,
\Big[
\theta(z_{n+1}^+\!-\!z_n^+)\, \theta(p^+) - \theta(z_n^+\!-\!z_{n+1}^+)\, \theta(-p^+)
\Big]
\bigg]\,
\nonumber\\
&\, 
\,\times\;
 \int  dz^-\,   e^{-i q^+  (x^- -z^-)}\, e^{-i k^+  (z^- -y^-)}\,
\left[{\cal P}_n\, \prod_{n=1}^{N}\left(-i g\, A^{-}\left(\underline{z_n}, z^- \right)\right)\right]\,
+\textrm{NNEik}
\label{expansion_scalar_prop_pure_A_minus_2_app}
\, .
\end{align}
Moreover, one has
\begin{align}
\frac{(2p^+)}{(2q^+)(2k^+)}
= &\,  \frac{(2p^+)}{[2p^+ \!+\!(q^+\!-\!k^+)][2p^+ \!-\!(q^+\!-\!k^+)]}
=    \frac{1}{(2p^+)}\,  \left[1+
\frac{(q^+\!-\!k^+)^2}{[(2p^+)^2 \!-\!(q^+\!-\!k^+)^2]}
\right]
\label{ratio_pplus_qplus_kplus}
\, .
\end{align}
When substituting Eq.~\eqref{ratio_pplus_qplus_kplus} into Eq.~\eqref{expansion_scalar_prop_pure_A_minus_2_app}, the second term in the bracket in Eq.~\eqref{ratio_pplus_qplus_kplus} is a NNEik correction (see Eq.~\eqref{qplus_min_kplus_as_power_suppressed}). Hence, one obtains
\begin{align}
&\delta G_F(x,y)\bigg|_{\textrm{pure }A^{-}}
= 
\sum_{N=1}^{+\infty} \int \bigg[\prod_{n=1}^{N} d^{D-1} \underline{z_n}\bigg]\,
\int \bigg[\prod_{n=0}^{N} \frac{d^{D-2} \p_n}{(2\pi)^{D-2}}\;
e^{i \p_{n} \cdot (\z_{n+1}-\z_n)}
\bigg]\,
\nonumber\\
&\, 
\times\;
\int \frac{d q^+}{2\pi}\:
e^{-i\, \frac{(\q^2+m^2)}{2q^+}\, (x^+-z_N^+)}\,
\Big[
\theta(x^+\!-\!z_N^+)\, \theta(q^+) - \theta(z_N^+\!-\!x^+)\, \theta(-q^+)
\Big]
\nonumber\\
&\, 
\times\;
\int  \frac{d k^+}{2\pi}\:
e^{-i\, \frac{(\k^2+m^2)}{2k^+}\, (z_{1}^+-y^+)}\,
\Big[
\theta(z_{1}^+\!-\!y^+)\, \theta(k^+) - \theta(y^+\!-\!z_{1}^+)\, \theta(-k^+)
\Big]
\nonumber\\
&\, 
\times\;
\frac{1}{(q^+\!+\!k^+)}
\bigg[\prod_{n=1}^{N-1}
e^{-i\, \frac{(\p_n^2+m^2)}{2p^+}\, (z_{n+1}^+-z_n^+)}\,
\Big[
\theta(z_{n+1}^+\!-\!z_n^+)\, \theta(q^+\!+\!k^+) - \theta(z_n^+\!-\!z_{n+1}^+)\, \theta(-q^+\!-\!k^+)
\Big]
\bigg]\,
\nonumber\\
&\, 
\,\times\;
 \int  dz^-\,   e^{-i q^+  (x^- -z^-)}\, e^{-i k^+  (z^- -y^-)}\,
\left[{\cal P}_n\, \prod_{n=1}^{N}\left(-i g\, A^{-}\left(\underline{z_n}, z^- \right)\right)\right]\,
+\textrm{NNEik}
\label{expansion_scalar_prop_pure_A_minus_3_app}
\, .
\end{align}
In Eq.~\eqref{expansion_scalar_prop_pure_A_minus_3_app}, if one selects the term in $\theta(q^+)$ in the second line and the term in  $\theta(k^+)$ in the third line, then only the term in $\theta(q^++k^+)$ survives in the fourth line.
Similarly, if one selects the term in $\theta(-q^+)$ in the second line and the term in  $\theta(-k^+)$ in the third line, then only the term in $\theta(-q^+-k^+)$ survives in the fourth line. However, at the level of Eq.~\eqref{expansion_scalar_prop_pure_A_minus_3_app}, there are non-zero contributions from cross terms: if one picks  $\theta(q^+)$ and $\theta(-k^+)$, or $\theta(-q^+)$ and $\theta(k^+)$. By Taylor expanding the gauge field insertions around $z^-=0$ in the last line of Eq.~\eqref{expansion_scalar_prop_pure_A_minus_3_app}, it is possible to show that these cross terms are actually zero modes, with $p^+=0$. Hence, they are not relevant in our expansion beyond the Eikonal approximation. Thus, one obtains
\begin{align}
\delta G_F(x,y)\bigg|_{\textrm{pure }A^{-}}
= &\,
\sum_{N=1}^{+\infty} \int \bigg[\prod_{n=1}^{N} d^{D-1} \underline{z_n}\bigg]\,
\int \bigg[\prod_{n=0}^{N} \frac{d^{D-2} \p_n}{(2\pi)^{D-2}}\;
e^{i \p_{n} \cdot (\z_{n+1}-\z_n)}
\bigg]\,
\int \frac{d q^+}{2\pi}\:
\int  \frac{d k^+}{2\pi}\:
\frac{1}{(q^+\!+\!k^+)}
\nonumber\\
&\, 
\times\;
e^{-i\, \frac{(\q^2+m^2)}{2q^+}\, (x^+-z_N^+)}\; 
\left[\prod_{n=1}^{N-1}
e^{-i\, \frac{(\p_n^2+m^2)}{2p^+}\, (z_{n+1}^+-z_n^+)}\,
\right]\,
e^{-i\, \frac{(\k^2+m^2)}{2k^+}\, (z_{1}^+-y^+)}\,
\nonumber\\
&\, 
\times\;
\bigg[
 \theta(q^+)\theta(k^+) \bigg[\prod_{n=0}^{N}
\theta(z_{n+1}^+\!-\!z_n^+)\bigg]\,
+ (-1)^{N+1} \theta(-q^+)\theta(-k^+) \bigg[\prod_{n=0}^{N}
\theta(z_n^+\!-\!z_{n+1}^+)\bigg]
\bigg]
\nonumber\\
&\, 
\,\times\;
 \int  dz^-\,   e^{-i q^+  (x^- -z^-)}\, e^{-i k^+  (z^- -y^-)}\,
\left[{\cal P}_n\, \prod_{n=1}^{N}\left(-i g\, A^{-}\left(\underline{z_n}, z^- \right)\right)\right]\,
+\textrm{NNEik}
\label{expansion_scalar_prop_pure_A_minus_4_app}
\, .
\end{align}


\section{Comparison of our result for the scalar propagator with the one of Ref.~\cite{Chirilli:2018kkw}\label{app:scalar_comparison_Giovanni}}


The scalar propagator at NEik accuracy has been previously studied in Ref.~\cite{Chirilli:2018kkw}. In particular, in Eq.~(3.7) there, a general expression has been given, which do not assume that the endpoints $x$ and $y$ are outside of the target, and which do not assume that the gauge field vanishes outside of the target. The aim of this appendix is to compare our results (corresponding to the sum of contributions from Eqs.~\eqref{expansion_scalar_prop_pure_A_minus_brownian_3}, \eqref{single_A_perp_scalar_prop_8} and \eqref{single_A_plus_scalar_prop_5}) and the ones of Ref.~\cite{Chirilli:2018kkw} for the NEik scalar propagator in that general situation.

Thanks to the symmetry of the Feynman scalar propagators with respect to its endpoints, we can focus without loss of generality on the regime $x^+>y^+$. In that regime, Eq.~(3.7) from Ref.~\cite{Chirilli:2018kkw} can be rewritten in our notations as\footnote{Ref.~\cite{Chirilli:2018kkw} uses a different convention than our Eqs.~\eqref{Cov_deriv} and \eqref{Cov_deriv_back} to define the covariant derivatives. It amounts to flipping the sign of the coupling $g$ everywhere. This sign change of $g$ has already been performed in Eq.~\eqref{NEik_scalar_generic_Giovanni} with respect to Eq.~(3.7) from Ref.~\cite{Chirilli:2018kkw}, in order to allow for a clearer comparison with our results.}
\begin{align}
&G_F(x,y)
=
\int \frac{dk^+}{2\pi}\: \frac{\theta(k^+)}{2k^+}\;  e^{-i k^+(x^--y^-)}
\int \frac{d^{D-2} \q}{(2\pi)^{D-2}}\:
\int \frac{d^{D-2} \k}{(2\pi)^{D-2}}\:
e^{-i x^+ \frac{\q^2}{2k+}}\, e^{i y^+\frac{\k^2}{2k+}}
\int d^{D-2}\z\,\, e^{i \q\cdot (\x-\z)}
\Bigg\{
{\cal U}_F(x^+,y^+)
\nonumber\\
&\,
-\frac{i}{2k^+}\, x^+
\left(\{\hat{P}_i, g A^i(x^+)\} +g^2 A_i(x^+)A^i(x^+) \right)
{\cal U}_F(x^+,y^+)
+\frac{i}{2k^+}\, {\cal U}_F(x^+,y^+)\, y^+
\left(\{\hat{P}_i, g A^i(y^+)\} +g^2 A_i(y^+)A^i(y^+) \right)
\nonumber\\
&\,
-\frac{i}{2k^+}\, \int_{y^+}^{x^+} dz^+\, z^+\, \Big\{\hat{P}^i,\,
  {\cal U}_F(x^+,z^+) g {F_i}^-(z^+)
{\cal U}_F(z^+,y^+)\Big\}
\nonumber\\
&\,
-\frac{i}{2k^+}\, \int_{y^+}^{x^+} dz^+\, \int_{z^+}^{x^+} dw^+\, (w^+\!-\!z^+)\,
{\cal U}_F(x^+,w^+) g F^{i-}(w^+){\cal U}_F(w^+,z^+) g {F_i}^-(z^+){\cal U}_F(z^+,y^+)
\Bigg\}
e^{i \k\cdot (\z-\y)}
+\textrm{NNEik}
\label{NEik_scalar_generic_Giovanni}
\, ,
\end{align}
where all Wilson lines, field strengths and gauge fields are implicitly functions of the same transverse coordinate $\z$, and are assumed to have no dependence on $z^-$.
In Eq.~\eqref{NEik_scalar_generic_Giovanni}, the canonical momentum operator $\hat{P}_i$ acts as
\begin{align}
\hat{P}_i=& i \overrightarrow{D}_i = i \overrightarrow{\partial}_i - g A_i
= -i \overleftarrow{D}_i =  -i \overleftarrow{\partial}_i - g A_i
\label{def_P_operator}
\, ,
\end{align}
where the transverse derivatives, with respect to $\z$, act not only on the fields and Wilson lines, but also on the $\z$ dependent phases in Eq.~\eqref{NEik_scalar_generic_Giovanni}.

Using the relations
\begin{align}
\{\hat{P}_i, g A^i(x^+)\} +g^2 A_i(x^+)A^i(x^+)
=&
2 (-i \overleftarrow{\partial}_i)g A^i(x^+) - [i \overrightarrow{\partial}_i,\, g A^i(x^+)]
-g^2 A_i(x^+)A^i(x^+)
\nonumber\\
\{\hat{P}_i, g A^i(y^+)\} +g^2 A_i(y^+)A^i(y^+)
=&
2 g A^i(y^+)(i \overrightarrow{\partial}_i) + [i \overrightarrow{\partial}_i,\, g A^i(y^+)]
-g^2 A_i(y^+)A^i(y^+)
\end{align}
one can rewrite Eq.~\eqref{NEik_scalar_generic_Giovanni} as
\begin{align}
&G_F(x,y)
=
\int \frac{dk^+}{2\pi}\: \frac{\theta(k^+)}{2k^+}\;  e^{-i k^+(x^--y^-)}
\int \frac{d^{D-2} \q}{(2\pi)^{D-2}}\:
\int \frac{d^{D-2} \k}{(2\pi)^{D-2}}\:
e^{-i x^+ \frac{\q^2}{2k+}}\, e^{i y^+\frac{\k^2}{2k+}}
\int d^{D-2}\z\,\, e^{i \q\cdot (\x-\z)}
\Bigg\{
{\cal U}_F(x^+,y^+)
\nonumber\\
&\,
+\frac{i}{2k^+}\, x^+
\left(-2 \q^i\, g A_i(x^+) + [i \overrightarrow{\partial}_i,\, g A^i(x^+)]  +g^2 A_i(x^+)A^i(x^+) \right)
{\cal U}_F(x^+,y^+)
\nonumber\\
&\,
-\frac{i}{2k^+}\, {\cal U}_F(x^+,y^+)\, y^+
\left(-2\k^i\, g A_i(y^+) - [i \overrightarrow{\partial}_i,\, g A^i(y^+)] +g^2 A_i(y^+)A^i(y^+) \right)
\nonumber\\
&\,
+\frac{i}{2k^+}\, \int_{y^+}^{x^+} dz^+\, z^+\, \Big(g A^i(x^+)
  {\cal U}_F(x^+,z^+) g {F_i}^-(z^+)
{\cal U}_F(z^+,y^+)
+
  {\cal U}_F(x^+,z^+) g {F_i}^-(z^+)
{\cal U}_F(z^+,y^+) g A^i(y^+)
\Big)
\nonumber\\
&\,
-\frac{i}{2k^+}\,(\q^i+\k^i) \int_{y^+}^{x^+} dz^+\, z^+\,
  {\cal U}_F(x^+,z^+) g {F_i}^-(z^+)
{\cal U}_F(z^+,y^+)
\nonumber\\
&\,
-\frac{i}{2k^+}\, \int_{y^+}^{x^+} dz^+\, \int_{z^+}^{x^+} dw^+\, (w^+\!-\!z^+)\,
{\cal U}_F(x^+,w^+) g F^{i-}(w^+){\cal U}_F(w^+,z^+) g {F_i}^-(z^+){\cal U}_F(z^+,y^+)
\Bigg\}
e^{i \k\cdot (\z-\y)}
+\textrm{NNEik}
\label{NEik_scalar_generic_Giovanni_2}
\, .
\end{align}

In order to make contact with our results, it is convenient to expand the decorated Wilson line appearing in the fourth and fifth lines of Eq.~\eqref{NEik_scalar_generic_Giovanni_2} as
\begin{align}
& i\int_{y^+}^{x^+} dz^+\, z^+\,
  {\cal U}_F(x^+,z^+) g {F_i}^-(z^+)
{\cal U}_F(z^+,y^+)
\nonumber\\
= &
\int_{y^+}^{x^+} dz^+\, z^+\,
  {\cal U}_F(x^+,z^+)
  \Big(
   [i\partial_j, g A^-(z^+)]
   +ig A_j(z^+)\, \overrightarrow{D}_{z^+}
   +ig \overleftarrow{D}_{z^+}\, A_j(z^+)
  \Big)
{\cal U}_F(z^+,y^+)
\nonumber\\
= &
\int_{y^+}^{x^+} dz^+\,
  {\cal U}_F(x^+,z^+)
  \Big(
   i g\, z^+ \left(\partial_j  A^-(z^+)\right)
   +ig A_j(z^+)
  \Big)
{\cal U}_F(z^+,y^+)
\nonumber\\
 & \hspace{1cm}
+y^+\, {\cal U}_F(x^+,y^+)\, ig A_j(y^+)
-x^+\, ig A_j(x^+)\, {\cal U}_F(x^+,y^+)
\, .
\label{single_F_min_perp_insert}
\end{align}
In the same way, one finds as well
\begin{align}
&\int_{y^+}^{x^+} dv^+\, \int_{v^+}^{x^+} dw^+\, (w^+\!-\!v^+)\,
  {\cal U}_F(x^+,w^+) g {F}^{i-}(w^+)
{\cal U}_F(w^+,v^+) g {F_i}^-(v^+)
{\cal U}_F(v^+,y^+)
\nonumber\\
= &
-\int_{y^+}^{x^+} dv^+\, \int_{v^+}^{x^+} dw^+\, (w^+\!-\!v^+)\,
  {\cal U}_F(x^+,w^+) g \left(\partial_i A^-(w^+)\right)
{\cal U}_F(w^+,v^+) g \left(\partial_i A^-(v^+)\right)
{\cal U}_F(v^+,y^+)
\nonumber\\
 &
 +i\int_{y^+}^{x^+} dv^+\, [\partial_i,\, {\cal U}_F(x^+,v^+)]\, g  A_i(v^+)\, {\cal U}_F(v^+,y^+)
 -i\int_{y^+}^{x^+} dv^+\, {\cal U}_F(x^+,v^+)\, g  A_i(v^+)\, [\partial_i,\, {\cal U}_F(v^+,y^+)]
\nonumber\\
 &
 +\int_{y^+}^{x^+} dv^+\, {\cal U}_F(x^+,v^+)\, g  A_i(v^+)\,g  A_i(v^+)\, {\cal U}_F(v^+,y^+)
 +(x^+\!-\!y^+)\, g  A_i(x^+)\, {\cal U}_F(x^+,y^+)\,g  A_i(y^+)\,
\nonumber\\
 &
 +g  A_i(x^+)\int_{y^+}^{x^+} dv^+\, {\cal U}_F(x^+,v^+)\,\Big(
     (x^+\!-\!v^+)\, g \left(\partial_i A^-(v^+)\right) - g  A_i(v^+)
 \Big)\, {\cal U}_F(v^+,y^+)
\nonumber\\
 &
 -\int_{y^+}^{x^+} dv^+\, {\cal U}_F(x^+,v^+)\,\Big(
     (v^+\!-\!y^+)\, g \left(\partial_i A^-(v^+)\right) + g  A_i(v^+)
 \Big)\, {\cal U}_F(v^+,y^+)\, g  A_i(y^+)
\nonumber\\
= &
-\int_{y^+}^{x^+} dv^+\, \int_{v^+}^{x^+} dw^+\, (w^+\!-\!v^+)\,
  {\cal U}_F(x^+,w^+) g \left(\partial_i A^-(w^+)\right)
{\cal U}_F(w^+,v^+) g \left(\partial_i A^-(v^+)\right)
{\cal U}_F(v^+,y^+)
\nonumber\\
 &
 +i\int_{y^+}^{x^+} dv^+\, [\partial_i,\, {\cal U}_F(x^+,v^+)]\, g  A_i(v^+)\, {\cal U}_F(v^+,y^+)
 -i\int_{y^+}^{x^+} dv^+\, {\cal U}_F(x^+,v^+)\, g  A_i(v^+)\, [\partial_i,\, {\cal U}_F(v^+,y^+)]
\nonumber\\
 &
 +\int_{y^+}^{x^+} dv^+\, {\cal U}_F(x^+,v^+)\, g  A_i(v^+)\,g  A_i(v^+)\, {\cal U}_F(v^+,y^+)
 \nonumber\\
 &
  -\int_{y^+}^{x^+} dz^+\, z^+\, \Big(g A_i(x^+)
  {\cal U}_F(x^+,z^+) g {F_i}^-(z^+)
{\cal U}_F(z^+,y^+)
+
  {\cal U}_F(x^+,z^+) g {F_i}^-(z^+)
{\cal U}_F(z^+,y^+) g A_i(y^+)
\Big)
 \nonumber\\
 &
 +x^+\, g A_i(x^+)\, \Big(
    -g A_i(x^+)\, {\cal U}_F(x^+,y^+) +i [\partial_i,\, {\cal U}_F(x^+,y^+)]
 \Big)
+y^+\, \Big(
    {\cal U}_F(x^+,y^+)\, g A_i(y^+)  +i [\partial_i,\, {\cal U}_F(x^+,y^+)]
 \Big)\, g A_i(y^+)
\, ,
\label{double_F_min_perp_insert}
\end{align}
using Eq.~\eqref{single_F_min_perp_insert} in the last step. Note that Eqs.~\eqref{single_F_min_perp_insert} and \eqref{double_F_min_perp_insert} are valid both if the background field is independent of $z^-$, or if all of the field insertions and Wilson lines depend on a common $z^-$.

On the other hand, our full result for the NEik scalar propagator assuming only $x^+>y^+$ is given by the sum of
Eqs.~\eqref{expansion_scalar_prop_pure_A_minus_brownian_3}, \eqref{single_A_perp_scalar_prop_8} and \eqref{single_A_plus_scalar_prop_5} and of the vacuum propagator as
\begin{align}
&G_F(x,y)
=  \int \frac{d^{D-1} \underline{q}}{(2\pi)^{D-1}}\:
\int \frac{d^{D-1} \underline{k}}{(2\pi)^{D-1}}\:
\frac{\theta(q^+)\theta(k^+)}{(q^+\!+\!k^+)}\: e^{-i \check{q}\cdot x}\, e^{i \check{k}\cdot y}\,
\int dz^-\, e^{i z^-   (q^+ -k^+)}\,
\int d^{D-2}\z\,\, e^{-i \z \cdot (\q -\k)}\,
\nonumber\\
&\, \times \; 
\Bigg\{{\cal U}_F(x^+,y^+)
+\frac{(\q^j+\k^j)}{(q^+\!+\!k^+)}
\int_{y^+}^{x^+}\! dv^+\, v^+\,
{\cal U}_F(x^+,v^+)  \big[-ig\, \d_j A^{-}(v^+)\big]
{\cal U}_F(v^+,y^+)
\nonumber\\
&\, \hspace{0.8cm}
+\frac{i}{(q^+\!+\!k^+)}
\int_{y^+}^{x^+}\! dv^+\int_{v^+}^{x^+}\! dw^+\, (w^+\!-\!v^+)\,
{\cal U}_F(x^+,w^+)  \big[g\, \d_j A^{-}(w^+)\big]
{\cal U}_F(w^+,v^+)
 \big[g\, \d_j A^{-}(v^+)\big]
{\cal U}_F(v^+,y^+)
\nonumber\\
&\, \hspace{0.8cm}
+\frac{1}{(q^++k^+)}
\int_{y^+}^{x^+}\! dz^+\,
\bigg[ {\cal U}_F(x^+,z^+)
\Big(-g\, A_{j}(z^+)\, \overrightarrow{\d}_{z^{j}}
+ \overleftarrow{\d}_{z^{j}}\,  g\, A_{j}(z^+)
-ig^2\, A_{j}(z^+)\, A_{j}(z^+)
\Big)
{\cal U}_F(z^+,y^+)\bigg]
\nonumber\\
&\, \hspace{0.8cm}
-\frac{(\q^j\!+\!\k^j)}{(q^++k^+)}
\int_{y^+}^{x^+}\! dz^+\,
{\cal U}_F(x^+,z^+)
\, ig A_j(z^+)\,
{\cal U}_F(z^+,y^+)
\nonumber\\
&\, \hspace{0.8cm}
+\frac{1}{(q^++k^+)}\, \Big(g A^+(x^+)\, {\cal U}_F(x^+,y^+)
+{\cal U}_F(x^+,y^+)\, g A^+(y^+)
\Big)
\Bigg\}
\label{NEik_scalar_generic_us_1}
\end{align}
keeping the $\z$ and $z^-$ dependence of the fields and Wilson lines implicit.
Inserting the relations \eqref{single_F_min_perp_insert} and \eqref{double_F_min_perp_insert}, into Eq.~\eqref{NEik_scalar_generic_us_1}, one finds
\begin{align}
&G_F(x,y)
=  \int \frac{d^{D-1} \underline{q}}{(2\pi)^{D-1}}\:
\int \frac{d^{D-1} \underline{k}}{(2\pi)^{D-1}}\:
\frac{\theta(q^+)\theta(k^+)}{(q^+\!+\!k^+)}\: e^{-i \check{q}\cdot x}\, e^{i \check{k}\cdot y}\,
\int dz^-\, e^{i z^-   (q^+ -k^+)}\,
\int d^{D-2}\z\,\, e^{-i \z \cdot (\q -\k)}\,
\nonumber\\
&\, \times \; 
\Bigg\{{\cal U}_F(x^+,y^+)
-\frac{i}{(q^+\!+\!k^+)}
\int_{y^+}^{x^+} dv^+\, \int_{v^+}^{x^+} dw^+\, (w^+\!-\!v^+)\,
  {\cal U}_F(x^+,w^+) g {F}^{j-}(w^+)
{\cal U}_F(w^+,v^+) g {F_j}^-(v^+)
{\cal U}_F(v^+,y^+)
\nonumber\\
&\, \hspace{0.8cm}
-\frac{i}{(q^+\!+\!k^+)}\int_{y^+}^{x^+} dz^+\, z^+\, \Big(g A_j(x^+)
  {\cal U}_F(x^+,z^+) g {F_j}^-(z^+)
{\cal U}_F(z^+,y^+)
+
{\cal U}_F(x^+,z^+) g {F_j}^-(z^+)
{\cal U}_F(z^+,y^+) g A_j(y^+)
\Big)
 \nonumber\\
 &\, \hspace{0.8cm}
 +\frac{i\, x^+}{(q^+\!+\!k^+)}\, g A_j(x^+)\, \Big(
    -g A_j(x^+)\, {\cal U}_F(x^+,y^+) +i [\partial_j,\, {\cal U}_F(x^+,y^+)]
 \Big)
  \nonumber\\
 &\, \hspace{0.8cm}
+\frac{i\, y^+}{(q^+\!+\!k^+)}\, \Big(
    {\cal U}_F(x^+,y^+)\, g A_j(y^+)  +i [\partial_j,\, {\cal U}_F(x^+,y^+)]
 \Big)\, g A_j(y^+)
\nonumber\\
&\, \hspace{0.8cm}
-\frac{i(\q^j\!+\!\k^j)}{(q^++k^+)}\bigg[
\int_{y^+}^{x^+}\! dz^+\, z^+
{\cal U}_F(x^+,z^+)
\, g {F_j}^-(z^+)\,
{\cal U}_F(z^+,y^+)
+x^+ g\, A_{j}(x^+) {\cal U}_F(x^+,y^+)
-y^+  {\cal U}_F(x^+,y^+) g\, A_{j}(y^+)
\bigg]
\nonumber\\
&\, \hspace{0.8cm}
+\frac{1}{(q^++k^+)}\, \Big(g A^+(x^+)\, {\cal U}_F(x^+,y^+)
+{\cal U}_F(x^+,y^+)\, g A^+(y^+)
\Big)
\Bigg\}
\, .
\label{NEik_scalar_generic_us_2}
\end{align}
Integrating by parts the transverse derivatives which act on Wilson lines in Eq.~\eqref{NEik_scalar_generic_us_2} and regrouping the terms, one finally arrives at
\begin{align}
&G_F(x,y)
=  \int \frac{d^{D-1} \underline{q}}{(2\pi)^{D-1}}\:
\int \frac{d^{D-1} \underline{k}}{(2\pi)^{D-1}}\:
\frac{\theta(q^+)\theta(k^+)}{(q^+\!+\!k^+)}\: e^{-i \check{q}\cdot x}\, e^{i \check{k}\cdot y}\,
\int dz^-\, e^{i z^-   (q^+ -k^+)}\,
\int d^{D-2}\z\,\, e^{-i \z \cdot (\q -\k)}\,
\nonumber\\
&\, \times \; 
\Bigg\{{\cal U}_F(x^+,y^+)
-\frac{i}{(q^+\!+\!k^+)}
\int_{y^+}^{x^+} dv^+\, \int_{v^+}^{x^+} dw^+\, (w^+\!-\!v^+)\,
  {\cal U}_F(x^+,w^+) g {F}^{j-}(w^+)
{\cal U}_F(w^+,v^+) g {F_j}^-(v^+)
{\cal U}_F(v^+,y^+)
\nonumber\\
&\, \hspace{0.8cm}
-\frac{i(\q^j\!+\!\k^j)}{(q^++k^+)}
\int_{y^+}^{x^+}\! dz^+\, z^+
{\cal U}_F(x^+,z^+)
\, g {F_j}^-(z^+)\,
{\cal U}_F(z^+,y^+)
\nonumber\\
&\, \hspace{0.8cm}
+\frac{i}{(q^+\!+\!k^+)}\int_{y^+}^{x^+} dz^+\, z^+\, \Big(g A^j(x^+)
  {\cal U}_F(x^+,z^+) g {F_j}^-(z^+)
{\cal U}_F(z^+,y^+)
+
{\cal U}_F(x^+,z^+) g {F_j}^-(z^+)
{\cal U}_F(z^+,y^+) g A^j(y^+)
\Big)
 \nonumber\\
 &\, \hspace{0.8cm}
 +\frac{i\, x^+}{(q^+\!+\!k^+)}\, \Big(
     g^2 A^j(x^+)\, A_j(x^+)
    -2\q^j\,  g A_j(x^+)
    +i [\partial_j,\, g A^j(x^+)]
 \Big)\, {\cal U}_F(x^+,y^+)
  \nonumber\\
 &\, \hspace{0.8cm}
+\frac{i\, y^+}{(q^+\!+\!k^+)}\, {\cal U}_F(x^+,y^+)\, \Big(
    - g^2 A_j(y^+) \, A^j(y^+)
    +2\k^j g A_j(y^+)
    +i [\partial_j,\, g A^j(y^+)]\,
 \Big)
\nonumber\\
&\, \hspace{0.8cm}
+\frac{1}{(q^++k^+)}\, \Big(g A^+(x^+)\, {\cal U}_F(x^+,y^+)
+{\cal U}_F(x^+,y^+)\, g A^+(y^+)
\Big)
\Bigg\}
\label{NEik_scalar_generic_us_3}
\, .
\end{align}
Neglecting the $z^-$ dependence of the background field as well as its $A^+$ component in Eq.~\eqref{NEik_scalar_generic_us_3}, one recovers the expression \eqref{NEik_scalar_generic_Giovanni_2} equivalent to Eq.~(3.7) from Ref.~\cite{Chirilli:2018kkw}. The last line of our result Eq.~\eqref{NEik_scalar_generic_us_3} involves terms with $A^+$ at the endpoints of the propagators $x^+$ and $y^+$. Such NEik contributions to the scalar propagators are absent from Eq.~(3.7) of Ref.~\cite{Chirilli:2018kkw} but were discussed in the Appendix C of Ref.~\cite{Chirilli:2018kkw}. Hence, the only difference between the results of Ref.~\cite{Chirilli:2018kkw} and ours for the scalar propagator at NEik accuracy is that we have taken into account the $z^-$ dependence of the background field, which is a source of extra NEik corrections, as illustrated by Eq.~\eqref{expansion_Wilson_line_zminus_equal_0}.


\bibliography{NEikBib}
\end{document}